\documentclass[a4paper,11pt]{article}
\pdfoutput=1 

\usepackage{jheppub_mod} 

\usepackage[T1]{fontenc} 

\usepackage{amsthm} 

\newtheorem{mytheorem}{Property}

\newcommand{\g}{g}
\newcommand{\dd}{\gamma}

\usepackage{tikz}
\usetikzlibrary{math}
\usetikzlibrary{arrows,shapes,positioning}
\usetikzlibrary{decorations.markings}
\tikzstyle arrowstyle=[scale=1]
\tikzstyle directed=[postaction={decorate,decoration={markings,
    mark=at position .65 with {\arrow[arrowstyle]{stealth}}}}]
\tikzstyle reverse directed=[postaction={decorate,decoration={markings,
    mark=at position .65 with {\arrowreversed[arrowstyle]{stealth};}}}]

\newlength{\mywidth}
\usepackage[colorlinks=true
,urlcolor=blue
,anchorcolor=blue
,citecolor=blue
,filecolor=blue
,linkcolor=blue
,menucolor=blue
,pagecolor=blue
,linktocpage=true
,pdfproducer=medialab
,pdfa=true
]{hyperref}

\title{\boldmath One-loop inelastic amplitudes from tree-level elasticity in 2d}

\author[a,b,c]{Davide Polvara}

\affiliation[a]{Department of Mathematical Sciences, Durham University, South Road, Durham DH1 3LE, United Kingdom }
\affiliation[b]{Dipartimento di Fisica e Astronomia,
Universita degli Studi di Padova, via Marzolo 8, 35131 Padova, Italy.}
\affiliation[c]{INFN,
Sezione di Padova, via Marzolo 8, 35131 Padova, Italy.}

\emailAdd{davide.polvara@unipd.it}

\abstract{
We investigate the perturbative integrability of different quantum field theories in 1+1 dimensions at one loop. 
Starting from massive bosonic Lagrangians with polynomial-like potentials and absence of inelastic processes at the tree level,
we derive a formula reproducing one-loop inelastic amplitudes for arbitrary numbers of external legs. 
We show that any one-loop inelastic amplitude is equal to its tree-level version, in which the masses of particles and propagators are corrected by one-loop bubble diagrams. 
These amplitudes are nonzero in general and counterterms need to be added to the Lagrangian to restore the integrability at one loop.
For the class of simply-laced affine Toda theories, we show that the necessary counterterms are obtained by scaling the potential with an overall multiplicative factor, 
proving in this way the one-loop integrability of these models.
Even though we focus on bosonic theories with polynomial-like interactions,
we expect that the on-shell techniques used in this paper to compute amplitudes can be applied to 
several other models.
}

\begin{document} 
\maketitle
\flushbottom

\section{Introduction}
\label{Introductory_section_on_conventions}

Starting from the pioneering paper~\cite{Zamolodchikov:1978xm}, the axiomatic S-matrix bootstrap program was applied in the past decades to conjecture analytical expressions for the S-matrices of several integrable models in 1+1 dimensions. 
However, the bootstrap philosophy is based on the assumption that the symmetries of a given classically integrable theory
survive the quantisation and the integrable structure extends to the full quantum theory. Unfortunately, this is not always the case. There are examples of classically integrable Lagrangians that require counterterms to preserve integrability at the quantum level~\cite{deVega:1981ka,Bonneau:1984pj}; in other cases, the masses of the particles, necessary to establishing the fusing angles entering the bootstrap relations, renormalise in a bad way (as it happens for the class of non-simply-laced affine Toda theories~\cite{Braden:1989bu}) making the bootstrap construction much more convoluted~\cite{Delius:1991kt}. Moreover,
in most cases, the S-matrix cannot be completely constrained through the bootstrap. Recent examples are provided by the class of non-linear sigma models defined on the worldsheet of superstrings propagating in the background $AdS_3 \times S^3 \times T^4$, which are known to be classically integrable~\cite{Cagnazzo:2012se}. Quantum S-matrices for these models were proposed in~\cite{Borsato:2013qpa,Hoare:2013pma,Borsato:2014hja,Lloyd:2014bsa}. 
These S-matrices were determined up to five dressing phases which are not completely fixed by symmetries and for which there exist conflicting proposals in the literature~\cite{Borsato:2013hoa,Frolov:2021fmj}. 
Differently from the proposal advanced in~\cite{Borsato:2013hoa}, the dressing phases conjectured in~\cite{Frolov:2021fmj} have good analytic properties and behave well under bound-state fusion.
However, these phases disagree with existing one-loop perturbative computations~\cite{Roiban:2014cia,Sundin:2016gqe}.
In~\cite{Frolov:2021fmj} it was argued that one possible reason for the disagreement comes from the necessity to add certain counterterms to the Lagrangian in the renormalization procedure.
These open problems lead to the following natural question: `Starting with a classically integrable theory, there is any chance to universally understand if integrability is preserved in loop computations and to determine higher-loop S-matrices?'.

In this paper, we address this question at one-loop order in perturbation theory for real Lagrangians of type
\begin{equation}
\label{eq0_1}
\mathcal{L}_0=\sum_{a=1}^r \biggl( \frac{1}{2} \partial_\mu \phi_a  \partial^\mu \phi_{\bar{a}} - \frac{1}{2} m_a^2 \phi_a \phi_{\bar{a}}\biggr) - \sum_{n=3}^{+\infty} \frac{1}{n!}\sum_{a_1,\dots, a_n=1}^r C^{(n)}_{a_1 \ldots a_n} \phi_{a_1} \ldots \phi_{a_n}.
\end{equation}
The Lagrangian in~\eqref{eq0_1} describes the interaction of $r$ massive bosonic scalar fields $(\phi_1,\ldots, \phi_r)$. 
To keep into account both real and complex fields we follow the same convention of~\cite{Dorey:2021hub} and we define $\phi_{\bar{a}} \equiv \phi_a^*$: in the case in which $\phi_a$ is real then $a=\bar{a} \in \{1,\ldots,r\}$, while in the case in which $\phi_a$ is complex then $a$ and $\bar{a}$ are different indices in $\{1,\ldots,r\}$. 
This convention allows taking any set of real or complex fields $(\phi_1,\ldots, \phi_r)$.
The masses and couplings in~\eqref{eq0_1} are defined in such a way that the properties below are satisfied.
\begin{mytheorem}\label{Condition_tree_level_elasticity_introduction}
All the on-shell amplitudes are purely elastic, with diagonal scattering and null production processes, 
at the tree level (i.e. the only tree-level amplitudes different from zero are those
in which the incoming and outgoing particles are of the same type and carry the same set of momenta).
\end{mytheorem} 
\begin{mytheorem}\label{Non_degenerate_mass_condition}
Given three types of particles $a, b$ and $c$ $\in \{1, \dots r\}$ satisfying $a\ne b$ and $m_a=m_b$, the tree-level inelastic amplitude $M^{(0)}_{ac \to bc}$ associated with the process
\begin{equation}
\label{tree_level_degenerate_elastic_process_in_property}
a+c \to b+c
\end{equation}
is zero for all on-shell and off-shell values of the momenta of the scattered particles.
\end{mytheorem} 
Even though it remains an open problem to classify all the sets of masses and couplings satisfying property~\ref{Condition_tree_level_elasticity_introduction} for arbitrary $r$, it is known that these sets are not empty.
Indeed, in~\cite{Dorey:2021hub} it was proven that all the bosonic affine Toda field theories verify property~\ref{Condition_tree_level_elasticity_introduction}\footnote{It is worth mentioning that the proof presented in~\cite{Dorey:2021hub} relies on previous observations of~\cite{Dorey:1996gd,Gabai:2018tmm,Bercini:2018ysh}. In particular in~\cite{Gabai:2018tmm,Bercini:2018ysh}
it was explained how to obtain higher order couplings in terms of the masses and lower order couplings for Lagrangians of the type in~\eqref{eq0_1} by imposing recursively absence of production at the tree level.}. 
While we require property~\ref{Condition_tree_level_elasticity_introduction} to be verified for on-shell values of the scattered particles, we impose that property~\ref{Non_degenerate_mass_condition} is satisfied also if the momenta of the particles are off-shell. 
Note indeed that if the particles $a$, $b$ and $c$ were on-shell in~\eqref{tree_level_degenerate_elastic_process_in_property}, property~\ref{Non_degenerate_mass_condition} would be just a consequence of property~\ref{Condition_tree_level_elasticity_introduction}.
This second property will be necessary to have compatibility between the renormalization procedure and integrability at one loop order in perturbation theory. This property is also satisfied by all the bosonic affine Toda field theories and we will prove this fact in this paper.

If we move from the tree level to the one-loop order
certain counterterms need to be added to~\eqref{eq0_1}. They are required by the standard renormalization procedure and are necessary to avoid ill-defined Feynman diagrams at one loop. We will introduce these counterterms by imposing the following conditions: 
\begin{itemize}
    \item all amplitudes need to be free of ultraviolet divergences at one loop;
    \item  $\forall$ pair of index $a, b \in \{1, \dots, r\}$ we define by $G_{ab}(p^2)$ the propagator (truncated at one-loop order in perturbation theory), having as incoming leg $a$ and outgoing leg $b$, both carrying momentum $p$.
    The only possible poles of this propagator need to be of order one in $p^2$ and need to be located at
     $p^2=m^2_a$ and $p^2=m^2_b$; we require
\begin{equation}
\label{renormalisation_condition_on_propagators}
\text{Res} \ G_{ab}(p^2)\Bigl|_{p^2=m^2_a}= \text{Res} \ G_{ab}(p^2)\Bigl|_{p^2=m^2_b}=i \delta_{ab}.
\end{equation}
\end{itemize}
These conditions allow determining which counterterms need to be added to the Lagrangian~\eqref{eq0_1}. 
After these counterterms are added, we will focus on one-loop inelastic processes of the form
\begin{equation}
\label{2_to_nminus2_production_process_Introduction}
    a_1(p_1)+a_2(p_2) \to a_3(p_3) \ldots a_n(p_n),
\end{equation}
where two particles of types $a_1$ and $a_2$ and momenta $p_1$ and $p_2$ collide generating $n-2$ outgoing particles of types $a_3, \ldots, a_{n}$ and associated momenta $p_3, \dots, p_n$ (with $n\ge4$). In the case $n=4$, we assume $\{a_1, a_2\}\ne \{a_3,a_4\}$.
We can write the renormalized amplitude $M_{a_1 a_2\to a_3 \dots a_n}$ associated with the process in~\eqref{2_to_nminus2_production_process_Introduction} (truncated at the one-loop order in the perturbative expansion) as
\begin{equation}
\label{amplitude_truncated_at_one_loop_2_to_n_minus_2_processes_Introduction}
M_{a_1 a_2\to a_3 \dots a_n}=M^{(0)}_{a_1 a_2\to a_3 \dots a_n}+M^{(1)}_{a_1 a_2\to a_3 \dots a_n}+M^{(\text{count.})}_{a_1 a_2\to a_3 \dots a_n}.
\end{equation}
$M^{(0)}_{a_1 a_2\to a_3 \dots a_n}$ and $M^{(1)}_{a_1 a_2\to a_3 \dots a_n}$ are obtained by summing over tree-level and one-loop Feynman diagrams respectively, as obtained by applying Feynman rules to the Lagrangian in~\eqref{eq0_1}. $M^{(\text{count.})}_{a_1 a_2\to a_3 \dots a_n}$ is instead obtained by summing over tree-level Feynman diagrams containing counterterms introduced by imposing the conditions in the bullet points listed above. 

In the following sections, we will prove that, if the amplitude~\eqref{amplitude_truncated_at_one_loop_2_to_n_minus_2_processes_Introduction} is obtained from a starting Lagrangian of type~\eqref{eq0_1} satisfying properties~\ref{Condition_tree_level_elasticity_introduction} and~\ref{Non_degenerate_mass_condition}, then it holds that
\begin{equation}
\label{final_formula_we_need_to_prove_first_formulation}
    M_{a_1 a_2\to a_3 \dots a_n}=-\sum_{j \in \{\text{prop, ext}\}} \Sigma^{(0)}_{jj} \frac{\partial}{\partial m^2_j} M^{(0)}_{a_1 a_2\to a_3 \dots a_n},
\end{equation}
where the mass corrections $\Sigma^{(0)}_{jj}$ will be defined in one moment.
The index $j$ in the sum~\eqref{final_formula_we_need_to_prove_first_formulation} runs over the virtual particles propagating inside Feynman diagrams and over the external particles. 
Since, by property~\ref{Condition_tree_level_elasticity_introduction}, for on-shell external particles it holds that $M^{(0)}_{a_1 a_2\to a_3 \dots a_n}=0$, then equation~\eqref{final_formula_we_need_to_prove_first_formulation} is equivalent to
\begin{equation}
\label{final_formula_we_need_to_prove_second_formulation}
    M_{a_1 a_2\to a_3 \dots a_n}= M^{(0)}_{a_1 a_2\to a_3 \dots a_n}\Bigl|_{m_j^2-\Sigma^{(0)}_{jj}} + \ O\Bigl((\Sigma^{(0)}_{jj})^2 M^{(0)}_{a_1 a_2\to a_3 \dots a_n}\Bigl|_{m_j^2}  \Bigr).
\end{equation}
This means that the renormalized amplitude, evaluated to one loop order in perturbation theory, is equal to the associated tree-level amplitude in which all the masses of the particles are shifted by
\begin{equation}
\label{mass_deformations_in_inelastic_amplitude}
    m_j^2 \to m^2_j-\Sigma^{(0)}_{jj}.
\end{equation}
The quantity $\Sigma^{(0)}_{jj}$ is defined as follows. 
We consider the propagator to one loop order obtained from the Lagrangian~\eqref{eq0_1} having as insertions an incoming leg of type $a$ and an outgoing leg of type $b$, both carrying momentum $p$. If we omit tadpole diagrams, before introducing counterterms to~\eqref{eq0_1}, the propagator is given by
\begin{equation}
\frac{i \delta_{ab}}{p^2 -m_a^2}+\frac{i}{p^2 -m_a^2} \bigl( -i \Sigma_{ab}(p^2) \bigr) \frac{i}{p^2 -m_b^2}.
\end{equation}
The quantity $-i\Sigma_{ab}(p^2)$ is obtained by summing over all possible bubble diagrams having $a$ and $b$ as external legs and is pictorially represented in figure~\ref{Image_changing_flavour_and_mass_corrections}.
\begin{figure}
\begin{center}
\begin{tikzpicture}
\tikzmath{\y=1.9;}


\filldraw[black] (-2.8*\y,0.1*\y)  node[anchor=west] {$-i \Sigma_{ab}(p^2) \equiv $};
\filldraw[black] (-1.45*\y,0.1*\y)  node[anchor=west] {\Large{$\sum_{c, d}$}};
\draw[directed] (-0.6*\y,0.1*\y) -- (-0.1*\y,0.1*\y);
\draw[directed] (0.3*\y,0.1*\y) arc(0:180:0.2*\y);
\draw[directed] (-0.1*\y,0.1*\y) arc(180:360:0.2*\y);
\draw[directed] (0.3*\y,0.1*\y) -- (0.8*\y,0.1*\y);
\filldraw[black] (-0.8*\y,0.2*\y)  node[anchor=west] {\scriptsize{$a(p)$}};
\filldraw[black] (0.8*\y,0.2*\y)  node[anchor=west] {\scriptsize{$b(p)$}};
\filldraw[black] (0*\y,0.4*\y)  node[anchor=west] {\scriptsize{$c(k)$}};
\filldraw[black] (-0.1*\y,-0.25*\y)  node[anchor=west] {\scriptsize{$d(p+k)$}};

\end{tikzpicture}
\caption{Sum of bubble diagrams with external legs of types $a$ and $b$.}
\label{Image_changing_flavour_and_mass_corrections}
\end{center}
\end{figure}
We define the expansion of $\Sigma_{ab}(p^2)$ around the on-shell value $p^2=m_a^2$ as
\begin{equation}
\label{expansion_bubble_diagram_one_loop_Sigma_ab}
\Sigma_{ab}(p^2)= \Sigma_{ab}^{(0)}+\Sigma_{ab}^{(1)} (p^2-m^2_a)+\ldots
\end{equation}
with
\begin{equation}
\label{derivatives_bubble_correction_to_propagator}
    \Sigma_{ab}^{(0)}\equiv \Sigma_{ab}(m^2_a) \hspace{4mm} \text{and} \hspace{4mm} \Sigma_{ab}^{(1)}\equiv  \frac{\partial}{\partial p^2}\Sigma_{ab}(p^2)\Bigl|_{p^2=m^2_a}.
\end{equation}
The quantities $\Sigma^{(0)}_{jj}$ in~\eqref{final_formula_we_need_to_prove_first_formulation} and~\eqref{final_formula_we_need_to_prove_second_formulation}, one for each mass associated with a virtual or external particle of type $j$, correspond to the corrections to the one-loop diagonal propagators evaluated at $p_j^2=m^2_j$. In the following sections, we will prove the validity of relations~\eqref{final_formula_we_need_to_prove_first_formulation} and~\eqref{final_formula_we_need_to_prove_second_formulation} and we will discuss their consequences on loop-level integrability. 

The paper is  organised as follows. In section~\ref{section_to_explain_the_mass_and_field_renormalization} we determine the counterterms necessary for equation~\eqref{renormalisation_condition_on_propagators} to be satisfied and we compute their contribution to one-loop inelastic amplitudes in theories satisfying properties~\ref{Condition_tree_level_elasticity_introduction} and~\ref{Non_degenerate_mass_condition}. In section~\ref{section_on_the_cutting_method} we describe a technique to write one-loop amplitudes as products and integrals of tree-level amplitudes: using this technique on theories satisfying properties~\ref{Condition_tree_level_elasticity_introduction} and~\ref{Non_degenerate_mass_condition} we derive the universal expressions~(\eqref{final_formula_we_need_to_prove_first_formulation}, \eqref{final_formula_we_need_to_prove_second_formulation}) for inelastic amplitudes. We discuss that these amplitudes are nonzero in general and the original Lagrangian needs to receive quantum corrections to preserve the integrability at one loop. 
We determine these quantum corrections in theories in which the ratios between $\Sigma_{aa}^{(0)}$ and $m_a^2$ do not depend on $a$ (i.e. all the masses, at one loop, renormalise scaling with a common multiplicative factor).
In section~\ref{Section_on_affine_Toda_field_theories} we focus on the class of affine Toda field theories: we show that properties~\ref{Condition_tree_level_elasticity_introduction} and~\ref{Non_degenerate_mass_condition} are universally satisfied in these theories and therefore their inelastic amplitudes are reproduced by formulas~(\eqref{final_formula_we_need_to_prove_first_formulation}, \eqref{final_formula_we_need_to_prove_second_formulation}). In simply-laced models, we provide universal proof that the ratios between $\Sigma_{aa}^{(0)}$ and $m_a^2$ do not depend on $a$, generalising results that were proven in~\cite{Braden:1989bu} on a case-by-case study. Through this consideration, we show that in simply-laced affine Toda theories, all inelastic amplitudes can be cancelled by a simple renormalization of the Lagrangian. In this way, we prove that these models are one-loop integrable. In section~\ref{Concluion_section} we present our conclusion, discussing our results and giving an overview of the open problems.
Finally, we present two appendices: in appendix~\ref{Appendix_on_counterterms_for_different_particles_with_equal_masses} we show how formulas~(\eqref{final_formula_we_need_to_prove_first_formulation}, \eqref{final_formula_we_need_to_prove_second_formulation}) are affected in theories not satisfying property~\ref{Non_degenerate_mass_condition}; in appendix~\ref{Potential_ill_defined_contributions_in_single_cuts} we give evidence that the only ill-defined contributions in amplitudes 
are due to diagonal bubbles in the external legs (which are removed by a standard renormalization procedure), while the cutting method described in section~\ref{section_on_the_cutting_method} does not produce additional ill-defined terms.

\section{Renormalization of masses and fields}
\label{section_to_explain_the_mass_and_field_renormalization}

In the following, we discuss what counterterms are needed for the two bullet points listed in section~\ref{Introductory_section_on_conventions} to be satisfied.
For Lagrangian theories of the form in~\eqref{eq0_1}, the only ultraviolet divergences at one loop come from diagrams containing propagators that
connect back to their originating vertices.
With a small abuse of notation, we call these diagrams `tadpoles'. 
It is always possible to remove all tadpole diagrams by adding suitable counterterms to~\eqref{eq0_1}. However, since~\eqref{eq0_1} contains infinitely many couplings\footnote{This is a requirement to have absence of inelastic processes at the tree level for any number of external particles, as it was observed in~\cite{Dorey:1996gd,Gabai:2018tmm,Bercini:2018ysh}.}, the number of counterterms needed to avoid tadpoles has to be infinite. Despite this fact, in section~\ref{Section_on_affine_Toda_field_theories} we will show that, for the class of bosonic affine Toda field theories, all tadpoles are cancelled by renormalising a single parameter in the Lagrangian. This renormalisation will introduce the infinite set of couplings necessary to cancel all one-loop tadpoles at a time.
We expect that something similar happens for all the Lagrangians of type~\eqref{eq0_1} satisfying properties~\ref{Condition_tree_level_elasticity_introduction} and~\ref{Non_degenerate_mass_condition}. 
Even though we cannot prove this in general, in the following analysis we do not consider tadpole diagrams and we assume the first bullet point in section~\ref{Introductory_section_on_conventions} to be satisfied. In any case, this point can always be imposed, eventually at a price to add infinitely many unrelated counterterms to~\eqref{eq0_1}. For this reason
we will just focus on the conditions necessary to satisfy the second bullet point listed in the introduction.

Before moving into the computations we set up the conventions that we will follow for the rest of this paper. The momenta in 1+1 dimensions can be equivalently parameterized using energy and spatial momentum components, $p_0$ and $p_1$, or using light-cone components, $p$ and $\bar{p}$. The choices are equivalent and connected by
\begin{equation}
    \label{standard_energy_momentum_components_PhD_thesis}
p_0=\frac{p+\bar{p}}{2} \hspace{5mm}, \hspace{5mm}p_1=\frac{p-\bar{p}}{2}.
\end{equation}
A particle of mass $m_j$, carrying momentum $p$, is on-shell if $p\bar{p}=p_0^2-p_1^2=m_j^2$. In this case, we can write the light-cone components of the momentum as
\begin{equation}
    \label{light_cone_components_PhD_thesis}
    p=m_j a_p=m_j e^{\theta_p} \hspace{5mm}, \hspace{5mm} \bar{p}=m_j \frac{1}{a_p}=m_j e^{-\theta_p}.
\end{equation}
The parameter $a_p\equiv e^{\theta_p}$ in~\eqref{light_cone_components_PhD_thesis} is the only degree of freedom associated with the on-shell particle and $\theta_p$ is called the `rapidity'. A particle is physical if the light-cone components of its momentum are positive, and therefore its rapidity is real.

\subsection{Counterterms and one-loop propagators}
\label{subsection_on_definition_of_counterterms}

In this section, reviewing the method used in~\cite{Braden:1991vz},
we prove that condition~\eqref{renormalisation_condition_on_propagators} is satisfied if we shift the masses and fields in~\eqref{eq0_1} as follows
\begin{equation}
\label{mass_renormalization_tab_definition}
\begin{split}
     &m^2_a \to m^2_a+ \delta m_a^2,\\
     &\phi_a \to \Bigl(1+\frac{1}{2} \Sigma^{(1)}_{aa} \Bigr) \phi_a - \sum_{\substack{b=1 \\ b\neq a}}^r t_{ba} \phi_b,
\end{split}
\end{equation}
where
\begin{equation}
\label{definition_of_mass_renormalization_and_tab_to_diagonalise_the_mass_matrix}
\delta m^2_a \equiv - \Sigma^{(0)}_{aa} \hspace{3mm} \text{and} \hspace{3mm}
    t_{ab}\equiv \begin{cases}
\frac{\Sigma^{(0)}_{ab}}{m^2_b - m^2_a} \ \ \text{if} \ \ m_a \ne m_b\\
0 \hspace{12mm} \text{if} \ \ m_a = m_b.
\end{cases}
\end{equation}

After substituting~\eqref{mass_renormalization_tab_definition} into~\eqref{eq0_1} the Lagrangian becomes
\begin{equation}
\label{renormalised_Lagrangian_due_to_standard_renormalisation_procedure}
\begin{split}
    \mathcal{L}&=\sum_{a=1}^r \biggl( \frac{1}{2} \partial_\mu \phi_a  \partial^\mu \phi_{\bar{a}} - \frac{1}{2} m_a^2 \phi_a \phi_{\bar{a}}\biggr) - \sum_{n=3}^{+\infty} \frac{1}{n!}\sum_{a_1,\dots, a_n=1}^r C^{(n)}_{a_1 \ldots a_n} \phi_{a_1} \ldots \phi_{a_n}\\
    &\begingroup\color{blue} +\sum_{a=1}^r\frac{1}{2}\Sigma^{(1)}_{aa} \Bigl(\partial_\mu \phi_a  \partial^\mu \phi_{\bar{a}}-  m^2_a \phi_a \phi_{\bar{a}}\Bigr)\endgroup \begingroup\color{red}+\sum_{\substack{a,b=1 \\ b\neq a}}^r t_{ba} \Bigl(-\partial_\mu \phi_{\bar{a}} \partial^{\mu} \phi_b+m^2_a  \phi_{\bar{a}} \phi_b \Bigr)\endgroup\\
    &\begingroup\color{blue}- \sum_{n=3}^{+\infty} \frac{1}{n!}\sum_{a_1,\dots, a_n=1}^r C^{(n)}_{a_1 \ldots a_n} \frac{1}{2}\Bigl( \Sigma^{(1)}_{a_1a_1}+\Sigma^{(1)}_{a_2a_2} + \ldots \Sigma^{(1)}_{a_n a_n}\Bigr) \phi_{a_1} \ldots \phi_{a_n} \endgroup\\
    &\begingroup\color{red}+\sum_{n=3}^{+\infty} \frac{1}{(n-1)!}\sum_{a_1,\dots, a_n=1}^r C^{(n)}_{a_1 \ldots a_n} \sum_{\substack{b=1 \\ b\neq a_1}}^r t_{b a_1} \phi_{b} \phi_{a_2} \ldots \phi_{a_n} \endgroup \begingroup\color{purple}-\sum_{a=1}^r\frac{1}{2} \delta m^2_a \phi_a \phi_{\bar{a}}\endgroup.
\end{split}
\end{equation}
The black terms in~\eqref{renormalised_Lagrangian_due_to_standard_renormalisation_procedure} are the same contributions appearing in ~\eqref{eq0_1}, while the remaining coloured terms are introduced in the renormalization procedure. The propagators and vertices associated with the renormalized Lagrangian~\eqref{renormalised_Lagrangian_due_to_standard_renormalisation_procedure}  are represented in figures~\ref{Image_all_propagators_and_two_point_vertices_of_renormalized_theory} and~\ref{Image_all_propagators_and_multiple_point_vertices_of_renormalized_theory} and are coloured as the Lagrangian terms from which they are generated.
\begin{figure}
\begin{center}
\begin{tikzpicture}
\tikzmath{\y=1.9;}

\filldraw[black] (0.3*\y,0.1*\y)  node[anchor=west] {\tiny{$a(p)$}};
\draw[directed] (0*\y,0*\y) -- (1*\y,0*\y);
\filldraw[black] (1.1*\y,0*\y)  node[anchor=west] {$= \frac{i}{p^2-m_a^2 + i \epsilon}$};

\filldraw[purple] (-0.1*\y,0.1*\y-1*\y)  node[anchor=west] {\tiny{$a(p)$}};
\filldraw[purple] (0.6*\y,0.1*\y-1*\y)  node[anchor=west] {\tiny{$a(p)$}};
\draw[directed][purple] (0*\y,-1*\y) -- (0.4*\y,-1*\y);
\draw[directed][purple] (0.5*\y,-1*\y) -- (0.9*\y,-1*\y);
\filldraw[purple] (0.3*\y,-1*\y)  node[anchor=west] {$\odot$};

\filldraw[purple] (1.1*\y,-1*\y)  node[anchor=west] {$= -i \delta m_a^2$};

\filldraw[blue] (-0.1*\y+4*\y,0.1*\y)  node[anchor=west] {\tiny{$a(p)$}};
\filldraw[blue] (0.6*\y+4*\y,0.1*\y)  node[anchor=west] {\tiny{$a(p)$}};
\draw[directed][blue] (4*\y,0*\y) -- (0.4*\y+4*\y,0*\y);
\draw[directed][blue] (0.4*\y+4*\y,0*\y) -- (0.9*\y+4*\y,0*\y);
\filldraw[blue] (0.3*\y+4*\y,0*\y)  node[anchor=west] {$\bullet$};
\filldraw[blue] (1.1*\y+4*\y,0*\y)  node[anchor=west] {$= i \Sigma_{aa}^{(1)} (p^2-m_a^2)$};

\filldraw[red] (-0.1*\y+4*\y,0.1*\y-1*\y)  node[anchor=west] {\tiny{$a(p)$}};
\filldraw[red] (0.6*\y+4*\y,0.1*\y-1*\y)  node[anchor=west] {\tiny{$b(p)$}};
\draw[directed][red] (0*\y+4*\y,-1*\y) -- (0.4*\y+4*\y,-1*\y);
\draw[directed][red] (0.5*\y+4*\y,-1*\y) -- (0.9*\y+4*\y,-1*\y);
\filldraw[red] (0.3*\y+4*\y,-1*\y)  node[anchor=west] {$\oslash$};
\filldraw[red] (1.1*\y+4*\y,-1*\y)  node[anchor=west] {$= -i t_{\bar{b} \bar{a}} \bigl(p^2 -m^2_a \bigr)- i t_{a b} \bigl( p^2 -m^2_b \bigr)$};
\filldraw[red] (0.25*\y+4*\y,-1.2*\y)  node[anchor=west] {\tiny{$a\ne b$}};

\end{tikzpicture}
\caption{Propagator and two-point vertices.}
\label{Image_all_propagators_and_two_point_vertices_of_renormalized_theory}
\end{center}
\end{figure}
For each $n$-point coupling $C^{(n)}_{a_1 \dots a_n}$ receiving quantum corrections, we use the convention to introduce $n$ distinct counterterms (one for each leg attached to the coupling). 
For example, instead of writing a vertex generated from the third line of~\eqref{renormalised_Lagrangian_due_to_standard_renormalisation_procedure} as
$$
-\frac{i}{2}C^{(n)}_{a_1 \dots a_n} \bigl( \Sigma^{(1)}_{a_1 a_1}+ \dots + \Sigma^{(1)}_{a_n a_n} \bigr).
$$
we split it into $n$ distinct vertices
$$
-\frac{i}{2}C^{(n)}_{a_1 \dots a_n} \Sigma^{(1)}_{a_1 a_1} \hspace{3mm}, \hspace{3mm} -\frac{i}{2}C^{(n)}_{a_1 \dots a_n} \Sigma^{(1)}_{a_2 a_2} \hspace{3mm},\hspace{3mm} \dots \hspace{3mm},\hspace{3mm} -\frac{i}{2}C^{(n)}_{a_1 \dots a_n} \Sigma^{(1)}_{a_n a_n}.
$$
Each of these vertices is pictorially represented with a bullet on the leg $j$ carrying the factor $\Sigma^{(1)}_{a_j a_j}$ ($j=1, \dots, n$), as it is shown in the blue picture in figure~\ref{Image_all_propagators_and_multiple_point_vertices_of_renormalized_theory}. 
\begin{figure}
\begin{center}
\begin{tikzpicture}
\tikzmath{\y=1.9;}

\draw[directed] (5*\y,0*\y) -- (5.6*\y,0*\y);
\draw[directed] (5.2*\y,0.4*\y) -- (5.6*\y,0*\y);
\draw[directed] (6*\y,0.4*\y) -- (5.6*\y,0*\y);
\draw[directed] (6.2*\y,0*\y) -- (5.6*\y,0*\y);
\draw[directed] (6*\y,-0.4*\y) -- (5.6*\y,0*\y);
\filldraw[black] (5.4*\y,-0.3*\y)  node[anchor=west] {$\ldots$};
\filldraw[black] (5.4*\y,0.3*\y)  node[anchor=west] {$\ldots$};
\draw[directed] (5.2*\y,-0.4*\y) -- (5.6*\y,0*\y);

\filldraw[black] (4.8*\y,0.1*\y)  node[anchor=west] {\tiny{$a_1$}};
\filldraw[black] (5*\y,0.5*\y)  node[anchor=west] {\tiny{$a_2$}};
\filldraw[black] (5.8*\y,0.5*\y)  node[anchor=west] {\tiny{$a_{j-1}$}};
\filldraw[black] (6*\y,0.1*\y)  node[anchor=west] {\tiny{$a_j$}};
\filldraw[black] (6*\y,-0.4*\y)  node[anchor=west] {\tiny{$a_{j+1}$}};
\filldraw[black] (4.9*\y,-0.4*\y)  node[anchor=west] {\tiny{$a_n$}};
\filldraw[black] (6.2*\y,0*\y)  node[anchor=west] {$= -i C^{(n)}_{a_1 \ldots a_n}$};

\draw[directed][blue] (9*\y,0*\y) -- (9.6*\y,0*\y);
\draw[directed][blue] (9.2*\y,0.4*\y) -- (9.6*\y,0*\y);
\draw[directed][blue] (10*\y,0.4*\y) -- (9.6*\y,0*\y);
\draw[directed][blue] (10.2*\y,0*\y) -- (9.6*\y,0*\y);
\filldraw[blue] (9.54*\y,-0.01*\y)  node[anchor=west] {$\bullet$};
\draw[directed][blue] (10*\y,-0.4*\y) -- (9.6*\y,0*\y);
\filldraw[blue] (9.4*\y,-0.3*\y)  node[anchor=west] {$\ldots$};
\filldraw[blue] (9.4*\y,0.3*\y)  node[anchor=west] {$\ldots$};
\draw[directed][blue] (9.2*\y,-0.4*\y) -- (9.6*\y,0*\y);

\filldraw[blue] (8.8*\y,0.1*\y)  node[anchor=west] {\tiny{$a_1$}};
\filldraw[blue] (9*\y,0.5*\y)  node[anchor=west] {\tiny{$a_2$}};
\filldraw[blue] (9.8*\y,0.5*\y)  node[anchor=west] {\tiny{$a_{j-1}$}};
\filldraw[blue] (10*\y,0.1*\y)  node[anchor=west] {\tiny{$a_j$}};
\filldraw[blue] (10*\y,-0.4*\y)  node[anchor=west] {\tiny{$a_{j+1}$}};
\filldraw[blue] (8.9*\y,-0.4*\y)  node[anchor=west] {\tiny{$a_n$}};
\filldraw[blue] (10.2*\y,0*\y)  node[anchor=west] {$= -\frac{i}{2} \Sigma_{a_j a_j}^{(1)} C^{(n)}_{a_1 \ldots a_n}$};

\draw[directed][red] (5*\y,0*\y-1.5*\y) -- (5.6*\y,-1.5*\y);
\draw[directed][red] (5.2*\y,0.4*\y-1.5*\y) -- (5.6*\y,0*\y-1.5*\y);
\draw[directed][red] (6*\y,0.4*\y-1.5*\y) -- (5.6*\y,-1.5*\y);
\draw[directed][red] (6.2*\y,-1.5*\y) -- (5.75*\y,-1.5*\y);
\filldraw[red] (5.54*\y,-0.01*\y-1.5*\y)  node[anchor=west] {\small{$\oslash$}};
\draw[directed][red] (6*\y,-0.4*\y-1.5*\y) -- (5.6*\y,-1.5*\y);
\filldraw[red] (5.4*\y,-0.3*\y-1.5*\y)  node[anchor=west] {$\ldots$};
\filldraw[red] (5.4*\y,0.3*\y-1.5*\y)  node[anchor=west] {$\ldots$};
\draw[directed][red] (5.2*\y,-0.4*\y-1.5*\y) -- (5.6*\y,-1.5*\y);

\filldraw[red] (4.8*\y,0.1*\y-1.5*\y)  node[anchor=west] {\tiny{$a_1$}};
\filldraw[red] (5*\y,0.5*\y-1.5*\y)  node[anchor=west] {\tiny{$a_2$}};
\filldraw[red] (5.8*\y,0.5*\y-1.5*\y)  node[anchor=west] {\tiny{$a_{j-1}$}};
\filldraw[red] (6*\y,0.1*\y-1.5*\y)  node[anchor=west] {\tiny{$a_j$}};
\filldraw[red] (6*\y,-0.4*\y-1.5*\y)  node[anchor=west] {\tiny{$a_{j+1}$}};
\filldraw[red] (4.9*\y,-0.4*\y-1.5*\y)  node[anchor=west] {\tiny{$a_n$}};
\filldraw[red] (6.2*\y,-1.5*\y)  node[anchor=west] {$= i\sum_{\substack{b=1 \\ b\neq a_j}}^r t_{a_j b} C^{(n)}_{a_1 \dots a_{j-1} b a_{j+1} \dots a_n}$};

\end{tikzpicture}
\caption{Multiple-point vertices.}
\label{Image_all_propagators_and_multiple_point_vertices_of_renormalized_theory}
\end{center}
\end{figure}
The same argument applies to vertices generated from the red term
in the last row of~\eqref{renormalised_Lagrangian_due_to_standard_renormalisation_procedure}; in this case, instead of a bullet, we represent the quantum correction with a circle containing a slash (see the red vertex in figure~\ref{Image_all_propagators_and_multiple_point_vertices_of_renormalized_theory}).

After having introduced the counterterms into the Lagrangian, the propagator associated with an incoming leg of type $a$ and a (possibly different) outgoing leg of type $b$ is
\begin{equation}
\label{Most_general_propagator_Gab}
\begin{split}
    G_{ab}(p^2)&=\frac{i \delta_{ab}}{p^2 -m_a^2} \bigl( 1 \begingroup\color{blue}-\Sigma^{(1)}_{aa} \endgroup \bigr)+\frac{i}{p^2 -m_a^2} \bigl( -i \Sigma_{ab}(p^2) \bigr) \frac{i}{p^2 -m_b^2} \begingroup\color{purple}+\frac{i}{p^2 -m_a^2} \bigl( -i \delta m_a^2 \delta_{ab} \bigr)  \frac{i}{p^2 -m_a^2}\endgroup\\
    &\begingroup\color{red}+ \frac{i t_{\bar{b} \bar{a}}}{p^2-m^2_b} + \frac{i t_{a b}}{p^2-m^2_a}. \endgroup
\end{split}
\end{equation}
The last two terms are null in the particular case in which $m_a=m_b$. Once again we used different colours to match the corrections to the propagator with the Lagrangian counterterms. 

Now we show that~\eqref{renormalisation_condition_on_propagators} is satisfied.
For $b\ne a$ and $m_b \ne m_a$, due to the second definition in~\eqref{definition_of_mass_renormalization_and_tab_to_diagonalise_the_mass_matrix} it holds that
\begin{equation}
\begin{split}
   &\text{Res} \ G_{ab}(p^2)\Bigl|_{p^2=m^2_a}= \frac{i \Sigma_{ab}(m_a^2)}{m_a^2 - m_b^2}  +i t_{ab}=0,\\
   &\text{Res} \ G_{ab}(p^2)\Bigl|_{p^2=m^2_b}= \frac{i \Sigma_{ab}(m_b^2)}{m_b^2 - m_a^2}  +i t_{\bar{b} \bar{a}}=0.
   \end{split}
\end{equation}
In the second relation above we used
\begin{equation}
\label{crossing_simmetry_on_loop_propagator}
    \Sigma_{ab}(p^2)= \Sigma_{\bar{b} \bar{a}}(p^2).
\end{equation}
This implies that the propagator does not contain poles whenever the incoming and outgoing legs are of different types and have different masses. For $b\ne a$ and $m_b = m_a$, if we plug~\eqref{definition_of_mass_renormalization_and_tab_to_diagonalise_the_mass_matrix} into~\eqref{Most_general_propagator_Gab} and we expand $\Sigma_{ab}(p^2)$ around $p^2=m^2_a$ we obtain
\begin{equation}
\label{expansion_of_propagator_different_particles_with_equal_masses}
   G_{ab}(p^2)= \frac{i \Sigma_{ab}^{(0)}}{(p^2 - m_a^2)^2}+ \frac{i \Sigma_{ab}^{(1)}}{p^2 - m_a^2} + \dots
\end{equation}
where in the ellipses are contained terms which are finite at $p^2=m^2_a$. In this case, by property~\ref{Non_degenerate_mass_condition} it holds that $\Sigma_{ab}^{(0)}=\Sigma_{ab}^{(1)}=0$ and~\eqref{renormalisation_condition_on_propagators} is satisfied.
We will prove this fact in section~\ref{section_to_explain_the_cut_method}.

Instead, when the incoming and outgoing legs are of the same type, we obtain
\begin{equation}
G_{aa}(p^2)= \frac{i}{p^2-m^2_a} (1- \Sigma_{aa}^{(1)}) + \frac{i}{(p^2-m^2_a)^2}  \bigl(\Sigma_{aa}(p^2)+ \delta m^2_a \bigr)
\end{equation}
Expanding $\Sigma_{aa}(p^2)$ around $p^2=m^2_a$ and using the first relation in~\eqref{definition_of_mass_renormalization_and_tab_to_diagonalise_the_mass_matrix} it is immediate to verify that
\begin{equation}
\text{Res} \ G_{aa}(p^2)\Bigl|_{p^2=m^2_a}= i.
\end{equation}
Since, after introducing the counterterms, \eqref{renormalisation_condition_on_propagators} is verified, all one-particle-reducible Feynman diagrams containing one-loop corrections in external legs can be omitted. This is because external legs in amplitudes need to be amputated: for each external leg of type $a$ we need to multiply the amplitude by
$$
\Bigl(\frac{i}{p^2-m^2_a}\Bigl)^{-1}.
$$
This corresponds, up to a factor $i$, to taking the residue of the propagator associated with the external leg $a$. As we just verified, this residue is null any time the propagator connects two particles $a$ and $b$ of different types, while in the case $a=b$ it holds that
$$
\Bigl[\Bigl(\frac{i}{p^2-m^2_a}\Bigr)^{-1} G_{aa}(p^2)\Bigl]\Bigl|_{p^2=m^2_a}=1.
$$
For a process of the type in~\eqref{2_to_nminus2_production_process_Introduction}, this implies that the sum of all the tree-level diagrams obtained from Lagrangian~\eqref{eq0_1} and the diagrams with corrections in external legs (these corrections include both one-loop bubbles and counterterms in the external legs) returns a tree-level amplitude
$M^{(0)}_{a_1 a_2\to a_3 \dots a_n}$. Therefore, in the computation of~\eqref{amplitude_truncated_at_one_loop_2_to_n_minus_2_processes_Introduction},
we can omit all the one-particle-reducible diagrams with corrections in the external legs. 
However, diagrams containing counterterms in vertices and internal propagators
cannot be omitted; this introduces some difficulties in understanding the universal behaviour of one-loop amplitudes. To avoid these difficulties we will follow the apparently more naive approach to compute the sums of one-loop diagrams and tree-level diagrams containing counterterms separately. 
These sums will generate $M^{(1)}_{a_1 a_2\to a_3 \dots a_n}$ and $M^{(\text{count.})}_{a_1 a_2\to a_3 \dots a_n}$ in~\eqref{amplitude_truncated_at_one_loop_2_to_n_minus_2_processes_Introduction}. For this reason, we do not omit Feynman diagrams containing corrections in external legs but rather we will compute them explicitly remarking when they simplify.

For the remaining part of this section, we will focus on the set of Feynman diagrams containing counterterms. We will show that many simplifications happen within this set and it will be possible to compute a universal formula for $M^{(\text{count.})}_{a_1 a_2\to a_3 \dots a_n}$.

\subsection{Off-diagonal field renormalization contributions}
\label{subsection_on_off_diagonal_field_renormalization_contributions}

We start by computing the sum of Feynman diagrams containing counterterms coloured red in figures~\ref{Image_all_propagators_and_two_point_vertices_of_renormalized_theory} and~\ref{Image_all_propagators_and_multiple_point_vertices_of_renormalized_theory}. These counterterms can appear both in internal or external legs. 
In both cases, we show that their sum is null. 

Let us consider the three Feynman diagrams in figure~\ref{Image_cancellation_of_red_contributions_in_diagrams}, where the legs $a_1, \dots, a_n$ and $b_1 \dots, b_m$ can be both external particles or internal propagators; in the latter case, the diagrams need to be thought as part of bigger Feynman diagrams in which we omit to write all vertices connected with $a_1, \dots, a_n$ and $b_1 \dots, b_m$. 
We sum over all the types of intermediate propagators $j$ and $k$ (with $j\ne k$ in diagram $(2)$), carrying momentum $p$.
Plugging the red vertices depicted in figures~\ref{Image_all_propagators_and_two_point_vertices_of_renormalized_theory} and~\ref{Image_all_propagators_and_multiple_point_vertices_of_renormalized_theory} into these Feynman diagrams, we obtain
\begin{equation}
\label{3_contributions_with_tab_in_internal_legs}
    \begin{split}
        &D^{(1)}=i \sum_{\substack{j, k=1 \\ j\neq k}}^r t_{jk} C^{(n+1)}_{a_1 a_2 \ldots a_n k} \frac{1}{p^2 - m^2_j}  C^{(m+1)}_{b_1 b_2 \ldots b_m \bar{j}},\\
        &D^{(2)}=-i \sum_{\substack{j, k=1 \\ j\neq k}}^r C^{(n+1)}_{a_1 a_2 \ldots a_n k} \frac{1}{p^2 - m^2_k} \Bigl[ t_{\bar{k}\bar{j}} (p^2-m^2_j) + t_{j k} (p^2-m^2_k) \Bigr] \frac{1}{p^2 - m^2_j}   C^{(m+1)}_{b_1 b_2 \ldots b_m \bar{j}},\\
        &D^{(3)}= i \sum_{\substack{j, k=1 \\ j\neq k}}^r C^{(n+1)}_{a_1 a_2 \ldots a_n k} \frac{1}{p^2 - m^2_k} t_{\bar{k} \bar{j}} C^{(m+1)}_{b_1 b_2 \ldots b_m \bar{j}}.
    \end{split}
\end{equation}
The superscript numbers $(1)$, $(2)$ and $(3)$ in the expressions above match the enumeration of the Feynman diagrams in figure~\ref{Image_cancellation_of_red_contributions_in_diagrams}.
\begin{figure}
\begin{center}
\begin{tikzpicture}
\tikzmath{\y=1;}

\filldraw[red] (0.4*\y,0.8*\y)  node[anchor=west] {\scriptsize{$(1)$}};

\draw[directed][red] (-0.6*\y,0.6*\y) -- (0*\y,0*\y);
\filldraw[red] (-0.6*\y,0.12*\y)  node[anchor=west] {$\vdots$};
\draw[directed][red] (-0.6*\y,-0.6*\y) -- (0*\y,0*\y);
\draw[directed][red] (0*\y,-0.9*\y) -- (0*\y,0*\y);
\filldraw[red] (-0.12*\y,0*\y)  node[anchor=west] {\small{$\oslash$}};
\draw[directed][red] (1.2*\y,0*\y) -- (0.3*\y,0*\y);
\draw[directed][red] (1.2*\y,-0.9*\y) -- (1.2*\y,0*\y);
\draw[directed][red] (1.8*\y,-0.6*\y) -- (1.2*\y,0*\y);
\filldraw[red] (1.5*\y,0.12*\y)  node[anchor=west] {$\vdots$};
\draw[directed][red] (1.8*\y,0.6*\y) -- (1.2*\y,0*\y);

\filldraw[red] (0.4*\y,0.2*\y)  node[anchor=west] {\tiny{$j$}};
\filldraw[red] (-0.2*\y,-1*\y)  node[anchor=west] {\tiny{$a_1$}};
\filldraw[red] (-0.8*\y,-0.7*\y)  node[anchor=west] {\tiny{$a_2$}};
\filldraw[red] (-0.8*\y,0.8*\y)  node[anchor=west] {\tiny{$a_n$}};
\filldraw[red] (1.1*\y,-1*\y)  node[anchor=west] {\tiny{$b_1$}};
\filldraw[red] (1.4*\y,-0.7*\y)  node[anchor=west] {\tiny{$b_2$}};
\filldraw[red] (1.4*\y,0.8*\y)  node[anchor=west] {\tiny{$b_m$}};

\filldraw[red] (2.3*\y,0*\y)  node[anchor=west] {$+$};


\filldraw[red] (4.6*\y,0.8*\y)  node[anchor=west] {\scriptsize{$(2)$}};

\draw[directed][red] (-0.6*\y+4*\y,0.6*\y) -- (0*\y+4*\y,0*\y);
\filldraw[red] (-0.6*\y+4*\y,0.12*\y)  node[anchor=west] {$\vdots$};
\draw[directed][red] (-0.6*\y+4*\y,-0.6*\y) -- (0*\y+4*\y,0*\y);
\draw[directed][red] (0*\y+4*\y,-0.9*\y) -- (0*\y+4*\y,0*\y);
\draw[directed][red] (4.7*\y,0*\y) -- (0*\y+4*\y,0*\y);
\filldraw[red] (+4.6*\y,0*\y)  node[anchor=west] {\small{$\oslash$}};
\draw[directed][red] (5.7*\y,0*\y) -- (5*\y,0*\y);
\draw[directed][red] (5.7*\y,-0.9*\y) -- (5.7*\y,0*\y);
\filldraw[red] (6*\y,0.12*\y)  node[anchor=west] {$\vdots$};
\draw[directed][red] (6.3*\y,-0.6*\y) -- (5.7*\y,0*\y);
\draw[directed][red] (6.3*\y,0.6*\y) -- (5.7*\y,0*\y);

\filldraw[red] (4.2*\y,0.2*\y)  node[anchor=west] {\tiny{$k$}};
\filldraw[red] (5.3*\y,0.2*\y)  node[anchor=west] {\tiny{$j$}};
\filldraw[red] (3.8*\y,-1*\y)  node[anchor=west] {\tiny{$a_1$}};
\filldraw[red] (3.2*\y,-0.7*\y)  node[anchor=west] {\tiny{$a_2$}};
\filldraw[red] (3.2*\y,0.8*\y)  node[anchor=west] {\tiny{$a_n$}};
\filldraw[red] (1.1*\y+4.5*\y,-1*\y)  node[anchor=west] {\tiny{$b_1$}};
\filldraw[red] (1.4*\y+4.5*\y,-0.7*\y)  node[anchor=west] {\tiny{$b_2$}};
\filldraw[red] (1.4*\y+4.5*\y,0.8*\y)  node[anchor=west] {\tiny{$b_m$}};

\filldraw[red] (6.8*\y,0*\y)  node[anchor=west] {$+$};

\filldraw[red] (8.8*\y,0.8*\y)  node[anchor=west] {\scriptsize{$(3)$}};

\draw[directed][red] (-0.6*\y+8.5*\y,0.6*\y) -- (0*\y+8.5*\y,0*\y);
\filldraw[red] (-0.6*\y+8.5*\y,0.12*\y)  node[anchor=west] {$\vdots$};
\draw[directed][red] (-0.6*\y+8.5*\y,-0.6*\y) -- (0*\y+8.5*\y,0*\y);
\draw[directed][red] (0*\y+8.5*\y,-0.9*\y) -- (0*\y+8.5*\y,0*\y);
\draw[directed][red] (9.4*\y,0*\y) -- (+8.5*\y,0*\y);
\filldraw[red] (+9.3*\y,0*\y)  node[anchor=west] {\small{$\oslash$}};
\draw[directed][red] (+9.7*\y,-0.9*\y) -- (+9.7*\y,0*\y);
\draw[directed][red] (10.3*\y,-0.6*\y) -- (+9.7*\y,0*\y);
\filldraw[red] (10*\y,0.12*\y)  node[anchor=west] {$\vdots$};
\draw[directed][red] (10.3*\y,0.6*\y) -- (+9.7*\y,0*\y);

\filldraw[red] (9*\y,0.2*\y)  node[anchor=west] {\tiny{$k$}};
\filldraw[red] (3.8*\y+4.5*\y,-1*\y)  node[anchor=west] {\tiny{$a_1$}};
\filldraw[red] (3.2*\y+4.5*\y,-0.7*\y)  node[anchor=west] {\tiny{$a_2$}};
\filldraw[red] (3.2*\y+4.5*\y,0.8*\y)  node[anchor=west] {\tiny{$a_n$}};
\filldraw[red] (1.1*\y+8.5*\y,-1*\y)  node[anchor=west] {\tiny{$b_1$}};
\filldraw[red] (1.4*\y+8.5*\y,-0.7*\y)  node[anchor=west] {\tiny{$b_2$}};
\filldraw[red] (1.4*\y+8.5*\y,0.8*\y)  node[anchor=west] {\tiny{$b_m$}};

\filldraw[red] (11*\y,0*\y)  node[anchor=west] {$= \ \ \ 0$};

\end{tikzpicture}
\caption{Cancellation between diagrams containing off-diagonal field renormalization counterterms in internal legs. The legs $a_1, \dots, a_n$ and $b_1, \dots, b_m$ can be both external particles or propagators attached to other vertices.}
\label{Image_cancellation_of_red_contributions_in_diagrams}
\end{center}
\end{figure}
Summing the three contributions in~\eqref{3_contributions_with_tab_in_internal_legs} it is immediate to verify that
$$
D^{(1)}+D^{(2)}+D^{(3)}=0.
$$
The same consideration can be repeated for any scattering process and for all the Feynman diagrams with these types of corrections attached to internal propagators.
The next step is to check what is the effect of these counterterms on the external legs. To do that, we focus on the pair of diagrams shown in figure~\ref{Image_cancellation_of_red_contributions_in_external_legs}, where $j$ represents an external particle carrying momentum $p$ and we are summing over all the possible propagators of type $k\ne j$ in the second diagram.
In this case, the legs $a_1, \dots, a_n$ can be external particles or propagators attached to other vertices that are not shown in figure~\ref{Image_cancellation_of_red_contributions_in_external_legs}.
The two diagrams can be written as
\begin{subequations}
    \label{External_legs_off_diagonal_counterterms}
\begin{align}
\label{External_legs_off_diagonal_counterterms_diagram_one}
 &D^{(4)}= i\sum_{\substack{k=1 \\ k\neq j}}^r t_{j k} C^{(n+1)}_{a_1 \dots a_{n} k} ,\\
\label{External_legs_off_diagonal_counterterms_diagram_two}
&D^{(5)}= -i \sum_{\substack{k=1 \\ k\neq j}}^r C^{(n+1)}_{a_1 \dots a_{n} k} \frac{1}{p^2 -m^2_k} \Bigl( t_{\bar{k} \bar{j}} (p^2 - m^2_j)+ t_{j k} (p^2 - m^2_k) \Bigr).
\end{align}
\end{subequations}
Combining the fact that $j$ is an external particle (satisfying the on-shell condition $p^2=m^2_j$) and $t_{jk}=0$ any time $m_k=m_j$, we conclude that 
$$
D^{(4)}+D^{(5)}=0.
$$

Due to these considerations, the sum of all Feynman diagrams containing vertices coloured red in figures~\ref{Image_all_propagators_and_two_point_vertices_of_renormalized_theory} and~\ref{Image_all_propagators_and_multiple_point_vertices_of_renormalized_theory} is zero. 
\begin{figure}
\begin{center}
\begin{tikzpicture}
\tikzmath{\y=1;}

\filldraw[red] (0.4*\y,0.8*\y)  node[anchor=west] {\scriptsize{$(4)$}};

\draw[directed][red] (-0.6*\y,0.6*\y) -- (0*\y,0*\y);
\filldraw[red] (-0.6*\y,0.12*\y)  node[anchor=west] {$\vdots$};
\draw[directed][red] (-0.6*\y,-0.6*\y) -- (0*\y,0*\y);
\draw[directed][red] (0*\y,-0.9*\y) -- (0*\y,0*\y);
\filldraw[red] (-0.12*\y,0*\y)  node[anchor=west] {\small{$\oslash$}};
\draw[directed][red] (1.2*\y,0*\y) -- (0.3*\y,0*\y);

\filldraw[red] (0.4*\y,0.2*\y)  node[anchor=west] {\tiny{$j$}};
\filldraw[red] (-0.2*\y,-1*\y)  node[anchor=west] {\tiny{$a_1$}};
\filldraw[red] (-0.8*\y,-0.7*\y)  node[anchor=west] {\tiny{$a_2$}};
\filldraw[red] (-0.4*\y,0.5*\y)  node[anchor=west] {\tiny{$a_n$}};

\filldraw[red] (2*\y,0*\y)  node[anchor=west] {$+$};


\filldraw[red] (4.6*\y,0.8*\y)  node[anchor=west] {\scriptsize{$(5)$}};

\draw[directed][red] (-0.6*\y+4*\y,0.6*\y) -- (0*\y+4*\y,0*\y);
\filldraw[red] (-0.6*\y+4*\y,0.12*\y)  node[anchor=west] {$\vdots$};
\draw[directed][red] (-0.6*\y+4*\y,-0.6*\y) -- (0*\y+4*\y,0*\y);
\draw[directed][red] (0*\y+4*\y,-0.9*\y) -- (0*\y+4*\y,0*\y);
\draw[directed][red] (4.7*\y,0*\y) -- (0*\y+4*\y,0*\y);
\filldraw[red] (+4.6*\y,0*\y)  node[anchor=west] {\small{$\oslash$}};
\draw[directed][red] (5.7*\y,0*\y) -- (5*\y,0*\y);

\filldraw[red] (4.2*\y,0.2*\y)  node[anchor=west] {\tiny{$k$}};
\filldraw[red] (5.3*\y,0.2*\y)  node[anchor=west] {\tiny{$j$}};
\filldraw[red] (3.8*\y,-1*\y)  node[anchor=west] {\tiny{$a_1$}};
\filldraw[red] (3.2*\y,-0.7*\y)  node[anchor=west] {\tiny{$a_2$}};
\filldraw[red] (3.2*\y,0.8*\y)  node[anchor=west] {\tiny{$a_n$}};

\filldraw[red] (6.5*\y,0*\y)  node[anchor=west] {$= \ \ \ 0$};

\end{tikzpicture}
\caption{Cancellation between diagrams containing off-diagonal field renormalization counterterms in external legs; $j$ represents an external particle incoming into the diagrams, while $a_1, \ldots, a_n$ can be both external particles or propagators attached to other vertices.}
\label{Image_cancellation_of_red_contributions_in_external_legs}
\end{center}
\end{figure}
We can write their contribution to the amplitude associated with the inelastic process~\eqref{2_to_nminus2_production_process_Introduction} as
\begin{equation}
\label{red_contributions_to_one_loop_amplitude}
   M^{(\text{red})}_{a_1, a_2 \to a_3 \dots a_n}= 0.
\end{equation}

In appendix~\ref{Appendix_on_counterterms_for_different_particles_with_equal_masses} we consider the case in which property~\ref{Non_degenerate_mass_condition} is violated; in this case, the imposition of the renormalization condition~\eqref{renormalisation_condition_on_propagators} forces to set $t_{ab} \ne 0$ when $a\ne b$ and $m_a=m_b$. For this reason~\eqref{red_contributions_to_one_loop_amplitude} may not vanish in general and relations~(\eqref{final_formula_we_need_to_prove_first_formulation}, \eqref{final_formula_we_need_to_prove_second_formulation}) could be violated in certain inelastic processes.

\subsection{Diagonal field renormalization contributions}

Similar considerations can be applied to vertices coloured blue in figures~\ref{Image_all_propagators_and_two_point_vertices_of_renormalized_theory} and~\ref{Image_all_propagators_and_multiple_point_vertices_of_renormalized_theory}. The sum of Feynman diagrams in which these vertices appear in internal propagators is once again zero. An example is provided in figure~\ref{Image_cancellation_of_blue_contributions_in_diagrams}, where we are summing over all the possible types of particles $j$, propagating internally to the diagrams with momentum $p$, keeping fixed the legs $a_1, \dots, a_n$ and $b_1, \dots, b_m$. Each of these legs can correspond to an external particle or a propagator attached to a vertex.
\begin{figure}
\begin{center}
\begin{tikzpicture}
\tikzmath{\y=1;}

\filldraw[blue] (0.4*\y,0.8*\y)  node[anchor=west] {\scriptsize{$(1')$}};

\draw[directed][blue] (-0.6*\y,0.6*\y) -- (0*\y,0*\y);
\filldraw[blue] (-0.6*\y,0.12*\y)  node[anchor=west] {$\vdots$};
\draw[directed][blue] (-0.6*\y,-0.6*\y) -- (0*\y,0*\y);
\draw[directed][blue] (0*\y,-0.9*\y) -- (0*\y,0*\y);
\filldraw[blue] (-0.14*\y,-0.02*\y)  node[anchor=west] {\small{$\bullet$}};
\draw[directed][blue] (1.2*\y,0*\y) -- (0.1*\y,0*\y);
\draw[directed][blue] (1.2*\y,-0.9*\y) -- (1.2*\y,0*\y);
\draw[directed][blue] (1.8*\y,-0.6*\y) -- (1.2*\y,0*\y);
\filldraw[blue] (1.5*\y,0.12*\y)  node[anchor=west] {$\vdots$};
\draw[directed][blue] (1.8*\y,0.6*\y) -- (1.2*\y,0*\y);

\filldraw[blue] (0.4*\y,0.2*\y)  node[anchor=west] {\tiny{$j$}};
\filldraw[blue] (-0.2*\y,-1*\y)  node[anchor=west] {\tiny{$a_1$}};
\filldraw[blue] (-0.8*\y,-0.7*\y)  node[anchor=west] {\tiny{$a_2$}};
\filldraw[blue] (-0.8*\y,0.8*\y)  node[anchor=west] {\tiny{$a_n$}};
\filldraw[blue] (1.1*\y,-1*\y)  node[anchor=west] {\tiny{$b_1$}};
\filldraw[blue] (1.4*\y,-0.7*\y)  node[anchor=west] {\tiny{$b_2$}};
\filldraw[blue] (1.4*\y,0.8*\y)  node[anchor=west] {\tiny{$b_m$}};

\filldraw[blue] (-2*\y,-2*\y)  node[anchor=west] {\tiny{$\Bigl(-\frac{i}{2}\Sigma^{(1)}_{jj} C^{(n+1)}_{a_1 \dots a_n j} \Bigr) \frac{1}{p^2-m^2_j} C^{(n+1)}_{b_1 \dots b_m \bar{j}}$}};

\filldraw[blue] (2.3*\y,0*\y)  node[anchor=west] {$+$};


\filldraw[blue] (4.6*\y,0.8*\y)  node[anchor=west] {\scriptsize{$(2')$}};

\draw[directed][blue] (-0.6*\y+4*\y,0.6*\y) -- (0*\y+4*\y,0*\y);
\filldraw[blue] (-0.6*\y+4*\y,0.12*\y)  node[anchor=west] {$\vdots$};
\draw[directed][blue] (-0.6*\y+4*\y,-0.6*\y) -- (0*\y+4*\y,0*\y);
\draw[directed][blue] (0*\y+4*\y,-0.9*\y) -- (0*\y+4*\y,0*\y);
\draw[directed][blue] (4.8*\y,0*\y) -- (0*\y+4*\y,0*\y);
\filldraw[blue] (+4.6*\y,-0.02*\y)  node[anchor=west] {\small{$\bullet$}};
\draw[directed][blue] (5.7*\y,0*\y) -- (4.9*\y,0*\y);
\draw[directed][blue] (5.7*\y,-0.9*\y) -- (5.7*\y,0*\y);
\filldraw[blue] (6*\y,0.12*\y)  node[anchor=west] {$\vdots$};
\draw[directed][blue] (6.3*\y,-0.6*\y) -- (5.7*\y,0*\y);
\draw[directed][blue] (6.3*\y,0.6*\y) -- (5.7*\y,0*\y);

\filldraw[blue] (4.2*\y,0.2*\y)  node[anchor=west] {\tiny{$j$}};
\filldraw[blue] (5.3*\y,0.2*\y)  node[anchor=west] {\tiny{$j$}};
\filldraw[blue] (3.8*\y,-1*\y)  node[anchor=west] {\tiny{$a_1$}};
\filldraw[blue] (3.2*\y,-0.7*\y)  node[anchor=west] {\tiny{$a_2$}};
\filldraw[blue] (3.2*\y,0.8*\y)  node[anchor=west] {\tiny{$a_n$}};
\filldraw[blue] (1.1*\y+4.5*\y,-1*\y)  node[anchor=west] {\tiny{$b_1$}};
\filldraw[blue] (1.4*\y+4.5*\y,-0.7*\y)  node[anchor=west] {\tiny{$b_2$}};
\filldraw[blue] (1.4*\y+4.5*\y,0.8*\y)  node[anchor=west] {\tiny{$b_m$}};

\filldraw[blue] (3*\y,-2*\y)  node[anchor=west] {\tiny{$i\Sigma^{(1)}_{jj} C^{(n+1)}_{a_1 \dots a_n j} \frac{1}{p^2-m^2_j}C^{(n+1)}_{b_1 \dots b_m \bar{j}}$}};

\filldraw[blue] (6.8*\y,0*\y)  node[anchor=west] {$+$};

\filldraw[blue] (8.8*\y,0.8*\y)  node[anchor=west] {\scriptsize{$(3')$}};

\draw[directed][blue] (-0.6*\y+8.5*\y,0.6*\y) -- (0*\y+8.5*\y,0*\y);
\filldraw[blue] (-0.6*\y+8.5*\y,0.12*\y)  node[anchor=west] {$\vdots$};
\draw[directed][blue] (-0.6*\y+8.5*\y,-0.6*\y) -- (0*\y+8.5*\y,0*\y);
\draw[directed][blue] (0*\y+8.5*\y,-0.9*\y) -- (0*\y+8.5*\y,0*\y);
\draw[directed][blue] (9.55*\y,0*\y) -- (+8.5*\y,0*\y);
\filldraw[blue] (+9.4*\y,-0.02*\y)  node[anchor=west] {\small{$\bullet$}};
\draw[directed][blue] (+9.7*\y,-0.9*\y) -- (+9.7*\y,0*\y);
\draw[directed][blue] (10.3*\y,-0.6*\y) -- (+9.7*\y,0*\y);
\filldraw[blue] (10*\y,0.12*\y)  node[anchor=west] {$\vdots$};
\draw[directed][blue] (10.3*\y,0.6*\y) -- (+9.7*\y,0*\y);

\filldraw[blue] (9*\y,0.2*\y)  node[anchor=west] {\tiny{$j$}};
\filldraw[blue] (3.8*\y+4.5*\y,-1*\y)  node[anchor=west] {\tiny{$a_1$}};
\filldraw[blue] (3.2*\y+4.5*\y,-0.7*\y)  node[anchor=west] {\tiny{$a_2$}};
\filldraw[blue] (3.2*\y+4.5*\y,0.8*\y)  node[anchor=west] {\tiny{$a_n$}};
\filldraw[blue] (1.1*\y+8.5*\y,-1*\y)  node[anchor=west] {\tiny{$b_1$}};
\filldraw[blue] (1.4*\y+8.5*\y,-0.7*\y)  node[anchor=west] {\tiny{$b_2$}};
\filldraw[blue] (1.4*\y+8.5*\y,0.8*\y)  node[anchor=west] {\tiny{$b_m$}};

\filldraw[blue] (8*\y,-2*\y)  node[anchor=west] {\tiny{$C^{(n+1)}_{a_1 \dots a_n j} \frac{1}{p^2-m^2_j} \Bigl( -\frac{i}{2}\Sigma^{(1)}_{jj} C^{(n+1)}_{b_1 \dots b_m \bar{j}} \Bigr)$}};

\filldraw[blue] (11*\y,0*\y)  node[anchor=west] {$= \ \ \ 0$};

\end{tikzpicture}
\caption{Cancellation between diagrams containing diagonal field renormalization counterterms in internal legs. As before $a_1, \dots, a_n$ and $b_1, \dots, b_m$ can label both external particles or propagators attached to other vertices. 
Under each diagram, its algebraic value is written.}
\label{Image_cancellation_of_blue_contributions_in_diagrams}
\end{center}
\end{figure}
As we can see from the algebraic values written under the diagrams, the sum of the three contributions in figure~\ref{Image_cancellation_of_blue_contributions_in_diagrams} is zero.

The case in which the counterterms are attached to the external legs is different and is shown in figure~\ref{Image_cancellation_of_blue_contributions_in_external_legs}.
\begin{figure}
\begin{center}
\begin{tikzpicture}
\tikzmath{\y=1;}

\filldraw[blue] (0.4*\y,0.8*\y)  node[anchor=west] {\scriptsize{$(4')$}};

\draw[directed][blue] (-0.6*\y,0.6*\y) -- (0*\y,0*\y);
\filldraw[blue] (-0.6*\y,0.12*\y)  node[anchor=west] {$\vdots$};
\draw[directed][blue] (-0.6*\y,-0.6*\y) -- (0*\y,0*\y);
\draw[directed][blue] (0*\y,-0.9*\y) -- (0*\y,0*\y);
\filldraw[blue] (-0.13*\y,-0.02*\y)  node[anchor=west] {\small{$\bullet$}};
\draw[directed][blue] (1.2*\y,0*\y) -- (0.1*\y,0*\y);

\filldraw[blue] (0.4*\y,0.2*\y)  node[anchor=west] {\tiny{$j$}};
\filldraw[blue] (-0.2*\y,-1*\y)  node[anchor=west] {\tiny{$a_1$}};
\filldraw[blue] (-0.8*\y,-0.7*\y)  node[anchor=west] {\tiny{$a_2$}};
\filldraw[blue] (-0.4*\y,0.5*\y)  node[anchor=west] {\tiny{$a_n$}};

\filldraw[blue] (2*\y,0*\y)  node[anchor=west] {$+$};

\filldraw[blue] (-1*\y,-2*\y)  node[anchor=west] {\scriptsize{$-\frac{i}{2}\Sigma^{(1)}_{jj} C^{(n+1)}_{a_1 \dots a_n j}$}};


\filldraw[blue] (4.6*\y,0.8*\y)  node[anchor=west] {\scriptsize{$(5')$}};

\draw[directed][blue] (-0.6*\y+4*\y,0.6*\y) -- (0*\y+4*\y,0*\y);
\filldraw[blue] (-0.6*\y+4*\y,0.12*\y)  node[anchor=west] {$\vdots$};
\draw[directed][blue] (-0.6*\y+4*\y,-0.6*\y) -- (0*\y+4*\y,0*\y);
\draw[directed][blue] (0*\y+4*\y,-0.9*\y) -- (0*\y+4*\y,0*\y);
\draw[directed][blue] (4.7*\y,0*\y) -- (0*\y+4*\y,0*\y);
\filldraw[blue] (+4.5*\y,-0.02*\y)  node[anchor=west] {\small{$\bullet$}};
\draw[directed][blue] (5.7*\y,0*\y) -- (4.8*\y,0*\y);

\filldraw[blue] (4.2*\y,0.2*\y)  node[anchor=west] {\tiny{$j$}};
\filldraw[blue] (5.3*\y,0.2*\y)  node[anchor=west] {\tiny{$j$}};
\filldraw[blue] (3.8*\y,-1*\y)  node[anchor=west] {\tiny{$a_1$}};
\filldraw[blue] (3.2*\y,-0.7*\y)  node[anchor=west] {\tiny{$a_2$}};
\filldraw[blue] (3.2*\y,0.8*\y)  node[anchor=west] {\tiny{$a_n$}};

\filldraw[blue] (3.5*\y,-2*\y)  node[anchor=west] {\scriptsize{$i\Sigma^{(1)}_{jj} C^{(n+1)}_{a_1 \dots a_n j}$}};

\filldraw[blue] (6.5*\y,0*\y)  node[anchor=west] {$\ne \ \ \ 0$};

\end{tikzpicture}
\caption{Contribution of diagrams containing diagonal field renormalization counterterms in external legs ($j$ corresponds to an external particle while $a_1, \dots, a_n$ can correspond to external particles or internal propagators attached to other vertices); differently from the diagrams in figure~\ref{Image_cancellation_of_red_contributions_in_external_legs} these diagrams do not sum to zero.}
\label{Image_cancellation_of_blue_contributions_in_external_legs}
\end{center}
\end{figure}
The two diagrams in figure~\ref{Image_cancellation_of_blue_contributions_in_external_legs} are similar to the ones in figure~\ref{Image_cancellation_of_red_contributions_in_external_legs}: as it happens in~\eqref{External_legs_off_diagonal_counterterms_diagram_two}, we can imagine diagram $(5')$ composed of two terms. However, in this case, due to the presence of a propagator of type $j$ which is of the same type of the external particle, both terms survive when we set $p^2=m^2_j$ (being $p$ the momentum carried by the particle $j$). For this reason, the sum of the two diagrams in figure~\ref{Image_cancellation_of_blue_contributions_in_external_legs} is nonzero and it is given by
\begin{equation}
    D^{(4')}+D^{(5')} = \frac{i}{2} \Sigma_{jj}^{(1)} C^{(n+1)}_{a_1 \dots a_n j} \times \bigl( \hspace{0.8mm} \dots \bigr)
\end{equation}
The ellipses in the expression above correspond to any tree-level diagram the legs $a_1 \dots a_n$ are attached to. Repeating this argument for all the external legs, we obtain that the contribution of the vertices coloured blue in figures~\ref{Image_all_propagators_and_two_point_vertices_of_renormalized_theory} and~\ref{Image_all_propagators_and_multiple_point_vertices_of_renormalized_theory} to the amplitude associated with the scattering process~\eqref{2_to_nminus2_production_process_Introduction} is
\begin{equation}
\label{blue_vertex_contributions_to_the_total_amplitude}
   M^{(\text{blue})}_{a_1, a_2 \to a_3 \dots a_n}= -\frac{1}{2} \sum_{j=1}^n \Sigma^{(1)}_{a_j a_j} M^{(0)}_{a_1, a_2 \to a_3 \dots a_n}.
\end{equation}
This contribution is nonzero in generic quantum field theories; however, since we are considering an inelastic process and the theory under consideration satisfies condition~\ref{Condition_tree_level_elasticity_introduction}, it has to hold that $M^{(0)}_{a_1, a_2 \to a_3 \dots a_n}=0$.
For this reason, we obtain
\begin{equation}
\label{blue_vertex_contributions_to_the_total_amplitude_zero_expression}
   M^{(\text{blue})}_{a_1, a_2 \to a_3 \dots a_n}= 0.
\end{equation}

\subsection{Mass renormalization contributions and diagonal bubbles in external legs}

Now we study diagrams containing the two-point counterterms coloured purple in figure~\ref{Image_all_propagators_and_two_point_vertices_of_renormalized_theory}. 
We focus on the case in which these counterterms appear on the external legs first.
An example, associated with the process in~\eqref{2_to_nminus2_production_process_Introduction}, is represented in figure~\ref{loop_on_external_leg_a_in_general_process_plus_purple_contribution}, where the big blob corresponds to a tree-level amplitude and a counterterm is attached to the external leg $a_1$. The algebraic value of the picture, corresponding to a particular combination of Feynman diagrams, is written under it.
\begin{figure}
\begin{center}
\begin{tikzpicture}
\tikzmath{\y=1.4;}


\draw[directed][purple] (8.9*\y,2.2*\y) -- (9.52*\y,1.58*\y);
\draw[directed][purple] (9.7*\y,1.4*\y) -- (10.3*\y,0.8*\y);
\filldraw[purple] (9.4*\y,1.5*\y)  node[anchor=west] {\Large{$\odot$}};
\draw[directed][purple] (9.9*\y,-0.3*\y) -- (10.3*\y,0.4*\y);
\draw[directed][purple] (10.6*\y,0.3*\y) -- (10.8*\y,-0.4*\y);
\draw[directed][purple] (10.8*\y,0.9*\y) -- (11.3*\y,1.4*\y);
\filldraw[color=purple!60, fill=purple!5, very thick](10.5*\y,0.6*\y) circle (0.4*\y);

\filldraw[purple] (8.6*\y,2.5*\y)  node[anchor=west] {\scriptsize{$a_1(p_1)$}};
\filldraw[purple] (9.9*\y,1.3*\y)  node[anchor=west] {\scriptsize{$a_1(p_1)$}};
\filldraw[purple] (9.5*\y,-0.5*\y)  node[anchor=west] {\scriptsize{$a_2$}};
\filldraw[purple] (10.6*\y,-0.5*\y)  node[anchor=west] {\scriptsize{$a_3$}};
\filldraw[purple] (11.3*\y,1.4*\y)  node[anchor=west] {\scriptsize{$a_n$}};
\filldraw[purple] (11.2*\y,0.6*\y)  node[anchor=west] {$\vdots$};

\filldraw[black] (9.4*\y,-1.4*\y)  node[anchor=west] {$\frac{\delta m^2_{a_1}}{p_1^2-m_{a_1}^2} M^{(0)}_{a_1 a_2 \to a_3 \dots a_n}$};

\end{tikzpicture}
\caption{Counterterm correction to the external leg $a_1$ and associated algebraic expression.}
\label{loop_on_external_leg_a_in_general_process_plus_purple_contribution}
\end{center}
\end{figure}
If we repeat the same argument for each external leg, 
we obtain that the
contribution to the amplitude returned by mass correction counterterms in external legs is given by
\begin{equation}
\label{purple_counterterms_in_external_legs}
M^{(\text{purple ext. legs})}_{a_1 a_2 \to a_3 \ldots a_n}= \sum_{j=1}^n \frac{\delta m^2_{a_j}}{p^2_{j} - m^2_{a_j}} M^{(0)}_{a_1 a_2 \to a_3 \dots a_n}.
\end{equation}
When the external particles are on-shell $M^{(0)}_{a_1 a_2 \to a_3 \dots a_n}=0$ but also the denominators on the r.h.s. of~\eqref{purple_counterterms_in_external_legs} are null. For this reason, \eqref{purple_counterterms_in_external_legs} is  nonzero in general. This contribution will be necessary to cancel certain ill-defined Feynman diagrams with loop corrections in external legs.

We now study the case in which the counterterms due to the renormalization of the masses appear in internal propagators.
This situation is shown in figure~\ref{counterterm_in_internal_propagator_for_random_n_point_amplitude_image}, where the counterterm is attached to a propagator of type $j$ carrying momentum $p$. The picture in figure~\ref{counterterm_in_internal_propagator_for_random_n_point_amplitude_image} corresponds to a particular combination of Feynman diagrams contributing to the process~\eqref{2_to_nminus2_production_process_Introduction}. The two blobs attached to the propagator $j$ are tree-level amplitudes and contain sums over all possible tree-level Feynman diagrams having as external particles $\{a_1, \dots , a_k ,j\}$ and $\{a_{k+1}, \dots , a_n ,j\}$. The particles can be incoming or outgoing depending on the directions of the associated arrows.
\begin{figure}
\begin{center}
\begin{tikzpicture}
\tikzmath{\y=1.4;}


\draw[directed][purple] (0.4*\y,0*\y) -- (1.4*\y,0*\y);
\filldraw[purple] (1.3*\y,0*\y)  node[anchor=west] {$\odot$};
\draw[directed][purple] (1.6*\y,0*\y) -- (2.6*\y,0*\y);

\draw[directed][purple] (-0.5*\y,1.3*\y) -- (-0.1*\y,0.2*\y);
\draw[directed][purple] (-1*\y,1*\y) -- (-0.2*\y,0.2*\y);
\draw[directed][purple] (-0.2*\y,0*\y) -- (-1.3*\y,0.5*\y);
\filldraw[purple] (-0.8*\y,-0.2*\y)  node[anchor=west] {$\vdots$};
\draw[directed][purple] (-0.1*\y,-0.2*\y) -- (-0.5*\y,-1.3*\y);
\draw[directed][purple] (-0.1*\y+3.2*\y,0.2*\y) -- (0.3*\y+3.2*\y,1.3*\y);
\draw[directed][purple] (-0.2*\y+3.5*\y,0.2*\y) -- (0.6*\y+3.5*\y,1*\y);
\filldraw[purple] (3.5*\y,-0.2*\y)  node[anchor=west] {$\vdots$};
\draw[directed][purple] (-0.1*\y+3.2*\y,-0.2*\y) -- (0.3*\y+3.2*\y,-1.3*\y);
\filldraw[color=purple!60, fill=purple!9, very thick](0*\y,0*\y) circle (0.4*\y);
\filldraw[color=purple!60, fill=purple!9, very thick](3*\y,0*\y) circle (0.4*\y);

\filldraw[purple] (-0.5*\y,1.3*\y)  node[anchor=west] {\scriptsize{$a_{1}$}};
\filldraw[purple] (-1.2*\y,1.1*\y)  node[anchor=west] {\scriptsize{$a_{2}$}};
\filldraw[purple] (-1.5*\y,0.6*\y)  node[anchor=west] {\scriptsize{$a_{3}$}};
\filldraw[purple] (-0.9*\y,-1.3*\y)  node[anchor=west] {\scriptsize{$a_{k}$}};
\filldraw[purple] (2.7*\y,1.3*\y)  node[anchor=west] {\scriptsize{$a_{k+1}$}};
\filldraw[purple] (3.7*\y,1.1*\y)  node[anchor=west] {\scriptsize{$a_{k+2}$}};
\filldraw[purple] (3.5*\y,-1.3*\y)  node[anchor=west] {\scriptsize{$a_{n}$}};

\filldraw[purple] (0.8*\y,0.3*\y)  node[anchor=west] {\scriptsize{$j$}};
\filldraw[purple] (2*\y,0.3*\y)  node[anchor=west] {\scriptsize{$j$}};

\end{tikzpicture}
\caption{Mass counterterm correction to a propagator of type $j$.}
\label{counterterm_in_internal_propagator_for_random_n_point_amplitude_image}
\end{center}
\end{figure}
The algebraic expression associated with the combination of diagrams in figure~\ref{counterterm_in_internal_propagator_for_random_n_point_amplitude_image} is
\begin{equation}
\label{mass_correction_in_internal_propagator_formulation_1}
     M^{(0)}_{a_1 a_2 \to a_3 \dots a_k j} \frac{i}{p^2 - m^2_j}   (-i \delta m^2_j) \frac{i}{p^2 - m^2_j}  M^{(0)}_{j \to a_{k+1} \dots a_n}
\end{equation}
While the particles $a_1 ,\dots a_n$ are on-shell (indeed they are the external particles entering process~\eqref{2_to_nminus2_production_process_Introduction}), $j$ is off-shell in general; for this reason, even though condition~\ref{Condition_tree_level_elasticity_introduction} holds, the tree-level amplitudes appearing in expression~\eqref{mass_correction_in_internal_propagator_formulation_1} are nonzero. We can write~\eqref{mass_correction_in_internal_propagator_formulation_1} as
\begin{equation}
\label{mass_correction_in_internal_propagator_formulation_2}
      \delta m^2_j \frac{\partial}{\partial m^2_j}  \Bigl(  M^{(0)}_{a_1 a_2 \to a_3 \dots a_k}   \frac{i}{p^2 - m^2_j} M^{(0)}_{j \to a_{k+1} \dots a_n} \Bigr).
\end{equation}
Repeating this argument for all the propagators, we see that the two-point counterterms appearing in internal propagators contribute to the amplitude associated with process~\eqref{2_to_nminus2_production_process_Introduction} with
\begin{equation}
\label{contribution_of_purple_terms_in_internal_propagators}
 M^{(\text{purple prop.})}_{a_1 a_2 \to a_3 \dots a_n}=\sum_{j \in \text{prop.}} \delta m^2_j \frac{\partial}{\partial m^2_j} M^{(0)}_{a_1 a_2 \to a_3 \dots a_n}.
\end{equation}
The sum in~\eqref{contribution_of_purple_terms_in_internal_propagators} is performed over the masses of all propagators appearing in $M^{(0)}_{a_1 a_2 \to a_3 \dots a_n}$.
The expression on the r.h.s. of~\eqref{contribution_of_purple_terms_in_internal_propagators} is nonzero.
Indeed condition~\ref{Condition_tree_level_elasticity_introduction} implies that 
$M^{(0)}_{a_1 a_2 \to a_3 \dots a_n}$ is null only if the particles $a_1, \dots, a_n$ are on-shell and the couplings and masses used to evaluate the amplitude are those appearing in the tree-level-integrable Lagrangian~\eqref{eq0_1}. In expression~\eqref{contribution_of_purple_terms_in_internal_propagators}, we should evaluate $M^{(0)}_{a_1 a_2 \to a_3 \dots a_n}$ for arbitrary values of the masses appearing in the tree-level propagators. Then we need to take the derivatives with respect to these masses and only in the end substitute these masses with their classical values. 
Even though $M^{(0)}_{a_1 a_2 \to a_3 \dots a_n}$ is null at these values, this does not imply that $\frac{\partial}{\partial m^2_j} M^{(0)}_{a_1 a_2 \to a_3 \dots a_n}$ is zero. This should be pretty obvious, indeed the condition that a generic function is zero at a certain point does not imply that the derivative of that function is zero at that point. 

Summing \eqref{red_contributions_to_one_loop_amplitude}, \eqref{blue_vertex_contributions_to_the_total_amplitude_zero_expression}, \eqref{purple_counterterms_in_external_legs} and~\eqref{contribution_of_purple_terms_in_internal_propagators}, we obtain that the contribution to the process~\eqref{2_to_nminus2_production_process_Introduction} generated by counterterms is given by
\begin{equation}
\label{total_counterterms_contribution_production_process}
    M^{(\text{count.})}_{a_1 a_2\to a_3 \dots a_n}=\sum_{j=1}^n \frac{\delta m^2_{a_j}}{p^2_{j} - m^2_{a_j}} M^{(0)}_{a_1 a_2 \to a_3 \dots a_n}+\sum_{j \in \text{prop.}} \delta m^2_j \frac{\partial}{\partial m^2_j} M^{(0)}_{a_1 a_2 \to a_3 \dots a_n}.
\end{equation}
We remark that the first sum runs over the external particles while the second sum runs over propagators appearing in all tree-level Feynman diagrams contributing to $M^{(0)}_{a_1 a_2 \to a_3 \dots a_n}$.

\section{The cutting method}
\label{section_on_the_cutting_method}

In this section, we show how one-loop amplitudes can be obtained in terms of particular combinations of tree-level amplitudes by splitting each propagator into the sum of a Dirac-delta function and a retarded propagator. This method is described in chapter 24 of~\cite{Schwartz_book} and we will apply it to integrable theories with Lagrangians of type~\eqref{eq0_1} and satisfying conditions~\ref{Condition_tree_level_elasticity_introduction} and~\ref{Non_degenerate_mass_condition}.
Using this split, we write one-loop amplitudes in terms of 
products and integrals of on-shell tree-level amplitudes. 

The goal of generating one-loop S-matrices from tree-level S-matrices of 1+1 dimensional integrable theories was already pursued in~\cite{Engelund:2013fja,Bianchi:2013nra,Bianchi:2014rfa} where unitarity methods~\cite{Bern:1994zx,Bern:1994cg} were used\footnote{Unitarity methods  were earlier applied also in~\cite{Arefeva:1974bk} to study 
the one-loop integrability of the sine-Gordon theory.}.
In particular, in~\cite{Bianchi:2014rfa} a formula reproducing one-loop two-to-two S-matrices of integrable theories was provided and tested on different models against symmetry considerations and one-loop computations. Up to a possible shift in the coupling, the authors found agreement with the expectations for the models that they considered. Even so, the formula obtained in~\cite{Bianchi:2014rfa} cannot be universal.
This is clear by the fact that the rational part of the one-loop S-matrix obtained in~\cite{Bianchi:2014rfa} is the sum of
the `square' of the tree-level S-matrix (expected by the optical theorem) plus a term which is linear in the tree-level S-matrix; this implies that any time the tree-level S-matrix has a pole of order one (corresponding to a bound state propagating particle) then the one-loop S-matrix contains a pole of order two. These higher-order poles, though can exist, are not common and require explanation in terms of Landau singularities in Feynman diagrams~\cite{Coleman:1978kk}. As a consequence of this fact, the formula obtained in~\cite{Bianchi:2014rfa} does not reproduce the correct expressions for one-loop amplitudes in general integrable theories. A simple example of a theory where it fails is the Bullough–Dodd model. A possible explanation for the failure of the formula proposed in~\cite{Bianchi:2014rfa} for certain integrable theories is that unitarity cut techniques can lead to an incomplete answer; indeed, it is possible to lose relevant rational terms in the cut procedure. Another reason can be that certain cuts are singular in two dimensions. While these singular contributions were avoided in~\cite{Bianchi:2014rfa} by using a particular prescription, it is not clear if the prescription used in that paper leads to the correct one-loop S-matrix for a general integrable theory.
For this reason, we propose a different approach to generate one-loop amplitudes.
Differently from unitarity methods, the technique described in the following sections is a different way to write sums of loop diagrams and returns complete answers for amplitudes. As it happened in~\cite{Bianchi:2014rfa} we also encounter potential singularities in particular degenerate inelastic processes: we will show that all potentially-singular contributions are finite and, if conditions~\ref{Condition_tree_level_elasticity_introduction} and~\ref{Non_degenerate_mass_condition} are satisfied, are zero in all inelastic processes.

\subsection{Cutting bubble integrals}
\label{section_to_explain_the_cut_method}

We start describing the method used to compute amplitudes by considering a simple example of a one-to-one process.
We define
$$
\omega_a(k) \equiv \sqrt{k_1^2 + m_a^2}
$$
to be the energy of an on-shell particle with mass $m_a$ and propagating with spatial momentum $k_1$. 
Keeping into account the following distribution relation
$$
\frac{1}{k_0-\omega_a(k)+i\epsilon}-\frac{1}{k_0-\omega_a(k)-i\epsilon}=-2\pi i \delta(k_0-\omega_a(k)),
$$
the Feynman propagator associated with the particle $a$ can be written as  
\begin{equation}
\label{splitting_propagators_into_retarded_part_and_delta_function}
\begin{split}
\Pi_a (k) &\equiv \frac{i}{k^2-m_a^2 + i \epsilon}= \frac{i}{2\omega_a(k)} \biggl( \frac{1}{k_0-\omega_a(k)+i\epsilon}-\frac{1}{k_0+\omega_a(k)-i\epsilon} \biggr)\\
&=\Pi_a^{(R)} (k) + \frac{\pi}{\omega_a(k)} \delta(k_0 - \omega_a(k))
\end{split}
\end{equation}
where
\begin{equation}
\label{definition_of_the_retarded_propagator}
\Pi_a^{(R)} (k) \equiv \frac{i}{k^2-m_a^2- i k_0 \ \epsilon}
\end{equation}
is called the `retarded propagator'.
The important point is that, while each Feynman propagator associated with a particle of type $a$ contains two poles at $k_0=\pm \omega_a (k) \mp i \epsilon$, one above and the other below the real axis of the $k_0$ complex plane, in the case of a retarded propagator both the poles lie on the upper half-plane and are located at $k_0=\pm \omega_a (k) + i \epsilon$. This implies that a loop integral involving only retarded propagators, in which all the loop momenta flow in the same direction, is equal to zero; indeed all the poles are contained in the same half-plane and it is possible to choose a contour not containing any poles.
As an example, let us consider a one-loop two-point process in which a particle of type $a$ changes its flavour transforming into a possibly-different particle $b$ with the same mass (in such a way that the event is kinematically allowed):
\begin{equation}
\label{one_to_one_flavour_changing_process}
    a(p) \to b(p)\hspace{5mm}\text{with}\hspace{5mm} p^2=m_a^2=m_b^2.
\end{equation}
We focus on the bubble diagram on the l.h.s. of the first row of figure \ref{Image_changing_flavour_one_loop_cuts}. 
\begin{figure}
\begin{center}
\begin{tikzpicture}
\tikzmath{\y=1.9;}

\draw[directed] (-0.6*\y,0.1*\y) -- (-0.1*\y,0.1*\y);
\draw[directed] (0.3*\y,0.1*\y) arc(0:180:0.2*\y);
\draw[directed] (-0.1*\y,0.1*\y) arc(180:360:0.2*\y);
\draw[directed] (0.3*\y,0.1*\y) -- (0.8*\y,0.1*\y);
\filldraw[black] (-0.8*\y,0.2*\y)  node[anchor=west] {\scriptsize{$a(p)$}};
\filldraw[black] (0.8*\y,0.2*\y)  node[anchor=west] {\scriptsize{$b(p)$}};
\filldraw[black] (0*\y,0.4*\y)  node[anchor=west] {\scriptsize{$c(k)$}};
\filldraw[black] (-0.1*\y,-0.25*\y)  node[anchor=west] {\scriptsize{$d(p+k)$}};

\draw[directed] (4.6*\y-2.4*\y,0*\y) -- (5.1*\y-2.4*\y,0*\y);
\draw[] (5.1*\y-2.4*\y,0*\y) -- (5.1*\y-2.4*\y,0.5*\y);
\draw[directed] (5.1*\y-2.4*\y,0.5*\y) arc(-90:270:0.2*\y);
\draw[directed] (5.1*\y-2.4*\y,0*\y) -- (5.6*\y-2.4*\y,0*\y);
\filldraw[black] (4.4*\y-2.4*\y,0.1*\y)  node[anchor=west] {\scriptsize{$a(p)$}};
\filldraw[black] (5.6*\y-2.4*\y,0.1*\y)  node[anchor=west] {\scriptsize{$b(p)$}};
\filldraw[black] (5*\y-2.4*\y,1*\y)  node[anchor=west] {\scriptsize{$c(k)$}};

\draw[directed] (2.2*\y+2.4*\y,0*\y) -- (2.7*\y+2.4*\y,0*\y);
\draw[directed] (2.7*\y+2.4*\y,0*\y) arc(-90:270:0.2*\y);
\draw[directed] (2.7*\y+2.4*\y,0*\y) -- (3.2*\y+2.4*\y,0*\y);
\filldraw[black] (2*\y+2.4*\y,0.1*\y)  node[anchor=west] {\scriptsize{$a(p)$}};
\filldraw[black] (3.2*\y+2.4*\y,0.1*\y)  node[anchor=west] {\scriptsize{$b(p)$}};
\filldraw[black] (2.6*\y+2.4*\y,0.5*\y)  node[anchor=west] {\scriptsize{$c(k)$}};

\draw[directed] (-0.6*\y-0.8*\y,0.1*\y-1.4*\y) -- (-0.1*\y-0.8*\y,0.1*\y-1.4*\y);
\draw[directed] (0.3*\y-0.8*\y,0.1*\y-1.4*\y) arc(0:180:0.2*\y);
\draw[directed] (0.3*\y-0.8*\y,0.1*\y-1.4*\y) -- (0.8*\y-0.8*\y,0.1*\y-1.4*\y);
\draw[directed] (-0.1*\y-0.8*\y,0.1*\y-1.4*\y) -- (0.*\y-0.8*\y,-0.2*\y-1.4*\y);
\draw[directed] (0.2*\y-0.8*\y,-0.2*\y-1.4*\y) -- (0.3*\y-0.8*\y,0.1*\y-1.4*\y);

\draw[red][dashed] (0.1*\y-0.8*\y,0*\y-1.4*\y) -- (0.1*\y-0.8*\y,-0.4*\y-1.4*\y);
\filldraw[black] (-0.6*\y-0.8*\y,0.3*\y-1.4*\y)  node[anchor=west] {\scriptsize{$a$}};
\filldraw[black] (0.6*\y-0.8*\y,0.3*\y-1.4*\y)  node[anchor=west] {\scriptsize{$b$}};
\filldraw[black] (0*\y-0.8*\y,-0.3*\y-0.7*\y)  node[anchor=west] {\scriptsize{$c$}};
\filldraw[black] (0.2*\y-0.8*\y,-0.3*\y-1.4*\y)  node[anchor=west] {\scriptsize{$d$}};
\filldraw[black] (-0.2*\y-0.8*\y,-0.3*\y-1.4*\y)  node[anchor=west] {\scriptsize{$d$}};

\draw[directed] (-0.6*\y+0.8*\y,0.1*\y-1.4*\y) -- (-0.1*\y+0.8*\y,0.1*\y-1.4*\y);
\draw[directed] (-0.1*\y+0.8*\y,0.1*\y-1.4*\y) arc(180:360:0.2*\y);
\draw[directed] (0.3*\y+0.8*\y,0.1*\y-1.4*\y) -- (0.8*\y+0.8*\y,0.1*\y-1.4*\y);
\draw[directed] (0.*\y+0.8*\y,0.4*\y-1.4*\y) -- (-0.1*\y+0.8*\y,0.1*\y-1.4*\y);
\draw[directed] (0.3*\y+0.8*\y,0.1*\y-1.4*\y) -- (0.2*\y+0.8*\y,0.4*\y-1.4*\y);

\draw[red][dashed] (0.1*\y+0.8*\y,0.6*\y-1.4*\y) -- (0.1*\y+0.8*\y,0.2*\y-1.4*\y);
\filldraw[black] (1*\y-0.8*\y,0.3*\y-1.4*\y)  node[anchor=west] {\scriptsize{$a$}};
\filldraw[black] (0.6*\y+0.8*\y,0.3*\y-1.4*\y)  node[anchor=west] {\scriptsize{$b$}};
\filldraw[black] (1.4*\y-0.8*\y,0.5*\y-1.4*\y)  node[anchor=west] {\scriptsize{$c$}};
\filldraw[black] (1.8*\y-0.8*\y,0.5*\y-1.4*\y)  node[anchor=west] {\scriptsize{$c$}};
\filldraw[black] (1.6*\y-0.8*\y,-0.2*\y-1.4*\y)  node[anchor=west] {\scriptsize{$d$}};

\draw[directed] (4.6*\y-2.4*\y,0*\y-1.4*\y) -- (5.1*\y-2.4*\y,0*\y-1.4*\y);
\draw[] (5.1*\y-2.4*\y,0*\y-1.4*\y) -- (5.1*\y-2.4*\y,0.5*\y-1.4*\y);
\draw[directed] (5.1*\y-2.55*\y,0.5*\y-1.05*\y) arc(130:270:0.2*\y);
\draw[directed] (5.1*\y-2.4*\y,0.5*\y-1.4*\y) arc(270:410:0.2*\y);
\draw[directed] (5.1*\y-2.4*\y,0*\y-1.4*\y) -- (5.6*\y-2.4*\y,0*\y-1.4*\y);

\draw[red][dashed] (5.1*\y-2.4*\y,1.1*\y-1.4*\y) -- (5.1*\y-2.4*\y,0.7*\y-1.4*\y);
\filldraw[black] (4.4*\y-2.4*\y,0*\y-1.4*\y)  node[anchor=west] {\scriptsize{$a$}};
\filldraw[black] (5.6*\y-2.4*\y,0*\y-1.4*\y)  node[anchor=west] {\scriptsize{$b$}};
\filldraw[black] (4.8*\y-2.4*\y,0.9*\y-1.4*\y)  node[anchor=west] {\scriptsize{$c$}};
\filldraw[black] (5.2*\y-2.4*\y,0.9*\y-1.4*\y)  node[anchor=west] {\scriptsize{$c$}};

\draw[directed] (2.2*\y+2.4*\y,0*\y-1.4*\y) -- (2.7*\y+2.4*\y,0*\y-1.4*\y);
\draw[directed] (2.7*\y+2.25*\y,0*\y-1.05*\y) arc(130:270:0.2*\y);
\draw[directed] (2.7*\y+2.4*\y,0*\y-1.4*\y) arc(270:410:0.2*\y);
\draw[directed] (2.7*\y+2.4*\y,0*\y-1.4*\y) -- (3.2*\y+2.4*\y,0*\y-1.4*\y);

\draw[red][dashed] (2.7*\y+2.4*\y,0.6*\y-1.4*\y) -- (2.7*\y+2.4*\y,0.2*\y-1.4*\y);
\filldraw[black] (2*\y+2.4*\y,0*\y-1.4*\y)  node[anchor=west] {\scriptsize{$a$}};
\filldraw[black] (3.2*\y+2.4*\y,0*\y-1.4*\y)  node[anchor=west] {\scriptsize{$b$}};
\filldraw[black] (2.4*\y+2.4*\y,0.4*\y-1.4*\y)  node[anchor=west] {\scriptsize{$c$}};
\filldraw[black] (2.8*\y+2.4*\y,0.4*\y-1.4*\y)  node[anchor=west] {\scriptsize{$c$}};

\end{tikzpicture}
\caption{Two-point one-loop Feynman diagrams (on the first line) and associated cuts (on the second line). The sum over one-loop Feynman diagrams becomes a sum of integrals of on-shell tree-level amplitudes. In the second row the $t$-, $s$-, $u$-channels and the $4$-point vertex contributions are reported from left to right.}
\label{Image_changing_flavour_one_loop_cuts}
\end{center}
\end{figure}
Using the notation \eqref{splitting_propagators_into_retarded_part_and_delta_function}, the diagram, omitting contributions from vertices, can be written as
\begin{equation}
\label{first_example_of_propagators_decomposition_inside_the_bubble}
   \int \frac{d^2 k}{(2 \pi)^2} \Bigl[ \Pi_c^{(R)} (k) + \frac{\pi}{\omega_c(k)} \delta(k_0 - \omega_c(k)) \Bigr] \Bigl[ \Pi_d^{(R)} (k+p) + \frac{\pi}{\omega_d(k+p)} \delta(k_0+p_0 - \omega_d(k+p)) \Bigr]
\end{equation}
where we have associated the loop-momentum $k$ with the $c$-particle propagating in the diagram.
We assume that the external particles are on-shell and physical; in this case, if all the particles are asymptotically stable at the tree level, only two among the four terms arising by computing the product in the equation above survive. Indeed, for physical values of the external momenta, it is never possible that both the $c$- and $d$-particle flowing in the loop are simultaneously mass-shell or the decay $d \to a+c$ would be allowed at the tree level and condition~\ref{Condition_tree_level_elasticity_introduction} would be violated. This implies that the product of the two delta functions in~\eqref{first_example_of_propagators_decomposition_inside_the_bubble} is equal to zero. The same is true for the product of the two retarded propagators: as explained previously they both have poles in the same half of the $k_0$ complex plane and therefore their integration does not contribute. The only surviving terms are given by
$$
 \int \frac{d^2 k}{(2 \pi)^2} \Bigl[ \Pi_c^{(R)} (k)  \frac{\pi}{\omega_d(k+p)} \delta(k_0+p_0 - \omega_d(k+p))  + \Pi_d^{(R)} (k + p)   \frac{\pi}{\omega_c(k)} \delta(k_0 - \omega_c(k))\Bigr].
$$
Each delta function constrains one particle propagating internally to the loop to be on-shell and transforms the sum of loops into the sum of integrals of tree-level amplitudes. Summing over all the possible propagators $c$ and $d$ corresponds to summing over different tree-level processes and all the possible forward and crossed channels entering the processes. If we insert also the remaining tadpole diagrams in figure \ref{Image_changing_flavour_one_loop_cuts} and we add factors coming from vertices, we obtain that the one-loop amplitude associated with the process~\eqref{one_to_one_flavour_changing_process} is
\begin{equation}
\label{changing_flavour_process_at_one_loop}
\frac{1}{8 \pi} \sum_{c=1}^r \int_{-\kappa}^{+\kappa} d\theta_k \ M^{(0)}_{a c \to b c}(p, k, p, k),
\end{equation}
where we labelled by $\theta_k$ the rapidity of particle $c$. Moreover, we used the change of integration variable $\frac{d k_1}{\omega_c(k)} = d \theta_k$.
A regulator $\kappa$ has been introduced in~\eqref{changing_flavour_process_at_one_loop} since in principle the integral can be divergent: this is the case if $M^{(0)}_{a c \to b c}$ is  non-zero for $\theta_{k} \to \pm \infty$. These non-zero contributions in the tree-level amplitude $M^{(0)}_{a c \to b c}$ are introduced by the tadpole diagrams in figure~\ref{Image_changing_flavour_one_loop_cuts}.
Before taking the limit $\kappa \to +\infty$, we should take care of removing possible divergences by adding suitable counterterms to the Lagrangian~\eqref{eq0_1}.
As already mentioned at the beginning of section~\ref{section_to_explain_the_mass_and_field_renormalization}, it is always possible to introduce suitable counterterms to avoid tadpoles.
After they are introduced we can take the limit $\kappa \to +\infty$ and, following the conventions used in the previous sections, we can write
\begin{equation}
\label{changing_flavour_process_at_one_loop_limit_k_infinity}
M^{(1)}_{a \to b} (p^2) =-i \Sigma_{ab}(p^2)= \frac{1}{8 \pi} \sum_{c=1}^r \int_{-\infty}^{+\infty} d\theta_k \ \hat{M}_{ac\to bc}^{(0)}(p, k, p, k),
\end{equation}
where we defined
\begin{equation}
\label{hat_convention_in_two_to_two_amplitude_tree_level}
    \hat{M}_{ac\to bc}^{(0)}(p, k, p, k) \equiv M_{ac\to bc}^{(0)}(p, k, p, k) -\lim_{\theta_{\tilde{k}} \to \infty}M_{ac\to bc}^{(0)}(p, \tilde{k}, p, \tilde{k}).
\end{equation}
We should remark that the integrand in~\eqref{changing_flavour_process_at_one_loop_limit_k_infinity} is not strictly speaking a tree-level amplitude since it contains retarded propagators instead of usual Feynman propagators. However, since the integrand never encounters poles on the integration line, in this case, there is no difference between the two types of propagators. 

Since the integrand in~\eqref{changing_flavour_process_at_one_loop_limit_k_infinity} is Lorentz invariant then $\Sigma_{ab}(p^2)$ depends only on the squared momentum of the incoming particle $a$. Even though we assumed the external particles on-shell, equation \eqref{changing_flavour_process_at_one_loop_limit_k_infinity} makes sense also if $p^2 \ne m_a^2$. 
In this case, the integrand must be thought of as the amplitude $\hat{M}^{(0)}_{a c \to b c}$ in which the mass $m_a$ (common to the particles $a$ and $b$) is deformed to $\sqrt{p^2}$. This makes sense if the deformation is small enough to avoid the possibility of having double cuts in the bubble diagrams in figure~\ref{Image_changing_flavour_one_loop_cuts}. This means that the particles of types $a$ and $b$, with a deformed mass $\sqrt{p^2}$, still need to be stable at the tree level. 
In this paper, we will consider theories satisfying property~\ref{Non_degenerate_mass_condition}. For this reason, any time $a$ and $b$ are different particles with the same mass, the integrand in~\eqref{changing_flavour_process_at_one_loop_limit_k_infinity} is zero at all values of $p^2$ and $\Sigma_{ab}(p^2)$ is null.
All the coefficients
$$
\Sigma^{(0)}_{ab}, \ \Sigma^{(1)}_{ab}, \dots
$$
of the expansion of $\Sigma_{ab} (p^2)$ around $p^2=m^2_a$ are therefore zero. This explains why equation~\eqref{expansion_of_propagator_different_particles_with_equal_masses} is null at the pole.

For $a=b$ the integrand of~\eqref{changing_flavour_process_at_one_loop_limit_k_infinity} is nonzero and it holds that
\begin{equation}
\label{definition_necessary_for_mass_renormalization_first_time_ga_appears}
\Sigma_{aa}(p^2)= \frac{i}{8 \pi} \sum_{c=1}^r \int_{-\infty}^{+\infty} d\theta_k \ \hat{M}^{(0)}_{a c \to a c}(p, k, p, k).
\end{equation}
At $p^2=m_a^2$, the integrand in~\eqref{definition_necessary_for_mass_renormalization_first_time_ga_appears} becomes a function of the rapidities (since the masses are fixed and on-shell). In this case, we obtain $\delta m_a^2$ expressed as the integral of an on-shell tree-level amplitude
\begin{equation}
\label{mass_renormalization_formula_on_shell}
\delta m_a^2=-\Sigma_{aa}(m_a^2)= -\frac{i}{8 \pi} \sum_{c=1}^r \int_{-\infty}^{+\infty} d\theta_k \ \hat{M}^{(0)}_{a c \to a c}(\theta_p,\theta_k,\theta_p, \theta_k).
\end{equation}

\subsection{Generalisation to two-to-two inelastic processes}
\label{section_to_explain_two_to_two_inelastic_processes}

Before moving to the study of one-loop inelastic processes of type~\eqref{2_to_nminus2_production_process_Introduction} for arbitrary number $n$ of external particles, we focus on the case $n=4$ and we consider the process
\begin{equation}
\label{generic_scattering_process_ab_into_cd}
a(p)+b(p')\to c(q)+d(q').
\end{equation}
We assume $\{a, b\} \ne \{c, d\}$ and we label the momenta carried by the different particles by $p$, $p'$, $q$ and $q'$. Using the cutting method explained previously, the sum of all one-loop Feynman diagrams can be written as
\begin{equation}
\label{sum_of_single_and_double_cut_contributions}
M_{ab\to cd}^{(1)}=M_{ab\to cd}^{(2\text{-cut})}+M_{ab\to cd}^{(1\text{-cut})}.
\end{equation}
The two contributions on the r.h.s. of the equality in~\eqref{sum_of_single_and_double_cut_contributions} are obtained by decomposing each propagator appearing inside a loop into the sum of a Dirac delta function and a retarded propagator, as written in~\eqref{splitting_propagators_into_retarded_part_and_delta_function}. 
The single-cut contribution $M_{ab\to cd}^{(1\text{-cut})}$ corresponds to a sum of terms, each containing a single Dirac delta function. The Dirac delta function removes one among the two integration variables appearing in the loop and similarly to what happened in~\eqref{changing_flavour_process_at_one_loop_limit_k_infinity} we obtain
\begin{equation}
\label{4_point_one_loop_from_tree_single_cut_contribution}
M_{ab\to cd}^{(1\text{-cut})} =\frac{1}{8\pi} \sum_{e=1}^r \int_{-\infty}^{+\infty} d\theta_k \hat{M}_{abe\to cde}^{(0)}(p,p',k,q,q',k).
\end{equation}
As before, the integration~\eqref{4_point_one_loop_from_tree_single_cut_contribution} is performed in the rapidity variable $\theta_k$ of the particle of type $e$ associated with the propagator that has been cut and we define
\begin{multline}
\label{removing_infinite_rapidity_of_particle_e_from_3_to_3_amplitude}
    \hat{M}_{abe\to cde}^{(0)}(p, p', k, q, q', k) \equiv \\ M_{abe\to cde}^{(0)}(p, p', k, q, q', k)- \lim_{\theta_{\tilde{k}} \to \infty}M_{abe\to cde}^{(0)}(p, p',\tilde{k}, q, q',\tilde{k})
\end{multline}
to be the integrand after having removed tadpoles by adding properly-tuned counterterms to the starting Lagrangian. 

Differently from one-to-one amplitudes, when we perform the substitution~\eqref{splitting_propagators_into_retarded_part_and_delta_function} in two-to-two one-loop diagrams certain terms containing two Dirac delta functions survive. These terms contribute to the amplitude through
\begin{equation}
\label{4_point_one_loop_from_tree_double_cut_contribution}
M_{ab\to cd}^{(2\text{-cut})} =M_{ab\to cd}^{(2\text{-cut},u)}+M_{ab\to cd}^{(2\text{-cut},t)}=\sum_{e,f=1}^r \frac{M_{ae\to cf}^{(0)}M_{bf\to de}^{(0)}}{8 m_e m_f |\sinh{\theta_{ef}}|} +\sum_{e,f=1}^r \frac{M_{ae\to df}^{(0)}M_{bf\to ce}^{(0)}}{8 m_e m_f |\sinh{\theta_{ef}}|}.
\end{equation}
In~\eqref{4_point_one_loop_from_tree_double_cut_contribution} we defined $\theta_{ef}\equiv \theta_e-\theta_f$ to be the difference between the rapidities of the particles associated to the cut propagators.
It is important to note that among the three different channels depicted in figure~\ref{All_possible_double_cuts_ab_into_cd} where the double-cuts can be performed, only the $u$- and $t$-channel generate a nonzero result. The cut in the $s$-channel, depicted in the centre of figure~\ref{All_possible_double_cuts_ab_into_cd} is zero. 
Indeed we are assuming that all the particles are stable at the tree level and is impossible that three on-shell particles, $a$, $b$ and $e$, with physical momenta, fuse to generate a physical on-shell particle $f$. This is why the double-cut contribution associated with the $s$-channel is missing in~\eqref{4_point_one_loop_from_tree_double_cut_contribution}.

We remark that the contributions on the r.h.s. of expressions~\eqref{4_point_one_loop_from_tree_single_cut_contribution} and~\eqref{4_point_one_loop_from_tree_double_cut_contribution} are tree-level amplitudes in which the propagators containing loop momentum variables
are of retarded type. 
\begin{figure}
\begin{center}
\begin{tikzpicture}
\tikzmath{\y=1;}


\draw[directed] (-7.4*\y,0.5*\y) -- (-6.8*\y,1.1*\y);
\draw[directed] (-6.8*\y,1.8*\y) -- (-7.4*\y,2.4*\y);
\draw[directed] (-5.5*\y,2.1*\y) -- (-6*\y,1.7*\y);
\draw[directed] (-4.6*\y,1.7*\y) -- (-5.1*\y,2.1*\y);
\draw[directed] (-6.2*\y,1.2*\y) -- (-5.8*\y,0.7*\y);
\draw[directed] (-5*\y,0.7*\y) -- (-4.6*\y,1.2*\y);
\draw[directed] (-3.2*\y,0.6*\y) -- (-3.8*\y,1.2*\y);
\draw[directed] (-3.8*\y,1.8*\y) -- (-3.2*\y,2.4*\y);

\filldraw[color=gray!60, fill=gray!5, very thick](-6.4*\y,1.5*\y) circle (0.4*\y);
\filldraw[color=gray!60, fill=gray!5, very thick](-4.2*\y,1.5*\y) circle (0.4*\y);
\filldraw[black] (-7.5*\y,0.3*\y)  node[anchor=west] {\tiny{$a(p)$}};
\filldraw[black] (-7.5*\y,2.5*\y)  node[anchor=west] {\tiny{$c(q)$}};
\filldraw[black] (-3.6*\y,0.3*\y)  node[anchor=west] {\tiny{$b(p')$}};
\filldraw[black] (-3.6*\y,2.5*\y)  node[anchor=west] {\tiny{$d(q')$}};
\filldraw[black] (-5.85*\y,2.2*\y)  node[anchor=west] {\tiny{$e$}};
\filldraw[black] (-5.25*\y,2.2*\y)  node[anchor=west] {\tiny{$e$}};
\filldraw[black] (-6*\y,0.5*\y)  node[anchor=west] {\tiny{$f$}};
\filldraw[black] (-5.3*\y,0.5*\y)  node[anchor=west] {\tiny{$f$}};

\filldraw[black] (-6.2*\y,-0.3*\y)  node[anchor=west] {\scriptsize{Allowed}};


\draw[directed] (-1.4*\y,0.5*\y) -- (-0.8*\y,1.1*\y);
\draw[directed] (-1.4*\y,2.4*\y) -- (-0.8*\y,1.8*\y);
\draw[directed] (0.5*\y,2.1*\y) -- (0*\y,1.7*\y);
\draw[directed] (1.4*\y,1.7*\y) -- (0.9*\y,2.1*\y);
\draw[directed] (-0.2*\y,1.2*\y) -- (0.2*\y,0.7*\y);
\draw[directed] (1*\y,0.7*\y) -- (1.4*\y,1.2*\y);
\draw[directed] (2.2*\y,1.2*\y) -- (2.8*\y,0.6*\y);
\draw[directed] (2.2*\y,1.8*\y) -- (2.8*\y,2.4*\y);

\filldraw[color=gray!60, fill=gray!5, very thick](-0.4*\y,1.5*\y) circle (0.4*\y);
\filldraw[color=gray!60, fill=gray!5, very thick](1.8*\y,1.5*\y) circle (0.4*\y);
\filldraw[black] (-1.5*\y,0.3*\y)  node[anchor=west] {\tiny{$a(p)$}};
\filldraw[black] (-1.5*\y,2.5*\y)  node[anchor=west] {\tiny{$b(p')$}};
\filldraw[black] (2.4*\y,0.3*\y)  node[anchor=west] {\tiny{$c(q)$}};
\filldraw[black] (2.4*\y,2.5*\y)  node[anchor=west] {\tiny{$d(q')$}};
\filldraw[black] (0.15*\y,2.2*\y)  node[anchor=west] {\tiny{$e$}};
\filldraw[black] (0.75*\y,2.2*\y)  node[anchor=west] {\tiny{$e$}};
\filldraw[black] (0*\y,0.5*\y)  node[anchor=west] {\tiny{$f$}};
\filldraw[black] (0.7*\y,0.5*\y)  node[anchor=west] {\tiny{$f$}};

\filldraw[black] (-0.2*\y,-0.3*\y)  node[anchor=west] {\scriptsize{Non-allowed}};


\draw[directed] (4.6*\y,0.5*\y) -- (5.2*\y,1.1*\y);
\draw[directed] (5.2*\y,1.8*\y) -- (4.6*\y,2.4*\y);
\draw[directed] (6.5*\y,2.1*\y) -- (6*\y,1.7*\y);
\draw[directed] (7.4*\y,1.7*\y) -- (6.9*\y,2.1*\y);
\draw[directed] (5.8*\y,1.2*\y) -- (6.2*\y,0.7*\y);
\draw[directed] (7*\y,0.7*\y) -- (7.4*\y,1.2*\y);
\draw[directed] (8.8*\y,0.6*\y) -- (8.2*\y,1.2*\y);
\draw[directed] (8.2*\y,1.8*\y) -- (8.8*\y,2.4*\y);

\filldraw[color=gray!60, fill=gray!5, very thick](5.6*\y,1.5*\y) circle (0.4*\y);
\filldraw[color=gray!60, fill=gray!5, very thick](7.8*\y,1.5*\y) circle (0.4*\y);
\filldraw[black] (4.5*\y,0.3*\y)  node[anchor=west] {\tiny{$a(p)$}};
\filldraw[black] (4.5*\y,2.5*\y)  node[anchor=west] {\tiny{$d(q')$}};
\filldraw[black] (8.4*\y,0.3*\y)  node[anchor=west] {\tiny{$b(p')$}};
\filldraw[black] (8.4*\y,2.5*\y)  node[anchor=west] {\tiny{$c(q)$}};
\filldraw[black] (6.15*\y,2.2*\y)  node[anchor=west] {\tiny{$e$}};
\filldraw[black] (6.75*\y,2.2*\y)  node[anchor=west] {\tiny{$e$}};
\filldraw[black] (6*\y,0.5*\y)  node[anchor=west] {\tiny{$f$}};
\filldraw[black] (6.7*\y,0.5*\y)  node[anchor=west] {\tiny{$f$}};

\filldraw[black] (5.8*\y,-0.3*\y)  node[anchor=west] {\scriptsize{Allowed}};

\end{tikzpicture}
\caption{Double cut in the $u$-, $s$- and $t$-channel.}
\label{All_possible_double_cuts_ab_into_cd}
\end{center}
\end{figure}
Equations~\eqref{sum_of_single_and_double_cut_contributions}, \eqref{4_point_one_loop_from_tree_single_cut_contribution} and~\eqref{4_point_one_loop_from_tree_double_cut_contribution}  are valid for any bosonic theory with stable particles at the tree level. 
In the following, we will show how the assumption that the tree-level amplitudes are purely elastic can be used to simplify further these equations.

From formula~\eqref{4_point_one_loop_from_tree_double_cut_contribution} we see that double cuts never contribute to inelastic processes if condition~\ref{Condition_tree_level_elasticity_introduction} is satisfied.
Indeed, if $\{a,b\}\ne \{c,d\}$, each term in the sum on the r.h.s. of~\eqref{4_point_one_loop_from_tree_double_cut_contribution} contains at least one inelastic two-to-two amplitude and \eqref{4_point_one_loop_from_tree_double_cut_contribution} is therefore zero. 
However, particular care should be taken to the degenerate situation in which $m_a=m_c$ and $m_b=m_d$ (the case $\{m_a=m_d, m_b=m_c\}$ is analogous). 
In this situation all the terms with $m_e=m_f$ in the sum associated with the $u$-channel cut on the r.h.s. of~\eqref{4_point_one_loop_from_tree_double_cut_contribution} are ill-defined.
If $\{m_a=m_c, m_b = m_d \}$ one of the two branches of the solution satisfying the overall energy-momentum conservation is $\theta_p=\theta_{q}$ and $\theta_{p'}=\theta_{q'}$. In this case the energy-momentum conservation applied to $M^{(0)}_{ae\to cf}$ and $M^{(0)}_{bf\to de}$ leads to $\theta_e=\theta_f$ and does not fix the values of these rapidities. 
Moreover, for $\theta_e=\theta_f$, $|\sinh{\theta_{ef}}|= 0$ and the $u$-channel cut in~\eqref{4_point_one_loop_from_tree_double_cut_contribution} becomes divergent.
These ill-defined contributions are a consequence of the cutting procedure adopted to compute amplitudes and are not real ill-defined terms that require to be cancelled through renormalization. For this reason, we expect that these contributions can be avoided or need to cancel similar ill-defined terms in~\eqref{4_point_one_loop_from_tree_single_cut_contribution}. Similar contributions were encountered also in~\cite{Bianchi:2014rfa} in performing unitarity cuts on elastic processes.
To handle these contributions we define
\begin{equation}
\label{definition_of_my_regulator_x}
x=\mu(e^{\theta_x},e^{-\theta_x})
\end{equation}
to be a small vector written in light-cone components~\eqref{light_cone_components_PhD_thesis}, such that in the process~\eqref{generic_scattering_process_ab_into_cd} (with $m_a=m_c$ and $m_b=m_d$) we have
\begin{equation}
\label{regularization_through_x_introducing_outgoing_mass_deformations}
q-p=p'-q'=x.
\end{equation}
In~\eqref{definition_of_my_regulator_x} we consider $\mu>0$ and $\theta_x \in \mathbb{R}$.
We assume $p$ and $p'$ to be the on-shell momenta of the incoming particles, satisfying $p^2=m^2_a$ and $p'^2=m^2_b$.
The vectors $q=p+x$ and $q'=p'-x$ are off-shell and become on-shell only in the limit $\mu \to 0$. 
With the regulator $x$, all the terms in the sum associated with the $u$-channel cut in~\eqref{4_point_one_loop_from_tree_double_cut_contribution}, with $m_e=m_f$, become well defined.
Labelling 
\begin{equation}
 p_e=\tilde{m} (e^{\theta_e},e^{-\theta_e}) \hspace{3mm} \text{and} \hspace{3mm} p_f=\tilde{m} (e^{\theta_f},e^{-\theta_f})
\end{equation}
the momenta of the particles $e$ and $f$ (written in light-cone components) having common mass $\tilde{m}$, it holds that
\begin{equation}
\label{difference_between_hatk_and_tildek_returning_x}
p_e-p_f=x.
\end{equation}
At this point it is immediate to verify that~\eqref{difference_between_hatk_and_tildek_returning_x} cannot be verified for $\mu>0$ and $\theta_x \in \mathbb{R}$.
Indeed relation~\eqref{difference_between_hatk_and_tildek_returning_x} is equivalent to a decay process in which a particle of type $e$ decays into a particle $f$ of the same mass plus a particle $x$ of small (but finite) mass $\mu$. If we go in the rest frame of $e$, we immediately see that 
this process cannot be realized. This implies that, if we turn on a small positive parameter $\mu$, all the ill-defined contributions in~\eqref{4_point_one_loop_from_tree_double_cut_contribution} do not appear by kinematical reasons. 
The final amplitude, obtained by summing over all Feynman diagrams and counterterms, is expected to be an analytic function of the masses and rapidities of the external particles.
For this reason, it is expected to be smooth in the limit $\mu \to 0$ and no discontinuity should appear. Therefore, all contributions in~\eqref{4_point_one_loop_from_tree_double_cut_contribution} which are ill-defined when $\mu=0$ have to be omitted since they are null for $\mu>0$ and no discontinuity is expected at $\mu = 0$. 

To obtain a consistent result for the final amplitude, the same limit has to be taken also to remove possible ill-defined contributions appearing in~\eqref{4_point_one_loop_from_tree_single_cut_contribution}. We will discuss these contributions in appendix~\ref{Potential_ill_defined_contributions_in_single_cuts}. 
If condition~\ref{Condition_tree_level_elasticity_introduction} is satisfied, we can therefore conclude that
\begin{equation}
M_{ab\to cd}^{(2\text{-cut})}=0.
\end{equation}
It is tempting to say that also~\eqref{4_point_one_loop_from_tree_single_cut_contribution} is zero by condition~\ref{Condition_tree_level_elasticity_introduction} since the integrand is a tree-level inelastic amplitude. Actually, the problem is more subtle than that and this guess turns out to be wrong, as we will show in the next sections.

\subsection{On-shell limits in tree-level inelastic amplitudes}
\label{section_on_properties_of_tree_level_inelastic_amplitudes}

Before evaluating the single-cut contribution~\eqref{4_point_one_loop_from_tree_single_cut_contribution} it is important to understand the properties of the integrands appearing in~\eqref{4_point_one_loop_from_tree_single_cut_contribution}. These are tree-level amplitudes associated with processes of the following type
\begin{equation}
\label{3_to_3_tree_level_process_abe_into_cde}
    a(p)+b(p')+e(k) \to c(\tilde{q})+d(\tilde{q}')+e(\tilde{k}).
\end{equation}
A way to properly define a tree-level amplitude associated with the process~\eqref{3_to_3_tree_level_process_abe_into_cde} is to take the following \textit{on-shell limit}: we keep
$p$, $p'$ and $k$ fixed and on-shell and move $\tilde{k}$, which also has to be on-shell. The on-shell momenta $\tilde{q}$ and $\tilde{q}'$ are then determined in terms of $p$, $p'$, $k$ and $\tilde{k}$ by the overall energy-momentum conservation\footnote{More correctly we should say that there are 
two branches of solutions for $\tilde{q}$ and $\tilde{q}'$; however our discussion applies equally to both the branches.}. Keeping all the particles on-shell, we take the limit $\tilde{k} \to k$, for which $\tilde{q} \to q$ and $\tilde{q}' \to q'$.
We define the tree-level amplitude obtained by moving along this limit by $M_{abe\to cde}^{(0,\text{on})}$ and it has to be zero by 
condition~\ref{Condition_tree_level_elasticity_introduction}. We split this zero amplitude into the sum of two terms
\begin{equation}
\label{on_shell_limit_of_a_3_to_3_non_elastic_amplitude}
    M_{abe\to cde}^{(0,\text{on})}=V_{abe \to cde}^{(\text{on})}+R_{abe \to cde}=0.
\end{equation}
The superscript word `on' in the terms above means
that we moved to the configuration $\tilde{k}=k$ keeping all the external particles on-shell along the limit.
In $V_{abe \to cde}^{(\text{on})}$ are contained all Feynman diagrams in which the two particles of type $e$ cross one of the other legs: $a$, $b$, $c$ or $d$. 
A particular collection of these Feynman diagrams, where the two legs of type $e$ cross leg $a$, is shown in figure~\ref{loop_on_external_leg_a_in_non_elastic_process_off_shell_limit}; the two blobs in the figure correspond to tree-level amplitudes.
\begin{figure}
\begin{center}
\begin{tikzpicture}
\tikzmath{\y=1.4;}

\draw[directed] (4.9*\y,2.2*\y) -- (5.5*\y,1.6*\y);
\draw[directed] (5.7*\y,1.4*\y) -- (6.3*\y,0.8*\y);
\draw[directed] (6.3*\y,2*\y) -- (5.8*\y,1.6*\y);
\draw[directed] (5.5*\y,1.4*\y) -- (5*\y,1*\y);
\filldraw[color=gray!60, fill=gray!5, very thick](5.6*\y,1.5*\y) circle (0.25*\y);
\draw[directed] (6.6*\y,0.3*\y) -- (6.8*\y,-0.4*\y);
\draw[directed] (5.9*\y,-0.3*\y) -- (6.3*\y,0.4*\y);
\draw[directed] (6.5*\y,0.7*\y) -- (7.3*\y,1.4*\y);
\filldraw[color=gray!60, fill=gray!5, very thick](6.5*\y,0.6*\y) circle (0.4*\y);

\filldraw[black] (4.6*\y,2.5*\y)  node[anchor=west] {\scriptsize{$a(p)$}};
\filldraw[black] (5.5*\y,-0.5*\y)  node[anchor=west] {\scriptsize{$b(p')$}};
\filldraw[black] (6.6*\y,-0.5*\y)  node[anchor=west] {\scriptsize{$c(\tilde{q})$}};
\filldraw[black] (7.3*\y,1.4*\y)  node[anchor=west] {\scriptsize{$d(\tilde{q}')$}};
\filldraw[black] (6.15*\y,2.2*\y)  node[anchor=west] {\scriptsize{$e(k)$}};
\filldraw[black] (4.5*\y,1*\y)  node[anchor=west] {\scriptsize{$e(\tilde{k})$}};
\filldraw[black] (5.9*\y,1.3*\y)  node[anchor=west] {\scriptsize{$a$}};

\end{tikzpicture}
\caption{Combination of Feynman diagrams contributing to $V^{(\text{on})}_{abe \to cd e}$ in which the two legs of type $e$ cross the leg $a$.}
\label{loop_on_external_leg_a_in_non_elastic_process_off_shell_limit}
\end{center}
\end{figure}
Other contributions to $V_{abe \to cde}^{(\text{on})}$ come from diagrams in which the two particles of type $e$ intersect the legs $b$, $c$ and $d$. In $R_{abe \to cde}$ are contained all the other Feynman diagrams.

It may look that $V_{abe \to cde}^{(\text{on})}$ is ill-defined in the limit $\tilde{k} \to k$ since potential singularities can arise. In this limit the propagator of type $a$ connecting the two blobs in figure~\ref{loop_on_external_leg_a_in_non_elastic_process_off_shell_limit} is singular; however, as the intermediate propagating particle connecting the two blobs becomes on-shell, the blob on the r.h.s. becomes zero. Indeed, it is an inelastic tree-level amplitude and has to be null by condition~\ref{Condition_tree_level_elasticity_introduction} any time it is evaluated on-shell.
A similar argument can be repeated for any potential pole appearing in the tree-level amplitude associated with the process~\eqref{3_to_3_tree_level_process_abe_into_cde}: 
after having found an analytic solution for the amplitude by taking a well-defined choice of the kinematics, we take the limit to the poles and we discover that the residues at the poles are always zero. This has to be true because the theory is purely elastic at the tree level. For this reason, the amplitude is well defined in the limit $\tilde{k} \to k$. Below we compute the contribution $V_{abe \to cde}^{(\text{on})}$.

As already mentioned, $V_{abe \to cde}^{(\text{on})}$ contains sums of Feynman diagrams in which the two particles $e(k)$ and $e(\tilde{k})$ are attached to one of the four other external legs $a$, $b$, $c$ and $d$. If we sum all these contributions we obtain
\begin{multline}
\label{V_on_shell_external_leg_corrections}
     V_{abe \to cde}^{(\text{on})}= \lim_{\theta_{\tilde{k}} \to \theta_k} \Bigl(\frac{i M^{(0)}_{ae \to ae}}{(p+k-\tilde{k})^2-m^2_a}+\frac{i M^{(0)}_{be \to be}}{(p'+k-\tilde{k})^2-m^2_b} \\
     +\frac{i M^{(0)}_{ce \to ce}}{(q+\tilde{k}-k)^2-m^2_c}+\frac{i M^{(0)}_{de \to de}}{(q'+\tilde{k}-k)^2-m^2_d} \Bigr) \times M^{(0)}_{ab \to cd}.
\end{multline}
We compute the sum of Feynman diagrams corresponding to the picture in figure~\ref{loop_on_external_leg_a_in_non_elastic_process_off_shell_limit}, which corresponds to multiply $M^{(0)}_{ab \to cd}$ with the first term in the brackets in~\eqref{V_on_shell_external_leg_corrections}.
We parameterize the light-cone components of the momentum of the propagator connecting the two blobs as
$$
(p_a,\bar{p}_a)=\mu_a (e^{\theta_a},e^{-\theta_a}).
$$ 
Then, the contribution associated with the picture is given by 
\begin{equation}
\label{non_expanded_on_shell_combination_at_external_leg_correction_non_elastic_amplitude}
M^{(0)}_{ae \to ae}(p,k,p_a,\tilde{k}) \frac{i}{\mu_a^2-m_a^2} M^{(0)}_{ab\to cd}(p_a,p',\tilde{q},\tilde{q}')
\end{equation}
In the limit in which $\tilde{k}$ becomes equal to $k$, then $p_a\to p$ and therefore $\mu_a \to m_a$ and $\theta_a \to \theta_p$. 
The amplitude $M^{(0)}_{ab\to cd}(p_a,p',\tilde{q},\tilde{q}') $ is completely determined once $\mu_a$ and $\theta_a$ are known since $\tilde{q}$ and $\tilde{q}'$ are fixed in terms of these parameters by requiring the momentum conservation. Therefore, we can expand 
\begin{multline}
\label{expansion_in_rapidity_and_mass_of_non_elastic_on_shell_amplitude}
M^{(0)}_{ab\to cd}(p_a,p',\tilde{q},\tilde{q}') = M^{(0)}_{ab\to cd}(p,p',q,q')+(\theta_a-\theta_p) \frac{\partial}{\partial \theta_a}M^{(0)}_{ab\to cd}(p_a,p',\tilde{q},\tilde{q}')\Bigr|_{\mu_a=m_a,\theta_a=\theta_p} \\
+(\mu_a^2-m_a^2) \frac{\partial}{\partial \mu_a^2}M^{(0)}_{ab\to cd}(p_a,p',\tilde{q},\tilde{q}')\Bigr|_{\mu_a=m_a,\theta_a=\theta_p}.
\end{multline}
The parameters $\mu_a$ and $\theta_a$ are both functions of $\theta_{\tilde{k}}$; however, their explicit dependence is not required for this discussion.
The first two terms on the r.h.s. of~\eqref{expansion_in_rapidity_and_mass_of_non_elastic_on_shell_amplitude} are identically zero since are evaluated at the fixed value $\mu_a=m_a$ where $M^{(0)}_{ab\to cd}=0$. The last term is instead different from zero in general since is evaluated at the value $\mu_a^2 \ne m_a^2$ and the on-shell condition $\mu_a=m_a$ is restored only after having performed the derivative. Therefore, in the limit $\tilde{k}\to k$, we obtain 
\begin{equation}
M^{(0)}_{ab\to cd}(p_a,p',\tilde{q},\tilde{q}') = (\mu_a^2-m_a^2) \frac{\partial}{\partial \mu_a^2}M^{(0)}_{ab\to cd}(p_a,p',q,q')\Bigr|_{\mu_a=m_a,\theta_a=\theta_p}
\end{equation}
and equation~\eqref{non_expanded_on_shell_combination_at_external_leg_correction_non_elastic_amplitude} becomes
\begin{equation}
i M^{(0)}_{ae \to ae}(p,k,p,k)  \frac{\partial}{\partial \mu_a^2}M^{(0)}_{ab\to cd}(p_a,p',q,q')\Bigr|_{\mu_a=m_a,\theta_a=\theta_p}.
\end{equation}
Repeating the same argument for all the terms appearing in~\eqref{V_on_shell_external_leg_corrections} we obtain
\begin{equation}
\label{expression_for_V_on_with_mass_derivatives}
\begin{split}
     V^{(\text{on})}_{abe \to cd e} &= i M^{(0)}_{ae \to ae}(p,k,p,k) \frac{\partial}{\partial m_a^2}M^{(0)}_{ab\to cd}(p,p',q,q')\\
&+i M^{(0)}_{be \to be}(p',k,p',k) \frac{\partial}{\partial m_b^2}M^{(0)}_{ab\to cd}(p,p',q,q')\\
&+i M^{(0)}_{ce \to ce}(q,k,q,k) \frac{\partial}{\partial m_c^2}M^{(0)}_{ab\to cd}(p,p',q,q')\\
&+i M^{(0)}_{de \to de}(q',k,q',k) \frac{\partial}{\partial m_d^2}M^{(0)}_{ab\to cd}(p,p',q,q').
\end{split}
\end{equation}
$M^{(0)}_{ab\to cd}(p,p',q,q')$ has to be thought as a function of $p$, $p'$, $q$ and $q'$ satisfying $p+p'=q+q'$; each of these momenta carries its mass and rapidity. $M^{(0)}_{ab\to cd}(p,p',q,q')$ is therefore a function of six parameters: four masses and the two rapidities of the incoming particles.
Once these six parameters are given, the rapidities of the outgoing particles are determined by the overall energy-momentum conservation. Due to condition~\ref{Condition_tree_level_elasticity_introduction} this function is zero on the surface (of this six-parameter space) at which the masses take their on-shell values. However, it is in general nonzero outside this surface and the derivatives with respect to the squares of the masses in~\eqref{expression_for_V_on_with_mass_derivatives} make sense.

The discussion can be repeated identically if we remove all Feynman diagrams 
which survive in the limit in which the rapidities of the two particles of type $e$ are infinite ($\theta_k=\theta_{\tilde{k}}=\infty$).  These are diagrams in which the two particles of type $e$ are attached to the same vertex.
Since the inelastic amplitude in~\eqref{on_shell_limit_of_a_3_to_3_non_elastic_amplitude} has to be zero at all values of $\theta_k$ and $\theta_{\tilde{k}}$ then it holds that
\begin{equation}
\label{on_shell_limit_of_a_3_to_3_non_elastic_amplitude_infinities_removed}
    \hat{M}_{abe\to cde}^{(0,\text{on})} \equiv M_{abe\to cde}^{(0,\text{on})} - M_{abe\to cde}^{(0,\text{on})}\Bigl|_{\theta_k =\theta_{\tilde{k}}= \infty} =\hat{V}_{abe \to cde}^{(\text{on})}+\hat{R}_{abe \to cde}=0.
\end{equation}
In~\eqref{on_shell_limit_of_a_3_to_3_non_elastic_amplitude_infinities_removed} we defined 
\begin{equation}
     \hat{V}_{abe\to cde}^{(\text{on})}=V_{abe\to cde}^{(\text{on})} - V_{abe\to cde}^{(\text{on})}\Bigl|_{\theta_k =\theta_{\tilde{k}}= \infty}
\end{equation}
and
\begin{equation}
    \hat{R}_{abe\to cde}=R_{abe\to cde} - R_{abe\to cde}\Bigl|_{\theta_k =\theta_{\tilde{k}}= \infty}.
\end{equation}
In this case, we find that
\begin{equation}
\label{definition_of_Vhat_on}
\begin{split}
     \hat{V}^{(\text{on})}_{abe \to cd e} &= i \hat{M}^{(0)}_{ae \to ae}(p,k,p,k) \frac{\partial}{\partial m_a^2}M^{(0)}_{ab\to cd}(p,p',q,q')\\
&+i \hat{M}^{(0)}_{be \to be}(p',k,p',k) \frac{\partial}{\partial m_b^2}M^{(0)}_{ab\to cd}(p,p',q,q')\\
&+i \hat{M}^{(0)}_{ce \to ce}(q,k,q,k) \frac{\partial}{\partial m_c^2}M^{(0)}_{ab\to cd}(p,p',q,q')\\
&+i \hat{M}^{(0)}_{de \to de}(q',k,q',k) \frac{\partial}{\partial m_d^2}M^{(0)}_{ab\to cd}(p,p',q,q')=- \hat{R}_{abe\to cde}.
\end{split}
\end{equation}
Once again the hat on the different two-to-two tree-level amplitudes means that we are removing the contributions at $\theta_k=\infty$ as indicated in~\eqref{hat_convention_in_two_to_two_amplitude_tree_level}.

\subsection{Off-shell limits in tree-level inelastic amplitudes and single-cut contributions}
\label{section_to_explain_single_cut_contributions}

From the argument just presented, one may naively expect that the integrands in~\eqref{4_point_one_loop_from_tree_single_cut_contribution} are all zero.
However, this is not correct.
We should stress that the limit adopted to define the amplitude~\eqref{on_shell_limit_of_a_3_to_3_non_elastic_amplitude_infinities_removed} was an \textit{on-shell limit}: we defined an amplitude at $\tilde{k} \ne k$ and we moved to the configuration $\tilde{k}=k$ keeping all the particles on-shell along the limit. To evaluate the amplitude $\hat{M}_{abe\to cde}^{(0)}$ in~\eqref{4_point_one_loop_from_tree_single_cut_contribution}, we cannot use this limit. Indeed, the two particles $e(k)$ and $e(\tilde{k})$ come from the same propagator that has been cut and therefore it needs to hold that $\tilde{k}=k$ with $k$ on-shell.
In this case, the only possibility to avoid ill-defined Feynman diagrams is to assume the particles $a$, $b$, $c$ and $d$ off-shell. 
If we label the momenta associated with these particles by $\tilde{p}$, $\tilde{p}'$ $\tilde{q}$ and $\tilde{q}'$ then the process we need to study is
$$
a(\tilde{p})+b(\tilde{p}')+e(k) \to c(\tilde{q}) +d(\tilde{q}')+e(k),
$$
where $k$ is the only on-shell momentum. Then, keeping the overall energy and momentum conserved, we need to take the limit $\tilde{p} \to p$, $\tilde{p}' \to p'$, $\tilde{q} \to q$ and $\tilde{q}' \to q'$ at which all the particles become on-shell. Since particles $a$, $b$, $c$ and $d$ are off-shell along the limit we refer to this limit as an \textit{off-shell limit}.

After removing contributions at $\theta_k=\infty$, we can still split the amplitude into
\begin{equation}
\label{off_shell_limit_of_a_3_to_3_non_elastic_amplitude}
    \hat{M}_{abe\to cde}^{(0)}=\hat{V}_{abe \to cde}+\hat{R}_{abe \to cde}.
\end{equation}
For general kinematics, $\hat{R}_{abe \to cde}$ contains Feynman diagrams which are finite and therefore it is the same both in~\eqref{on_shell_limit_of_a_3_to_3_non_elastic_amplitude_infinities_removed} and~\eqref{off_shell_limit_of_a_3_to_3_non_elastic_amplitude} (i.e. it does not depend on the limit adopted to reach the on-shell configuration); on the contrast, $\hat{V}_{abe \to cde}$ in~\eqref{off_shell_limit_of_a_3_to_3_non_elastic_amplitude} is different from $\hat{V}_{abe \to cde}^{(\text{on})}$ appearing in~\eqref{on_shell_limit_of_a_3_to_3_non_elastic_amplitude_infinities_removed}; this is due to the fact that diagrams of the type depicted in figure~\ref{loop_on_external_leg_a_in_non_elastic_process_off_shell_limit} contain singular propagators and their values depend on the limit we follow to reach the on-shell configuration 
\begin{equation}
\label{on_shell_configuration_3_to_3}
    a(p)+b(p')+e(k) \to c(q) +d(q')+e(k).
\end{equation}
Due to this fact, the amplitude in~\eqref{off_shell_limit_of_a_3_to_3_non_elastic_amplitude} is nonzero in general. 

The contribution $\hat{V}_{abe \to cde}$ is given by
\begin{multline}
\label{V_off_shell_external_leg_corrections}
     \hat{V}_{abe \to cde}= \lim_{\substack{\tilde{p} \to p,\ \tilde{p}' \to p' \\ \tilde{q} \to q,\ \tilde{q}' \to q'}} \Bigl(\frac{i \hat{M}^{(0)}_{ae \to ae}}{\tilde{p}^2-m^2_a+ i\epsilon}+\frac{i \hat{M}^{(0)}_{be \to be}}{\tilde{p}'^2-m^2_b+ i\epsilon} \\
     +\frac{i \hat{M}^{(0)}_{ce \to ce}}{\tilde{q}^2-m^2_c+ i\epsilon}+\frac{i \hat{M}^{(0)}_{de \to de}}{\tilde{q}'^2-m^2_d+ i\epsilon} \Bigr) \times M^{(0)}_{ab \to cd}
\end{multline}
and is not well-defined; 
suppose to send $\tilde{p} \to p$ first and then take the limit for the remaining three momenta. In this case, since $p^2=m^2_a$, the first term in parenthesis in~\eqref{V_off_shell_external_leg_corrections} is divergent. However, $M^{(0)}_{ab \to cd}\ne 0$ since the momenta $\tilde{p}'$, $\tilde{q}$ and $\tilde{q}'$ are off-shell. This implies that this limit is ill-defined and it is not clear what direction we should take when we send the external momenta to their on-shell values. 
We will show later that these ill-defined contributions appearing in the integrands of~\eqref{4_point_one_loop_from_tree_single_cut_contribution} will be removed thanks to the counterterms introduced in the renormalization procedure.

We define
\begin{subequations}
    \label{V_integrand_V_integral_R_integrand_R_integral}
\begin{align}
\label{V_integrand_V_integral_R_integrand_R_integral_1}
 &\hat{\mathcal{V}}^{(\text{on})}_{ab \to cd}\equiv \frac{1}{8\pi} \sum_{e=1}^r \int_{-\infty}^{+\infty} d\theta_k \hat{V}_{abe \to cde}^{(\text{on})},\\
\label{V_integrand_V_integral_R_integrand_R_integral_2}
&\hat{\mathcal{V}}_{ab \to cd}\equiv \frac{1}{8\pi} \sum_{e=1}^r \int_{-\infty}^{+\infty} d\theta_k \hat{V}_{abe \to cde},\\
\label{V_integrand_V_integral_R_integrand_R_integral_3}
 &\hat{\mathcal{R}}_{ab \to cd}\equiv \frac{1}{8\pi} \sum_{e=1}^r \int_{-\infty}^{+\infty} d\theta_k \hat{R}_{abe \to cde},
\end{align}
\end{subequations}
where, as usual, $\theta_k$ is the common value of the rapidities of the two particles of type $e$.
Plugging~\eqref{off_shell_limit_of_a_3_to_3_non_elastic_amplitude} into~\eqref{4_point_one_loop_from_tree_single_cut_contribution}, and using (\eqref{V_integrand_V_integral_R_integrand_R_integral_2},\eqref{V_integrand_V_integral_R_integrand_R_integral_3}), the single-cut contribution can be written as
\begin{equation}
\label{off_shell_limit_of_a_3_to_3_non_elastic_amplitude_integral}
    M_{ab\to cd}^{(1\text{-cut})} = \hat{\mathcal{V}}_{ab \to cd}+\hat{\mathcal{R}}_{ab \to cd}=\hat{\mathcal{V}}_{ab \to cd}-\hat{\mathcal{V}}^{(\text{on})}_{ab \to cd} \ ,
\end{equation}
where in the last equality we used~\eqref{on_shell_limit_of_a_3_to_3_non_elastic_amplitude_infinities_removed}. 
Plugging~\eqref{V_off_shell_external_leg_corrections} into~\eqref{V_integrand_V_integral_R_integrand_R_integral_2} and using~\eqref{definition_necessary_for_mass_renormalization_first_time_ga_appears} we obtain 
\begin{equation}
\label{Voff_explicit_expression_with_all_terms}
\begin{split}
    \hat{\mathcal{V}}_{ab \to cd}&= \Bigl( \frac{\Sigma_{aa}(\tilde{p}^2)}{\tilde{p}^2-m^2_a} + \frac{\Sigma_{bb}(\tilde{p'}^2)}{\tilde{p}'^2-m^2_b}+\frac{\Sigma_{cc}(\tilde{q}^2)}{\tilde{q}^2-m^2_c}+
   \frac{\Sigma_{dd}(\tilde{q}'^2)}{\tilde{q}'^2-m^2_d} \Bigr)M^{(0)}_{ab\to cd}(p,p',q,q')\\
    &=\Bigl( \frac{\Sigma^{(0)}_{aa}}{\tilde{p}^2-m^2_a} + \frac{\Sigma^{(0)}_{bb}}{\tilde{p}'^2-m^2_b}+\frac{\Sigma^{(0)}_{cc}}{\tilde{q}^2-m^2_c}+
   \frac{\Sigma^{(0)}_{dd}}{\tilde{q}'^2-m^2_d} \Bigr)M^{(0)}_{ab\to cd}(p,p',q,q')\\
   &+\Bigl( \Sigma^{(1)}_{aa} + \Sigma^{(1)}_{bb}+\Sigma^{(1)}_{cc}+
   \Sigma^{(1)}_{dd} \Bigr)M^{(0)}_{ab\to cd}(p,p',q,q') ,
\end{split}
\end{equation}
where in the second equality we expanded the diagonal bubble corrections as shown in~\eqref{expansion_bubble_diagram_one_loop_Sigma_ab}. 
Similarly, if we substitute~\eqref{definition_of_Vhat_on} into~\eqref{V_integrand_V_integral_R_integrand_R_integral_1}, we obtain
\begin{equation}
\label{Von_explicit_expression_with_all_terms}
    \hat{\mathcal{V}}^{(\text{on})}_{ab \to cd}= \Bigl( \Sigma^{(0)}_{aa} \frac{\partial}{\partial m_a^2} + \Sigma^{(0)}_{bb} \frac{\partial}{\partial m_b^2}+\Sigma^{(0)}_{cc} \frac{\partial}{\partial m_c^2}+\Sigma^{(0)}_{dd} \frac{\partial}{\partial m_d^2} \Bigr)M^{(0)}_{ab\to cd}(p,p',q,q').
\end{equation}
Finally, combining~\eqref{Voff_explicit_expression_with_all_terms} and~\eqref{Von_explicit_expression_with_all_terms} into~\eqref{off_shell_limit_of_a_3_to_3_non_elastic_amplitude_integral}, and using the fact that $M^{(0)}_{ab\to cd}(p,p',q,q')=0$ on-shell, we find the contribution of single cuts:
\begin{equation}
\label{final_expression_for_single_cut_contribution_two_to_two}
\begin{split}
    M_{ab\to cd}^{(1\text{-cut})}&= \Bigl(\delta m_a^2 \frac{\partial}{\partial m_a^2} + \delta m_b^2 \frac{\partial}{\partial m_b^2}+\delta m_c^2 \frac{\partial}{\partial m_c^2}+\delta m_d^2 \frac{\partial}{\partial m_d^2} \Bigr)M^{(0)}_{ab\to cd}(p,p',q,q')\\
   &-\Bigl( \frac{\delta m_a^2}{\tilde{p}^2-m^2_a} + \frac{\delta m_b^2}{\tilde{p}'^2-m^2_b}+\frac{\delta m_c^2}{\tilde{q}^2-m^2_c}+
   \frac{\delta m_d^2}{\tilde{q}'^2-m^2_d} \Bigr)M^{(0)}_{ab\to cd}(p,p',q,q').
\end{split}
\end{equation}

\subsection{Universal expressions for one-loop inelastic amplitudes}

One-loop amplitudes associated with processes of the type in~\eqref{2_to_nminus2_production_process_Introduction}, with $n\ge4$, can be obtained by generalising the results of the previous sections.
The one-loop amplitude can again be written as a sum of a single- and a double-cut contribution
\begin{equation}
M_{a_1 a_2\to a_3 \dots a_n}^{(1)}=M_{a_1 a_2\to a_3 \dots a_n}^{(2\text{-cut})}+M_{a_1 a_2\to a_3 \dots a_n}^{(1\text{-cut})},
\end{equation}
where
\begin{equation}
\label{single_cut_contribution_production_amplitude_one_loop}
M_{a_1 a_2\to a_3 \dots a_n}^{(1\text{-cut})} =\frac{1}{8\pi} \sum_{e=1}^r \int_{-\infty}^{+\infty} d\theta_k \hat{M}_{a_1 a_2 e\to a_3 \dots a_n e}^{(0)}(p_1, p_2 , k ,p_3, \dots, p_n,k)
\end{equation}
and
\begin{equation}
\label{double_cut_contribution_production_amplitude_one_loop}
M_{a_1 a_2\to a_3 \dots a_n}^{(2\text{-cut})} = \sum_{k=2}^{n}\sum_{\sigma} \sum_{e,f=1}^r \frac{M_{a_1 e\to a_{\sigma(3)} \dots a_{\sigma(k)}f}^{(0)} \ M_{a_2 f\to a_{\sigma(k+1)}\dots a_{\sigma(n)}e}^{(0)}}{8 m_e m_f |\sinh{\theta_{ef}}|}.
\end{equation}
The integrand in~\eqref{single_cut_contribution_production_amplitude_one_loop} is defined following a similar convention to~\eqref{removing_infinite_rapidity_of_particle_e_from_3_to_3_amplitude}: it corresponds to a tree-level production amplitude in which the contribution at $\theta_k=\infty$ has been removed (being $\theta_k$ the rapidity of the particle of type $e$). 

For each $k$ in~\eqref{double_cut_contribution_production_amplitude_one_loop} we sum over all the possible configurations $\sigma$ splitting $\{a_3, \ldots, a_n \}$ into two sets composed of $k-2$ and $n-k$ elements respectively. For example, for $k=3$, there are $n-2$ different configurations $\sigma$ for the numerator in~\eqref{double_cut_contribution_production_amplitude_one_loop}:
\begin{equation}
    \begin{split}
        &M_{a_1 e\to a_{3} f}^{(0)}M_{a_2 f\to a_4 a_5\dots a_{n}e}^{(0)},\\
        &M_{a_1 e\to a_{4} f}^{(0)}M_{a_2 f\to a_{3} a_5\dots a_{n}e}^{(0)},\\
        &\vdots\\
        &M_{a_1 e\to a_{n} f}^{(0)}M_{a_2 f\to a_{3} a_4\dots a_{n-1}e}^{(0)}.
    \end{split}
\end{equation}
Tree-level amplitudes in which there is a single particle as an incoming or outgoing state are included in the cases $k=2$ and $k=n$. However, these are of course zero if all particles are stable at the tree level.
As expected from the study performed in the previous sections, also in this case all the tree-level amplitudes appearing on the r.h.s. of~\eqref{double_cut_contribution_production_amplitude_one_loop} are inelastic and therefore by condition~\ref{Condition_tree_level_elasticity_introduction} it holds that
\begin{equation}
\label{double_cut_contribution_is_zero_in_production_amplitudes}
M_{a_1 a_2\to a_3 \dots a_n}^{(2\text{-cut})} = 0.
\end{equation}

The single cut contribution~\eqref{single_cut_contribution_production_amplitude_one_loop} can be computed following the same method explained previously. As before, we can reach the on-shell configuration at which the cut is evaluated, following two different limits. If we adopt an on-shell limit similar to the one used in section~\ref{section_on_properties_of_tree_level_inelastic_amplitudes} we obtain
\begin{equation}
\label{on_shell_limit_generic_production_amplitude}
   \hat{M}_{a_1 a_2 e\to a_3 \dots a_n e}^{(0,\text{on})}= \hat{V}_{a_1 a_2 e\to a_3 \dots a_n e}^{(\text{on})} + \hat{R}_{a_1 a_2 e\to a_3 \dots a_n e} = 0. 
\end{equation}
As before, $\hat{V}_{a_1 a_2 e\to a_3 \dots a_n e}^{(\text{on})}$ contains all diagrams in which the two legs of type $e$ cross the same external leg and can be written as
\begin{equation}
\label{Von_explicit_expression_with_all_terms_production_amplitude}
    \hat{V}_{a_1 a_2 e\to a_3 \dots a_n e}^{(\text{on})}= i \sum_{j=1}^n M^{(0)}_{a_j e \to a_j e} \frac{\partial}{\partial m_{a_j}^2} M^{(0)}_{a_1 a_2\to a_3 \dots a_n}(p_1, \dots , p_n).
\end{equation}
$\hat{R}_{a_1 a_2 e\to a_3 \dots a_n e}$ contains all the other Feynman diagrams. 
For the same reason explained in the previous section, \eqref{on_shell_limit_generic_production_amplitude} is not an integrand of~\eqref{single_cut_contribution_production_amplitude_one_loop}. To evaluate the expressions appearing in~\eqref{single_cut_contribution_production_amplitude_one_loop} we should reach the on-shell configuration at which the cut is realised by moving along a limit in which the particles $a_1, a_2, \dots a_n$ are off-shell. If we do so we obtain
\begin{equation}
\label{off_shell_limit_generic_production_amplitude}
\begin{split}
       \hat{M}_{a_1 a_2 e\to a_3 \dots a_n e}^{(0)}&= \hat{V}_{a_1 a_2 e\to a_3 \dots a_n e} + \hat{R}_{a_1 a_2 e\to a_3 \dots a_n e}\\
       &=\hat{V}_{a_1 a_2 e\to a_3 \dots a_n e} - \hat{V}_{a_1 a_2 e \to a_3 \dots a_n e}^{(\text{on})},
\end{split}
\end{equation}
where in the last equality we used the key point that the contribution $\hat{R}_{a_1 a_2 e\to a_3 \dots a_n e}$ in~\eqref{off_shell_limit_generic_production_amplitude} is the same as the one appearing in~\eqref{on_shell_limit_generic_production_amplitude}. In other words $\hat{R}_{a_1 a_2 e\to a_3 \dots a_n e}$ does not depend on the limit adopted to reach the configuration at which all the particles $a_1,\dots a_n$ are on-shell and the two particles of type $e$ are both on-shell and carry the same momentum. 
$\hat{V}_{a_1 a_2 e\to a_3 \dots a_n e}$ is easily obtained and can be written as
\begin{equation}
\label{V_off_shell_external_leg_correction_multiple_particle_production}
     \hat{V}_{a_1 a_2 e\to a_3 \dots a_n e}= \lim_{p^2_j \to m^2_{a_j}} \sum_{j=1}^n \frac{i \hat{M}^{(0)}_{a_j e \to a_j e}}{p_j^2-m^2_{a_j}} \times M^{(0)}_{a_1 a_2\to a_3 \dots a_n}.
\end{equation}
An expression for the single cut contribution in~\eqref{single_cut_contribution_production_amplitude_one_loop} can be obtained by
substituting~\eqref{off_shell_limit_generic_production_amplitude} into~\eqref{single_cut_contribution_production_amplitude_one_loop}.
If we define
\begin{subequations}
    \label{V_integrand_V_integral_R_integrand_R_integral_production}
\begin{align}
\label{V_integrand_V_integral_R_integrand_R_integral_1_production}
 &\hat{\mathcal{V}}^{(\text{on})}_{a_1 a_2 \to a_3 \dots a_n}\equiv \frac{1}{8\pi} \sum_{e=1}^r \int_{-\infty}^{+\infty} d\theta_k \hat{V}_{a_1 a_2 e\to a_3 \dots a_n e}^{(\text{on})},\\
\label{V_integrand_V_integral_R_integrand_R_integral_2_production}
&\hat{\mathcal{V}}_{a_1 a_2\to a_3 \dots a_n}\equiv \frac{1}{8\pi} \sum_{e=1}^r \int_{-\infty}^{+\infty} d\theta_k \hat{V}_{a_1 a_2 e\to a_3 \dots a_n e},
\end{align}
\end{subequations}
and use~\eqref{Von_explicit_expression_with_all_terms_production_amplitude}, \eqref{V_off_shell_external_leg_correction_multiple_particle_production} and~\eqref{definition_necessary_for_mass_renormalization_first_time_ga_appears}, together with the fact that $M^{(0)}_{a_1 a_2\to a_3 \dots a_n}=0$ on-shell, we find
\begin{equation}
\label{single_cut_contribution_production_amplitude_final_result}
\begin{split}
    M_{a_1 a_2\to a_3 \dots a_n}^{(1\text{-cut})}&= \sum_{j=1}^r\delta m_{a_j}^2 \frac{\partial}{\partial m_{a_j}^2} M^{(0)}_{a_1 a_2\to a_3 \dots a_n} (p_1 \dots , p_n)\\
   &- \sum_{j=1}^r \frac{\delta m_{a_j}^2}{p_j^2-m^2_{a_j}} M^{(0)}_{a_1 a_2\to a_3 \dots a_n}(p_1 \dots , p_n).
\end{split}
\end{equation}

The final result for the renormalized amplitude associated with the process~\eqref{2_to_nminus2_production_process_Introduction}, truncated at the one-loop order in perturbation theory, is given by summing~\eqref{total_counterterms_contribution_production_process}, \eqref{double_cut_contribution_is_zero_in_production_amplitudes} and \eqref{single_cut_contribution_production_amplitude_final_result}. 
The ill-defined contributions in~\eqref{total_counterterms_contribution_production_process} and~\eqref{single_cut_contribution_production_amplitude_final_result} cancel in the sum and the final result is given by
\begin{equation}
\label{full_one_loop_amplitude_final_result}
\begin{split}
     M_{a_1 a_2\to a_3 \dots a_n}&=\sum_{{j \in \{\text{prop, ext}\}}} \delta m^2_j \frac{\partial}{\partial m^2_j} M^{(0)}_{a_1 a_2 \to a_3 \dots a_n}\\
     &=M^{(0)}_{a_1 a_2\to a_3 \dots a_n}\Bigl|_{m_j^2+\delta m^2_j} + O\bigl( (\delta m^2_j)^2 \bigl).
\end{split}
\end{equation}
This is precisely the result anticipated in~(\eqref{final_formula_we_need_to_prove_first_formulation},\eqref{final_formula_we_need_to_prove_second_formulation}), where the sum over $j$ is performed on the masses of the external particles and internal propagators.

\subsection{Sufficient conditions for absence of inelasticity at one-loop}
\label{Section_sufficient_conditions_absence_inelasticity_one_loop}

One-loop inelastic amplitudes associated with classically integrable Lagrangians satisfying conditions~\ref{Condition_tree_level_elasticity_introduction} and~\ref{Non_degenerate_mass_condition} are generally nonzero. This should be quite obvious: note that in~\eqref{full_one_loop_amplitude_final_result} we are shifting all the masses by certain quantities determined by one-loop corrections to the propagators. Even if the original tree-level amplitude $M^{(0)}_{a_1 a_2\to a_3 \dots a_n}$ is zero for on-shell values of the external momenta this does not mean that the amplitude remains zero after the deformation
$$
m^2_j \to m^2_j +\delta m^2_j.
$$
Indeed, in~\eqref{full_one_loop_amplitude_final_result} both the masses appearing inside propagators and the masses of the external particles do not correspond to the classical masses appearing in~\eqref{eq0_1}.
For this reason, Lagrangian~\eqref{renormalised_Lagrangian_due_to_standard_renormalisation_procedure}, obtained by modifying the tree-level integrable Lagrangian~\eqref{eq0_1} through a standard renormalization procedure, is not integrable at one loop. To restore integrability at one loop order in perturbation theory additional counterterms need to be introduced.
In this section, we show how these additional counterterms can be introduced in theories satisfying the following condition 
$$
\delta m^2_j = \dd  \cdot m^2_j,
$$
where $\dd$ is a common factor not depending on the type of particle $j \in \{1, \dots, r\}$.
If all the corrections to the squares of the masses scale with the same multiplicative factor then, defining $\lambda=1+\dd$, it holds that
\begin{equation}
\label{delta_m_square_j_does_not_depend_on_j}
    m^2_j+\delta m^2_j = (1+ \dd)m^2_j = \lambda \cdot m^2_j.
\end{equation}
If condition~\eqref{delta_m_square_j_does_not_depend_on_j} is satisfied it has to be possible to add suitable counterterms to the renormalized Lagrangian~\eqref{renormalised_Lagrangian_due_to_standard_renormalisation_procedure} to preserve integrability at the one-loop order. Since all masses are scaled by the same multiplicative factor,
the fusing angles do not change
and the flipping rule reviewed in~\cite{Dorey:2021hub} necessary for the cancellation of singularities in sums of Feynman diagrams contributing to inelastic processes still applies 
after the deformation~\eqref{delta_m_square_j_does_not_depend_on_j}. 
In particular, collections of diagrams that were singular simultaneously in tree-level amplitudes $M^{(0)}_{a_1 a_2\to a_3 \dots a_n}\bigl|_{m^2_j}$ continue to be singular simultaneously in amplitudes $M^{(0)}_{a_1 a_2\to a_3 \dots a_n}\bigl|_{\lambda m_j^2}$. However, while the fusing angles determine the location of the poles in amplitudes, the requirement that the residues at these poles are zero follows from particular conditions on the couplings. The scaling~\eqref{delta_m_square_j_does_not_depend_on_j} slightly modifies these conditions and the couplings require to be adjusted to ensure cancellation of inelastic processes.

Let us focus on the two-to-two inelastic process in~\eqref{generic_scattering_process_ab_into_cd} first.
Using~\eqref{full_one_loop_amplitude_final_result}, we obtain that the amplitude to one loop order is 
\begin{equation}
\label{renormalized_two_to_two_amplitude_with_all_masses_dilatated_in_the_same_manner}
\begin{split}
    &M_{ab\to cd}= M^{(0)}_{ab \to cd} \Bigl|_{\lambda m^2_j}=\\
    &-\frac{i}{\lambda} \sum_{i \in s} \frac{ C^{(3)}_{ab\bar{i}}  C^{(3)}_{i \bar{c} \bar{d}}}{s-m^2_i} -\frac{i}{\lambda} \sum_{j \in t} \frac{ C^{(3)}_{a\bar{c}\bar{j}}  C^{(3)}_{j b \bar{d}}}{t-m^2_j}-\frac{i}{\lambda} \sum_{k \in u} \frac{ C^{(3)}_{a\bar{d}\bar{k}}  C^{(3)}_{k b \bar{c}}}{u-m^2_k} -i C^{(4)}_{ab \bar{c} \bar{d}}.
\end{split}
\end{equation}
Note that also the Mandelstam variables $s$, $t$ and $u$ scale with $\lambda$ since the masses of the external particles are scaled.
Even though expression~\eqref{renormalized_two_to_two_amplitude_with_all_masses_dilatated_in_the_same_manner} is nonzero, we can scale the couplings of the theory in such a way to make~\eqref{renormalized_two_to_two_amplitude_with_all_masses_dilatated_in_the_same_manner} null. This can be done as follows
\begin{equation}
\label{scaling_of_couplings_necessary_for_cancellation_purposes}
    C_{a_1 a_2 a_3}^{(3)} \to \lambda^\rho C_{a_1 a_2 a_3}^{(3)} \hspace{4mm} \text{and} \hspace{4mm} C_{a_1 a_2 a_3 a_4}^{(4)} \to \lambda^{2 \rho -1} C_{a_1 a_2 a_3 a_4}^{(4)},
\end{equation}
with $\rho \in \mathbb{R}$, and leads to
\begin{equation}
\begin{split}
    M_{ab\to cd}&= \lambda^{2 \rho -1} \Bigl(-i \sum_{i \in s} \frac{ C^{(3)}_{ab\bar{i}}  C^{(3)}_{i \bar{c} \bar{d}}}{s-m^2_i} -i \sum_{j \in t} \frac{ C^{(3)}_{a\bar{c}\bar{j}}  C^{(3)}_{j b \bar{d}}}{t-m^2_j}-i \sum_{k \in u} \frac{ C^{(3)}_{a\bar{d}\bar{k}}  C^{(3)}_{k b \bar{c}}}{u-m^2_k} -i C^{(4)}_{ab \bar{c} \bar{d}}\Bigl)\\
    &=\lambda^{2 \rho -1} M^{(0)}_{ab\to cd}=0.
    \end{split}
\end{equation}
Equation~\eqref{scaling_of_couplings_necessary_for_cancellation_purposes} easily generalises to higher-order couplings. 
Necessary conditions to ensure the absence of production processes at the tree level for Lagrangians of type~\eqref{eq0_1} were determined in~\cite{Gabai:2018tmm} and can be written in the form of recursion relations on higher-order couplings
\begin{equation}
 \label{higher_order_couplings_tree_level_constraints}
\begin{split}
C^{(n)}_{a_1 \ldots a_n}&- \sum_l C^{(n-1)}_{a_1 \ldots a_{n-2} \bar{l}} \frac{1}{m^2_l}C^{(3)}_{l  a_{n-1}a_{n}}- \sum_s C^{(n-2)}_{a_1 \ldots a_{n-3} \bar{s}} \frac{1}{m^2_s} C_{s a_{n-2} a_{n-1} a_n}^{(4)}\\
&+ \sum_l C^{(n-2)}_{a_1 \ldots a_{n-3} \bar{s}} \frac{1}{m^2_s} C_{s a_{n-2} \bar{l}}^{(3)} \frac{1}{m_l^2} C_{l a_{n-1} a_n}^{(3)}=0,
\end{split}
\end{equation}
where $n\ge5$.
If all the masses are scaled as in~\eqref{delta_m_square_j_does_not_depend_on_j}, the l.h.s. of~\eqref{higher_order_couplings_tree_level_constraints} transforms as follows
\begin{equation}
 \label{higher_order_couplings_tree_level_constraints_lambda_rescaled}
\begin{split}
C^{(n)}_{a_1 \ldots a_n}&- \frac{1}{\lambda}\sum_l C^{(n-1)}_{a_1 \ldots a_{n-2} \bar{l}} \frac{1}{m^2_l}C^{(3)}_{l  a_{n-1}a_{n}}- \frac{1}{\lambda} \sum_s C^{(n-2)}_{a_1 \ldots a_{n-3} \bar{s}} \frac{1}{m^2_s} C_{s a_{n-2} a_{n-1} a_n}^{(4)}\\
&+\frac{1}{\lambda^2} \sum_l C^{(n-2)}_{a_1 \ldots a_{n-3} \bar{s}} \frac{1}{m^2_s} C_{s a_{n-2} \bar{l}}^{(3)} \frac{1}{m_l^2} C_{l a_{n-1} a_n}^{(3)}.
\end{split}
\end{equation}
Once again the quantity in~\eqref{higher_order_couplings_tree_level_constraints_lambda_rescaled} is nonzero unless we perform the following transformations on the couplings
\begin{equation}
\label{coupling_integrable_renormalization}
    C_{a_1 \dots a_n}^{(n)} \to \lambda^{n(\rho-1)+3-2 \rho} C_{a_1 \dots a_n}^{(n)} \hspace{4mm} \text{with} \ \rho \in \mathbb{R}
\end{equation}

While the Lagrangian~\eqref{renormalised_Lagrangian_due_to_standard_renormalisation_procedure} is not integrable at one loop, an integrable Lagrangian is given by
\begin{equation}
\label{renormalised_Lagrangian_due_to_standard_renormalisation_procedure_plus_integrable_corrections}
\begin{split}
    \mathcal{L}_{\text{int.}}&=\sum_{a=1}^r \biggl( \frac{1}{2} \partial_\mu \phi_a  \partial^\mu \phi_{\bar{a}} - \frac{1}{2} \lambda m_a^2 \phi_a \phi_{\bar{a}}\biggr) - \sum_{n=3}^{+\infty} \frac{1}{n!}\sum_{a_1,\dots, a_n=1}^r \lambda^{n(\rho-1)+3-2 \rho} C^{(n)}_{a_1 \ldots a_n} \phi_{a_1} \ldots \phi_{a_n}\\
    &\begingroup\color{blue} +\sum_{a=1}^r\frac{1}{2}\Sigma^{(1)}_{aa} \Bigl(\partial_\mu \phi_a  \partial^\mu \phi_{\bar{a}}-  m^2_a \phi_a \phi_{\bar{a}}\Bigr)\endgroup \begingroup\color{red}+\sum_{\substack{a,b=1 \\ b\neq a}}^r t_{ba} \Bigl(-\partial_\mu \phi_{\bar{a}} \partial^{\mu} \phi_b+m^2_a  \phi_{\bar{a}} \phi_b \Bigr)\endgroup\\
    &\begingroup\color{blue}- \sum_{n=3}^{+\infty} \frac{1}{n!}\sum_{a_1,\dots, a_n=1}^r C^{(n)}_{a_1 \ldots a_n} \frac{1}{2}\Bigl( \Sigma^{(1)}_{a_1a_1}+\Sigma^{(1)}_{a_2a_2} + \ldots \Sigma^{(1)}_{a_n a_n}\Bigr) \phi_{a_1} \ldots \phi_{a_n} \endgroup\\
    &\begingroup\color{red}+\sum_{n=3}^{+\infty} \frac{1}{(n-1)!}\sum_{a_1,\dots, a_n=1}^r C^{(n)}_{a_1 \ldots a_n} \sum_{\substack{b=1 \\ b\neq a_1}}^r t_{b a_1} \phi_{b} \phi_{a_2} \ldots \phi_{a_n}\endgroup.
\end{split}
\end{equation}
where the last term in~\eqref{renormalised_Lagrangian_due_to_standard_renormalisation_procedure} (coloured purple) has been absorbed in the factor $\lambda$ multiplying the masses.
If all masses scale as in~\eqref{delta_m_square_j_does_not_depend_on_j} under one-loop corrections then all tree-level and one-loop inelastic amplitudes obtained from the Lagrangian~\eqref{renormalised_Lagrangian_due_to_standard_renormalisation_procedure_plus_integrable_corrections} are zero. This Lagrangian depends on one free parameter $\rho$ which is not fixed by requiring the absence of inelasticity at one loop. 

While in this section we have renormalized the Lagrangian couplings in such a way as to kill all inelastic one-loop amplitudes, it is worth mentioning that in some cases the renormalization required to preserve integrability follows from integrating out certain auxiliary fields in apparently different theories.
This is for example the case for the class of generalised sine-Gordon models considered in~\cite{Hoare:2010fb}, whose quantum actions (comprising the counterterms) can be obtained by integrating out the unphysical modes in gauged Wess-Zumino-Witten models for a coset $G/H$ plus an integrable potential.

\section{One-loop integrability in affine Toda field theories}
\label{Section_on_affine_Toda_field_theories}

Affine Toda models are a class of 1+1 dimensional quantum field theories, describing the interaction of $r$ bosonic scalar fields $\phi_1, \dots, \phi_r$, in 1+1 dimensions. Their Lagrangian is defined as follows
\begin{equation}
\label{Toda_theory_lagrangian_defined_in_terms_of_roots}
\mathcal{L}^{(\text{Toda})}_0=\frac{1}{2} \partial_\mu \phi_a  \partial^\mu \phi_a - \frac{m^2}{\g^2} \Bigl( \sum_{i=0}^r n_i e^{\g \alpha^a_i  \phi_a} -h \Bigr),
\end{equation}
where $\{\alpha_i\}_{i=0}^r$ is a set of $r+1$ vectors in $\mathbb{R}^r$, having inner products encoded in one of the twisted or untwisted affine Dynkin diagram. The set of integers $\{n_i\}_{i=0}^r$ are characteristic for each algebra and satisfy
$$
\sum_{i=0}^r n_i \alpha_i=0
$$
so that $\phi_1=\ldots=\phi_r=0$ is a stationary point around which it is possible to perform perturbation theory. The integer $r$ is the rank of the Lie algebra, while the parameters $m$ and $\g$ provide the theory's mass scale and the coupling constant. In~\eqref{Toda_theory_lagrangian_defined_in_terms_of_roots} the potential has been translated by 
\begin{equation}
\label{Definition_of_h_in_general_Toda_model}
   h\equiv\sum_{i=0}^r n_i
\end{equation}
so that it is equal to zero at the point $\phi_1=\ldots=\phi_r=0$.
The classical integrability of these theories was proven by finding a Lax connection from which higher-spin conserved charges can be generated~\cite{a1,a2}. In the past, these theories were mainly studied through the S-matrix bootstrap approach~\cite{Braden:1989bu,Delius:1991kt,a16,a18, a19,a20,aa20,a21,a22,a25,Corrigan:1993xh,Dorey:1993np,Oota:1997un,a24,aa24,Fring:1991gh}, through which non-perturbative S-matrix elements were conjectured. 

These theories can also be studied through perturbation theory by expanding~\eqref{Toda_theory_lagrangian_defined_in_terms_of_roots} around the minimum $\phi_1=\ldots=\phi_r=0$ and performing a change of basis so to make the squared of the mass matrix diagonal. In this case, we obtain a Lagrangian of type~\eqref{eq0_1}, which is amenable to Feynman diagrams computations.
Using universal properties of affine Toda field theories 
the bootstrapped S-matrices  were confirmed at the tree level by standard perturbative computations in~\cite{Dorey:2021hub}.
In that paper, a complete proof of the tree-level perturbative integrability of these theories was provided by showing that combinations of Feynman diagrams contributing to non-elastic processes always sum to zero at the tree level.
At the loop level, partial results were obtained in~\cite{Braden:1991vz,Braden:1990wx,Braden:1990qa,Sasaki:1992sk,Braden:1992gh,Dorey:2022fvs,Second_loop_paper_sagex}: while many nice features of these models were shown,
the results were based on a case-by-case study performed over different theories and a proof of the vanishing of inelastic one-loop amplitudes has so far been missing.

In this section, we extend the results of~\cite{Dorey:2021hub} to the one-loop order in perturbation theory: using the techniques developed in the previous sections we show 
how integrability manifests itself in one-loop computations in affine Toda field theories. In particular, we prove the absence of one-loop inelastic amplitudes in all simply-laced affine Toda models.

\subsection{A quick review of the Coxeter geometry}

The integrability of the affine Toda theories is due to the Coxeter geometry of their underlying root systems. We briefly review some properties emerging from this geometry relevant to prove the one-loop integrability of these models. More detailed reviews of the topic can be found in~\cite{Dorey:2021hub,Corrigan:1994nd}.

We define by $\Phi \in \mathbb{R}^r$ the root system associated with a generic semisimple Lie algebra $\mathcal{G}$ of rank $r$.
For any root $\alpha \in \Phi$ 
we define the corresponding Weyl reflection $w_\alpha$ as
\begin{equation} \label{Weyl_reflection}
w_\alpha(x)=x-2 \frac{(x,\alpha)}{\alpha^2} \alpha \, ,
\end{equation}
where $( \ , \, )$ is the standard symmetric invariant bilinear form on $\mathcal{G}$ and $\alpha^2\equiv( \alpha , \alpha )$. $w_\alpha$ is a linear map acting as a reflection with respect to the hyperplane orthogonal to $\alpha$.
It is always possible to follow the Steinberg ordering~\cite{Steinberg_Paper} and split the simple roots of $\Phi$ into two sets, $\circ$ and $\bullet$ (white and black), both composed of roots mutually orthogonal to each other. 
An example is shown in figure~\ref{Example_of_Dynkin_diagrams_with_balck_and_white_roots} for the Dynkin diagram associated with $E_8$.
\begin{figure}
\begin{center}
\begin{tikzpicture}
\tikzmath{\y=1.1;}

\filldraw[color=black,fill=black, very thick](0*\y ,0*\y) circle (0.1*\y);
\draw[] (0.1*\y,0*\y) -- (0.9*\y,0*\y);
\filldraw[color=black,fill=white, very thick](1*\y ,0*\y) circle (0.1*\y);
\draw[] (1.1*\y,0*\y) -- (1.9*\y,0*\y);
\filldraw[color=black,fill=black, very thick](2*\y ,0*\y) circle (0.1*\y);
\draw[] (2.1*\y,0*\y) -- (2.9*\y,0*\y);
\filldraw[color=black,fill=white, very thick](3*\y ,0*\y) circle (0.1*\y);
\draw[] (3.1*\y,0*\y) -- (3.9*\y,0*\y);
\filldraw[color=black,fill=black, very thick](4*\y ,0*\y) circle (0.1*\y);
\draw[] (4.1*\y,0*\y) -- (4.9*\y,0*\y);
\filldraw[color=black,fill=white, very thick](5*\y ,0*\y) circle (0.1*\y);
\draw[] (5.1*\y,0*\y) -- (5.9*\y,0*\y);
\filldraw[color=black,fill=black, very thick](6*\y ,0*\y) circle (0.1*\y);
\draw[] (2*\y,0.1*\y) -- (2*\y,0.9*\y);
\filldraw[color=black,fill=white, very thick](2*\y,1*\y) circle (0.1*\y);
\end{tikzpicture}
\end{center}
\caption{$E_8$ Dynkin diagram.}
\label{Example_of_Dynkin_diagrams_with_balck_and_white_roots}
\end{figure}
We can then define the Coxeter element $w$ as
\begin{equation} \label{Coxeter_element}
w=w_{\bullet} w_{\circ}=\prod_{\alpha \in \bullet} w_{\alpha} \prod_{\beta \in \circ} w_{\beta} .
\end{equation}
Note that the order of $w_{\alpha \in \bullet}$ in the first product in~\eqref{Coxeter_element} is irrelevant since all the roots $\alpha \in \bullet$ are orthogonal to each other. The same argument applies to the second product.
It can be shown that the Coxeter element $w$ is diagonalised with respect to an orthonormal basis $\{ z_s \} \in \mathbb{C}^r$ satisfying
\begin{equation} 
\label{eq1_9}
w  z_{s} = e^{2 i \theta_s} z_s \hspace{3mm}, \hspace{3mm} \theta_s= \frac{s \pi}{h} \,
\end{equation}
where $h$ is called the Coxeter number and for untwisted affine Toda field theories is the same as the number defined in~\eqref{Definition_of_h_in_general_Toda_model}.
The integers $s$ in~\eqref{eq1_9} are the exponents of $\mathcal{G}$ and belong to the interval $1 \le s \le h$. Apart from the case $s=\frac{h}{2}$, these exponents appear always in pairs $\{ s, h-s \}$. The eigenvectors $\{z_s , z_{h-s} \}$ associated with each pair of exponents span the spin-$s$ eigenplane of $w$ and can be used to define
a projector $P_s$ onto this plane.
Then, for a generic root $\alpha$, it holds that
$$
P_s (w^{p}\alpha)=e^{2 i p \theta_s}P_s(\alpha) .
$$ 
From this relation, we see that the Coxeter element acts as a combination of rotations of different subspaces of $\mathbb{R}^r$: for each root $\alpha$ and for each exponent $s$, the projection $P_s (\alpha)$ is rotated by $w$ of an angle $\theta_s$ in the counterclockwise direction. Since $2 h \theta_s=2 \pi s$ then $w^h=1$.

In~\cite{B. Kostant1} it was shown how to organize the root system into $r$ Coxeter orbits $\{ \Gamma_a \}_{a=1}^r$, each composed of $h$ roots: first we label the fundamental weights by $\{ \lambda_a \}_{a=1}^r$, satisfying
\begin{equation}
\label{fundamental_weights_definition}
\frac{2}{\alpha^2_a }(\lambda_a , \alpha_b)=\delta_{ab} \, ,
\end{equation}
then we generate the orbits acting on the following roots
\begin{equation}
\label{wirting_roots_labeling_the_orbits_in_terms_of_fundamental_weights}
\gamma_a \equiv (1-w^{-1}) \lambda_a=\begin{cases}
&\alpha_a \hspace{20mm} \text{if} \ a \in \circ \, , \\
& - w^{-1} \alpha_a \hspace{10.5mm} \text{if} \ a \in \bullet \, .
\end{cases}
\end{equation}
\begin{figure}
\begin{center}
\begin{tikzpicture}
\tikzmath{\y=0.8;}

\draw[->,thick](0*\y,0*\y) -- (4.5*\y,0*\y);
\draw[->,thick](0*\y,0*\y) -- (-4*\y,-2*\y);
\draw[->,thick](0*\y,0*\y) -- (4*\y,-2*\y);
\draw[](0*\y,0*\y) -- (-4.5*\y,0*\y);
\draw[directed] (4.8*\y,-1.2*\y) arc(-30:60:0.4*\y);
\filldraw[black] (4.8*\y,-1.1*\y)  node[anchor=west] {\footnotesize{$w$}};

\filldraw[black] (-3.1*\y,-1.75*\y)  node[anchor=west] {\footnotesize{$P_{s}(\alpha_\bullet )$}};
\filldraw[black] (1.6*\y,-1.75*\y)  node[anchor=west] {\footnotesize{$P_{s}(\gamma_\bullet)$}};
\filldraw[black] (2.6*\y,0.35*\y)  node[anchor=west] {\footnotesize{$P_{s}(\gamma_\circ)=P_{s}(\alpha_\circ)$}};

\draw[][] (1.2*\y,0*\y) arc(0:-27:1.2*\y);
\draw[][] (-1.2*\y,0*\y) arc(180:207:1.2*\y);
\filldraw[black] (1.2*\y,-0.4*\y)  node[anchor=west] {\tiny{$ \theta_s$}};
\filldraw[black] (-1.8*\y,-0.4*\y)  node[anchor=west] {\tiny{$\theta_s$}};
\end{tikzpicture}
\end{center}
\caption{Projections of the roots $\gamma_\circ$, $\alpha_\circ$, $\gamma_\bullet$ and $\alpha_\bullet$ on the spin-$s$ eigenplane of the Coxeter element. By $\circ$ and $\bullet$ we refer to any index in $\circ$ and $\bullet$.}
\label{Coxeter_geometry_projections_of_roots_on_the_spin_s_plans_representations_of_abullet_and_acirc}
\end{figure}
The projection of a root $\gamma_{a}$ onto the spin-$s$ eigenplane of $w$ can be written using a complex number notation as
\begin{equation}
\label{projections_complex_numbers_notation}
\begin{split}
&P_s(\gamma_a)=\gamma_a^{(s)} \hspace{14mm} \text{if} \  a \in \circ \\
&P_s(\gamma_a)=\gamma_a^{(s)} e^{-i \theta_s} \hspace{6mm} \text{if} \  a \in \bullet\\
\end{split}
\end{equation}
The real number $\gamma_a^{(s)}$ depends on the orbit $\Gamma_a$ and on the exponent $s$; instead, the direction of the projection only depends on the colour of $a$. These projections are depicted in figure~\ref{Coxeter_geometry_projections_of_roots_on_the_spin_s_plans_representations_of_abullet_and_acirc}.

Some important properties follow from this geometrical construction of root systems: remarkably in~\cite{a24,Freeman:1991xw} it was shown that there is a one-to-one correspondence between Coxeter orbits $\{ \Gamma_a \}_{a=1}^r$ and particles of affine Toda field theories. In particular, in~\cite{Freeman:1991xw} it was shown that \footnote{More correctly we should write $\gamma_a^{(1)}= c \cdot m_a$ where the proportionality constant $c$ is the same for all the orbits.}
\begin{equation}
\label{root_projection_rule}
\gamma_a^{(1)}=m_a.
\end{equation}
Moreover, all $3$-point couplings are characterised by the following fusing rule~\cite{a24}:\\
`$C^{(3)}_{abc} \ne 0$ iff $\exists \ \alpha \in \Gamma_a , \, \beta \in \Gamma_b$ and $\gamma \in \Gamma_c$ such that $\alpha+\beta+\gamma=0$.'\\
Whenever this is the case then the $3$-point coupling is determined by~\cite{Braden:1989bu,Fring:1991me}
\begin{equation}
\label{Connection_among_three_point_couplings_and_areas}
|C^{(3)}_{abc}|=  4 \frac{\g}{\sqrt{h}} \Delta_{abc} |N_{\alpha , \beta}|,
\end{equation}
where $\Delta_{abc}$ is the area of the triangle having as sides the masses $m_a$, $m_b$ and $m_c$ (given by projecting three roots $\alpha \in \Gamma_a$, $\beta \in \Gamma_b$ and $\gamma \in \Gamma_c$ on the spin-$1$ eigenplane of $w$), and $N_{\alpha \beta}$ is the structure constant obtained by commuting the Lie algebra generators associated with the roots $\alpha$ and $\beta$. 
Once the masses and the $3$-point couplings of the theory are determined all the higher-order couplings can be obtained recursively through the following simple rule~\cite{Fring:1992tt}: $\forall$ integer $n \ge 4$ and $\forall$ integer $k \in [2,n-2]$ it holds that
\begin{equation}
\label{Higher_order_couplings}
C^{(n)}_{a_1 \ldots a_n}= \frac{\g^2 }{h \cdot m^2}C_{a_1 \ldots a_k}^{(k)} C_{a_{k+1} \ldots a_n}^{(n-k)} + \sum_{j} C_{a_1 \ldots a_k \bar{j}}^{(k+1)} \frac{1}{m_j^2} C_{j a_{k+1} \ldots a_n}^{(n+1-k)}
\end{equation}
 where we define
$$
C_{ab}^{(2)}=m_a^2 \delta_{a\bar{b}}.
$$
Starting from these universal features, in~\cite{Dorey:2021hub} it was proven that for the entire class of affine Toda field theories, the tree-level inelastic amplitudes are zero; therefore property~\ref{Condition_tree_level_elasticity_introduction} listed in the introduction is satisfied. An important observation to prove this fact is that the momenta of the scattered particles and internal propagator at the pole positions of amplitudes can be described by root projections onto the spin-$1$ eigenplane of $w$.

Property~\ref{Non_degenerate_mass_condition} can also be proven. 
In the next section, we show that, in all simply-laced affine Toda theories, there are no tree-level Feynman diagrams contributing to processes of type~\eqref{tree_level_degenerate_elastic_process_in_property} and therefore property~\ref{Non_degenerate_mass_condition} is universally satisfied in simply-laced Toda theories. This implies that property~\ref{Non_degenerate_mass_condition} is also satisfied by the non-simply-laced theories for the following reason.
Non-simply-laced models can be split into two sets: twisted and untwisted theories.
In~\cite{Braden:1989bu,Dorey:2021hub,Dorey:1992gr} it was argued that the masses and couplings of twisted theories are a subset of the masses and couplings of simply-laced theories; therefore, if in simply-laced models there are no tree-level Feynman diagrams contributing to the process~\eqref{tree_level_degenerate_elastic_process_in_property}, also in twisted models there are no tree-level diagrams contributing to this process and property~\ref{Non_degenerate_mass_condition} is valid also in twisted models.
On the other hand, the folding leading to untwisted non-simply-laced theories turns out to remove all mass degeneracies from the spectrum~\cite{Braden:1989bu} and property~\ref{Non_degenerate_mass_condition} trivially applies to these models. For this reason, we limit ourselves to considering only the class of simply-laced affine Toda theories.

\subsection{Proving property 2}

We consider the process~\eqref{tree_level_degenerate_elastic_process_in_property} in the kinematical configuration in which the incoming and outgoing momenta are the same:
$$
a(p)+c(p') \to b(p)+c(p') \,;
$$
$a$ and $b$ are particles with the same mass.
The condition to have at least one particle propagating in the $s$-channel is that there exists a particle $e$ such that $C^{(3)}_{ac \bar{e}}$ and $C^{(3)}_{\bar{b} \bar{c} e}$ are both nonzero.
If this is the case then on the kinematical configuration making the particle $e$ on-shell it holds that
$$
\alpha+\gamma= \epsilon = \beta+\gamma  \ \ \text{with} \ \alpha \in \Gamma_a , \, \beta \in \Gamma_b, \, \gamma \in \Gamma_c \ \text{and} \ \epsilon \in \Gamma_e.
$$
The projections of these roots onto the spin-$1$ eigenplane of $w$ close a tiled polygon corresponding to an on-shell Feynman diagram.
This on-shell diagram needs to be cancelled through the flipping rule~\cite{a24,Braden:1990wx}, which requires the existence of an on-shell diagram (with a particle, say $d$, propagating in the $t$-channel) obtained by projecting
$$
\alpha-\gamma= \delta = \beta-\gamma  \ \ \text{with} \ \delta \in \Gamma_d.
$$
By the fact that we are considering a simply-laced theory all the roots have the same length; therefore, $\alpha+\gamma$ and $\alpha-\gamma$ cannot be both roots, otherwise (normalising the root length by $\Lambda$) we obtain
$$
\Lambda^2= (\alpha+\gamma)^2= 2 \alpha^2 + 2 \gamma^2 - (\alpha - \gamma)^2 = 3 \Lambda^2
$$
which is clearly false. For this reason, it does not exist any particle $e$ such that $C^{(3)}_{ac \bar{e}}$ and $C^{(3)}_{\bar{b} \bar{c} e}$ are both nonzero. 
For the same argument there is no particle $d$ such that $C^{(3)}_{a\bar{c} \bar{d}}$ and $C^{(3)}_{c \bar{b} d}$ are both nonzero. This implies that there are no Feynman diagrams with particles propagating in the $s$ or $t$ channel in the process~\eqref{tree_level_degenerate_elastic_process_in_property}.

Also in the $u$ channel, no particles can propagate. 
The existence of a particle of type $k$ propagating in the $u$-channel would imply that $C^{(3)}_{a\bar{b} \bar{k}}$ and $C^{(3)}_{c\bar{c}k}$ are both nonzero. If this is the case, the following equalities are realized in the root system:
$$
\alpha-\beta= \kappa = \gamma - \tilde{\gamma}  \ \ \text{with} \ \alpha \in \Gamma_a , \, \beta \in \Gamma_b, \, \gamma \in \Gamma_c, \, \tilde{\gamma} \in \Gamma_c \ \text{and} \ \kappa \in \Gamma_k.
$$
The projection of these roots onto the spin-$1$ eigenplane of $w$ corresponds to an on-shell Feynman diagram associated with the process~\eqref{tree_level_degenerate_elastic_process_in_property} in the kinematical configuration in which the incoming- and outgoing-momenta are different. By the flipping rule,
this implies that $\alpha-\gamma=\beta-\tilde{\gamma}$ is a root or $\alpha+\tilde{\gamma}=\beta+\gamma$ is a root. In the first case, it has to exist a particle $e$ such that $C^{(3)}_{ac \bar{e}}$ and $C^{(3)}_{\bar{b} \bar{c} e}$ are both nonzeroes at a time; in the second case, it has to exist a particle $d$ such that $C^{(3)}_{a\bar{c} \bar{d}}$ and $C^{(3)}_{c \bar{b} d}$ are both nonzeroes at a time. Since the previous discussion forbids both situations, also in the $u$ channel no particles can propagate. We conclude that there are no Feynman diagrams containing $3$-point couplings contributing to the process~\eqref{tree_level_degenerate_elastic_process_in_property}. By relation~\eqref{Higher_order_couplings} also the $4$-point coupling $C^{(4)}_{a c \bar{b} \bar{c}}$ is then zero. For this reason, the tree-level amplitude associated with the process~\eqref{tree_level_degenerate_elastic_process_in_property} is null for all on-shell and off-shell values of the external particles since no diagrams contribute to this process.

\subsection{Cancellation of self-contracted vertices}

After having proven that properties~\ref{Condition_tree_level_elasticity_introduction} and~\ref{Non_degenerate_mass_condition} are satisfied by all the affine Toda theories, we show that all Feynman diagrams 
with contractions that connect back to their originating vertices are removed by renormalizing a single parameter in the Lagrangian. Part of this discussion is already known from~\cite{aa20}, however, it is worth reviewing it here using the cutting method described in the previous sections.
\begin{figure}
\begin{center}
\begin{tikzpicture}
\tikzmath{\y=1.9;}


\filldraw[black] (1.1*\y,0*\y)  node[anchor=west] {\LARGE{$\sum_{c,j}$}};
\draw[directed] (1.9*\y,0*\y) -- (2.5*\y,0*\y);
\draw[directed] (2*\y,-0.4*\y) -- (2.5*\y,0*\y);
\draw[directed] (3*\y,-0.4*\y) -- (2.5*\y,0*\y);
\draw[directed] (3.1*\y,0*\y) -- (2.5*\y,0*\y);
\filldraw[black] (2.35*\y,-0.3*\y)  node[anchor=west] {\scriptsize{$\ldots$}};
\filldraw[black] (1.8*\y,0.1*\y)  node[anchor=west] {\scriptsize{$a_1$}};
\filldraw[black] (1.8*\y,-0.3*\y)  node[anchor=west] {\scriptsize{$a_2$}};
\filldraw[black] (2.9*\y,-0.3*\y)  node[anchor=west] {\scriptsize{$a_{n-1}$}};
\filldraw[black] (2.9*\y,0.1*\y)  node[anchor=west] {\scriptsize{$a_n$}};
\filldraw[black] (2.5*\y,0.3*\y)  node[anchor=west] {\scriptsize{$j$}};
\draw[] (2.5*\y,0*\y) -- (2.5*\y,0.5*\y);
\draw[directed] (5.1*\y-2.6*\y,0.5*\y) arc(-90:270:0.2*\y);
\filldraw[black] (5*\y-2.6*\y,1*\y)  node[anchor=west] {\scriptsize{$c(k)$}};

\filldraw[black] (3.4*\y,0*\y)  node[anchor=west] {\scriptsize{$+$}};
\filldraw[black] (3.8*\y,0*\y)  node[anchor=west] {\LARGE{$\sum_{c}$}};

\draw[directed] (2.1*\y+2.4*\y,0*\y) -- (2.7*\y+2.4*\y,0*\y);
\draw[directed] (2.2*\y+2.4*\y,-0.4*\y) -- (2.7*\y+2.4*\y,0*\y);
\draw[directed] (3.2*\y+2.4*\y,-0.4*\y) -- (2.7*\y+2.4*\y,0*\y);
\draw[directed] (3.3*\y+2.4*\y,0*\y) -- (2.7*\y+2.4*\y,0*\y);
\draw[directed] (2.7*\y+2.4*\y,0*\y) arc(-90:270:0.2*\y);
\filldraw[black] (2*\y+2.4*\y,0.1*\y)  node[anchor=west] {\scriptsize{$a_1$}};
\filldraw[black] (2*\y+2.4*\y,-0.3*\y)  node[anchor=west] {\scriptsize{$a_2$}};
\filldraw[black] (2.55*\y+2.4*\y,-0.3*\y)  node[anchor=west] {\scriptsize{$\ldots$}};
\filldraw[black] (3.1*\y+2.4*\y,-0.3*\y)  node[anchor=west] {\scriptsize{$a_{n-1}$}};
\filldraw[black] (3.1*\y+2.4*\y,0.1*\y)  node[anchor=west] {\scriptsize{$a_n$}};
\filldraw[black] (2*\y+2.9*\y,0.55*\y)  node[anchor=west] {\scriptsize{$c(k)$}};

\filldraw[black] (3.8*\y+2.4*\y,0*\y)  node[anchor=west] {\scriptsize{$+$}};

\draw[directed] (6.9*\y,0*\y) -- (7.45*\y,0*\y);
\draw[directed] (7*\y,-0.4*\y) -- (7*\y+0.45*\y,-0.4*\y+0.36*\y);
\draw[directed] (8*\y,-0.4*\y) -- (7*\y+0.55*\y,-0.4*\y+0.36*\y);
\draw[directed] (8.1*\y,0*\y) -- (7.55*\y,0*\y);
\filldraw[black] (2.56*\y+4.8*\y,0*\y)  node[anchor=west] {\scriptsize{$\otimes$}};

\filldraw[black] (2*\y+4.8*\y,0.1*\y)  node[anchor=west] {\scriptsize{$a_1$}};
\filldraw[black] (2*\y+4.8*\y,-0.3*\y)  node[anchor=west] {\scriptsize{$a_2$}};
\filldraw[black] (2.55*\y+4.8*\y,-0.3*\y)  node[anchor=west] {\scriptsize{$\ldots$}};
\filldraw[black] (3.1*\y+4.8*\y,-0.3*\y)  node[anchor=west] {\scriptsize{$a_{n-1}$}};
\filldraw[black] (3.1*\y+4.8*\y,0.1*\y)  node[anchor=west] {\scriptsize{$a_n$}};

\filldraw[black] (3.6*\y+4.8*\y,0*\y)  node[anchor=west] {\scriptsize{$=$}};

\filldraw[black] (8.8*\y,0*\y)  node[anchor=west] {\scriptsize{$0$}};

\end{tikzpicture}
\caption{A counterterm proportional to $C^{(n)}_{a_1,\ldots, a_n}$ is needed to cancel tadpoles.}
\label{Image_for_tadpole_cancellation}
\end{center}
\end{figure}
These ultraviolet divergences are due to Feynman diagrams of the type depicted in the first two pictures on the l.h.s. of figure~\ref{Image_for_tadpole_cancellation}.
Using the cutting method described in the previous sections the sum of this pair of diagrams can be written as
\begin{equation}
\label{general_tadpole_diagrams}
    \frac{i}{8 \pi} \sum_c \int_{-\kappa}^\kappa d\theta_k \Bigl[ \sum_j C^{(n+1)}_{a_1 \ldots a_n \bar{j}} \frac{1}{m_j^2} C^{(3)}_{j c \bar{c}} - C^{(n+2)}_{a_1 \ldots a_n c \bar{c}} \Bigr].
\end{equation}
The loop propagator, carrying momentum $k$, has been cut and a cut-off $\kappa$ has been introduced. Using~\eqref{Higher_order_couplings} then equation~\eqref{general_tadpole_diagrams} becomes
\begin{equation}
\label{general_tadpole_diagrams_simplified}
    -\frac{i}{8 \pi} \frac{\g^2}{h} \sum_c \frac{m_c^2}{m^2}  C^{(n)}_{a_1 \ldots a_n} \int_{-\kappa}^\kappa d\theta_k.
\end{equation}
This quantity is cancelled by replacing
$$
C^{(n)}_{a_1 \ldots a_n} \to C^{(n)}_{a_1 \ldots a_n} + \delta C^{(n)}_{a_1 \ldots a_n}
$$
in the Lagrangian~\eqref{eq0_1}, where
\begin{equation}
    \delta C^{(n)}_{a_1 \ldots a_n} \equiv - \g^2 \frac{\delta m_{\text{tad}}^2}{m^2} C^{(n)}_{a_1 \ldots a_n}
\end{equation}
and
\begin{equation}
\label{definition_necessary_to_absorb_infinities_in_tadpoles}
    \delta m_{\text{tad}}^2 \equiv \frac{1}{8 \pi} \frac{1}{h} \sum_c m_c^2 \int_{-\kappa}^\kappa d\theta_k \, .
\end{equation}
The vertex associated with this counterterm is represented on the r.h.s. of figure~\ref{Image_for_tadpole_cancellation}. 
For any $n\ge 2$ and for any set of indices $\{a_1, \dots, a_n\}$ the counterterm necessary to cancel the divergent diagrams generating expression~\eqref{general_tadpole_diagrams_simplified} is obtained by modifying the Lagrangian~\eqref{Toda_theory_lagrangian_defined_in_terms_of_roots} as follows
\begin{equation}
\label{Toda_theory_lagrangian_defined_in_terms_of_roots_tadpoles_removed}
\mathcal{L}_0^{(\text{Toda})}=\frac{1}{2} \partial_\mu \phi_a  \partial^\mu \phi_a - \frac{m^2-\g^2 \delta m_{\text{tad}}^2}{\g^2} \Bigl( \sum_{i=0}^r n_i e^{\g \alpha^a_i  \phi_a} -h \Bigr).
\end{equation}
The renormalization of a single term in the Lagrangian of the affine Toda field theories is sufficient to remove all the UV divergences from the one-loop amplitudes. For this reason, when we compute amplitudes, we can omit all Feynman diagrams with contractions that connect back to their originating vertices, as we did in section~\ref{section_on_the_cutting_method}.

With this discussion we conclude that formulas~(\eqref{final_formula_we_need_to_prove_first_formulation},\eqref{final_formula_we_need_to_prove_second_formulation}) are valid in all the affine Toda fields theories; however, this does not imply that the one-loop inelastic amplitudes are zero. We still need to determine how to renormalize the couplings in such a way as to cancel the expressions in~(\eqref{final_formula_we_need_to_prove_first_formulation},\eqref{final_formula_we_need_to_prove_second_formulation}). 
While this problem is difficult in non-simply-laced theories, it simplifies considerably in simply-laced models. The reason is that in all simply-laced affine Toda theories, the masses renormalize scaling with the same multiplicative factor. Simply-laced models 
enter in the class of models discussed in section~\ref{Section_sufficient_conditions_absence_inelasticity_one_loop} and therefore are one-loop integrable. 
How the masses of simply-laced theories renormalize is known
from several years; a universal formula for their one-loop corrections was determined in~\cite{Braden:1989bu} based on a case-by-case study. 
However, a universal understanding of why all the masses scale with the same factor is unknown and will be covered in the next section.

\subsection{Mass renormalization in simply-laced affine Toda theories}

At the one-loop order, all the masses of simply-laced affine Toda theories are scaled with the same multiplicative factor. 
We provide universal proof of this fact. The relations discussed in the first part of this section were largely derived in~\cite{Dorey:1992tw}, though with slightly different conventions.
Instead, the universal proof of the renormalization of the masses is new.

Any root $\alpha$, whatever the orbit it belongs to, can be expressed as 
\begin{equation}
\label{Coxeter_geometry_general_root_alpha_written_as_a_sum_over_rcirc_and_rbullet}
\alpha=r^{(\alpha)}_\circ + r^{(\alpha)}_\bullet
\end{equation}
where $r^{(\alpha)}_\circ$ and $r^{(\alpha)}_\bullet$ depend on the root $\alpha$ we choose and are defined by
\begin{equation}
\label{definition_of_rcirc_and_rbullet}
r^{(\alpha)}_{\circ}=\sum_{i \in \circ} \frac{2}{\alpha_i^2}(\lambda_i, \alpha) \alpha_i \hspace{5mm} , \hspace{5mm} r^{(\alpha)}_{\bullet}=\sum_{i \in \bullet} \frac{2}{\alpha_i^2} (\lambda_i, \alpha) \alpha_i .
\end{equation}
From~\eqref{fundamental_weights_definition}, we see that this corresponds to writing  $\alpha$ as a linear combination of simple roots with integer coefficients.
The lengths of the projections of the vectors $r^{(\alpha)}_\circ$ and $r^{(\alpha)}_\bullet$ on the spin-$1$ eigenplane of $w$ depend on the root $\alpha$ we choose; however, the directions of these projections do not depend on $\alpha$. Indeed, from figure~\ref{Coxeter_geometry_projections_of_roots_on_the_spin_s_plans_representations_of_abullet_and_acirc} and equation~\eqref{projections_complex_numbers_notation}, we see that all the projections of the white roots have the same direction and therefore $r^{(\alpha)}_{\circ}$ (which is a combination of white roots) needs to have a projection $P_1(r^{(\alpha)}_{\circ})$ which is independent by $\alpha$. A similar argument applies to $r^{(\alpha)}_{\bullet}$.
This allows for finding additional constraints on the set of masses of the theory. Let us consider an integer $p \in [0,h-1]$ and a root $\gamma_c = \alpha_c \in \circ$; then projecting the relation
\begin{equation}
w^{-p}\alpha_c=r^{(w^{-p}\alpha_c)}_\circ + r^{(w^{-p}\alpha_c)}_\bullet
\end{equation}
onto the spin-$1$ eigenplane of $w$ and using simple trigonometric properties we obtain
\begin{equation}
\label{projection_rcirc_rbullet_relation_1}
\begin{split}
&|P_1(r^{(w^{-p} \alpha_c)}_\circ)|= \frac{\sin((1+2p) \theta_1)}{\sin \theta_1} |P_1(\alpha_c)| \, ,\\
&|P_1(r^{(w^{-p} \alpha_c)}_\bullet)|= \frac{\sin(2p\theta_1)}{\sin \theta_1} |P_1(\alpha_c)| \, .
\end{split}
\end{equation}
The angle $\theta_1$ is defined in~\eqref{eq1_9} and the projected vectors are represented in figure~\ref{Coxeter_geometry_projections_of_triangles_connecting_the_masses_through_rcirc_and_rbullet}.
\begin{figure}
\begin{center}
\begin{tikzpicture}
\tikzmath{\y=0.8;}

\draw[->,thick](0*\y,0*\y) -- (-6*\y,-3*\y);
\draw[->,thick](-6*\y,-3*\y) -- (-1*\y,-3*\y);
\draw[->,thick](0*\y,0*\y) -- (3.5*\y,0*\y);
\draw[->,thick](0*\y,0*\y) -- (-1*\y,-3*\y);
\draw[](0*\y,0*\y) -- (-4.5*\y,0*\y);

\draw[][] (0.6*\y,0*\y) arc(0:-105:0.6*\y);
\draw[][] (-0.6*\y,0*\y) arc(180:207:0.6*\y);
\draw[][] (-5.3*\y,-3*\y) arc(0:27:0.7*\y);
\draw[][] (-1.6*\y,-3*\y) arc(180:70:0.6*\y);
\filldraw[black] (0.4*\y,-0.6*\y)  node[anchor=west] {\tiny{$2 p \theta_1$}};
\filldraw[black] (-2.2*\y,-2.4*\y)  node[anchor=west] {\tiny{$2 p \theta_1$}};
\filldraw[black] (-1.3*\y,-0.2*\y)  node[anchor=west] {\tiny{$\theta_1$}};
\filldraw[black] (-5.3*\y,-2.8*\y)  node[anchor=west] {\tiny{$\theta_1$}};

\filldraw[black] (2*\y,0.35*\y)  node[anchor=west] {\footnotesize{$P_1(\alpha_c)$}};
\filldraw[black] (-0.8*\y,-2.4*\y)  node[anchor=west] {\footnotesize{$P_1(w^{-p}\alpha_c)$}};
\filldraw[black] (-3.7*\y,-3.4*\y)  node[anchor=west] {\footnotesize{$P_1(r^{(w^{-p} \alpha_c)}_\circ)$}};
\filldraw[black] (-5.2*\y,-1*\y)  node[anchor=west] {\footnotesize{$P_1(r^{(w^{-p} \alpha_c)}_\bullet)$}};

\filldraw[black] (2.6*\y,-2.4*\y)  node[anchor=west] {$c \in \circ$};

\end{tikzpicture}
\end{center}
\caption{Projections of $w^{-p} \alpha_c$, $r^{(w^{-p} \alpha_c)}_\circ$ and $r^{(w^{-p} \alpha_c)}_\bullet$ on the spin-$1$ eigenplane of the Coxeter element. }
\label{Coxeter_geometry_projections_of_triangles_connecting_the_masses_through_rcirc_and_rbullet}
\end{figure}
If we consider a simply-laced theory with roots normalized as follows
$$
\alpha^2=\Lambda^2 \ \forall \ \alpha \in \Phi \, ,
$$
then plugging~\eqref{definition_of_rcirc_and_rbullet} into~\eqref{projection_rcirc_rbullet_relation_1}, and using (\eqref{projections_complex_numbers_notation},\eqref{root_projection_rule}), we obtain that for any $c \in \circ$ it holds
\begin{equation}
\label{projection_rcirc_rbullet_masses_relation_2}
\begin{split}
\sum_{i \in \circ } (\lambda_i , w^{-p} \alpha_c) m_i &= \frac{\Lambda^2}{2}  \frac{\sin \bigl((1+2p) \theta_1 \bigr)}{\sin \theta_1} \ m_c \, ,\\
\sum_{i \in \bullet }  (\lambda_i , w^{-p} \alpha_c) m_i  &=  \frac{\Lambda^2}{2}\frac{\sin \bigl(2p \theta_1 \bigr)}{\sin \theta_1} \ m_c.
\end{split}\end{equation} 
From these relations, we note that the masses of the theory are connected to each other. If the length common to all the roots is $\Lambda=\sqrt{2}$ then the sum of the two equations in~\eqref{projection_rcirc_rbullet_masses_relation_2} returns a particular case of formula (3.5) of~\cite{Dorey:1992tw}.

Before performing the computation of the one-loop corrections to the masses we remind the following result from~\cite{Dorey:2021hub}: 
the tree-level amplitudes involving the scattering of a particle $a \in \circ$ with any other particle of the theory, after having removed infinite-rapidity contributions, as in~\eqref{hat_convention_in_two_to_two_amplitude_tree_level}, can be written as
\begin{equation}
\begin{split}
&\hat{M}^{(0)}_{ab \to ab}(\theta)=
\frac{2i \g^2}{h} m_a m_b \sum_{\substack{p \in \mathbb{N} \\ 0<1+2p< h}}(\lambda_b , w^{-p} \gamma_a) [2p+1]_\theta \hspace{8mm} \text{if} \ \ b \in \circ \, , \\
&\hat{M}^{(0)}_{ab \to ab}(\theta)=
\frac{2i \g^2}{h} m_a m_b \sum_{\substack{p \in \mathbb{N} \\ 0<2p< h}} (\lambda_b , w^{-p} \gamma_a) [2p]_\theta \hspace{19mm} \text{if} \ \ b \in \bullet \, ,
\end{split}
\end{equation}
where
\begin{equation}
    [x]_\theta \equiv - \frac{\sinh^2{ \Bigl( \frac{i \pi}{h} (x-1)\Bigr)}}{ \cosh{\theta} - \cos{ \bigl(\frac{\pi}{h} (x-1)} \bigl)} + \frac{\sinh^2{ \Bigl( \frac{i \pi}{h} (x+1)\Bigr)}}{\cosh{\theta} - \cos{ \bigl( \frac{\pi}{h} (x+1)} \bigl)} \, .
\end{equation}
$\theta$ is the difference between the rapidities of the particles $a$ and $b$.
Using this fact, together with~\eqref{mass_renormalization_formula_on_shell}, we can write
\begin{equation}
\begin{split}
\delta m_a^2= 
\frac{1}{4 \pi} \frac{\g^2}{h} m_a
& \biggl(\sum_{\substack{p \in \mathbb{N} \\ 0<2p+1< h}}  \sum_{b \in \circ} m_b (\lambda_a , w^{-p} \gamma_b) \int_{-\infty}^{+\infty} d\theta [2p+1]_\theta \\
+&\sum_{\substack{p \in \mathbb{N} \\ 0<2p< h}}  \sum_{b \in \bullet} m_b (\lambda_a , w^{-p} \gamma_b) \int_{-\infty}^{+\infty} d\theta [2p]_\theta \biggr) \, .
\end{split}
\end{equation}
Plugging~\eqref{projection_rcirc_rbullet_masses_relation_2} into this equation we obtain
\begin{equation}
\label{universal_mass_renormalization_simply_laced_theory_1}
\begin{split}
\delta m_a^2= 
\frac{\Lambda^2}{8 \pi} \frac{\g^2}{h} m^2_a \biggl(
&\sum_{\substack{p \in \mathbb{N} \\ 0<2p+1< h}} \frac{\sin{\bigl( (1+2p) \theta_1\bigr)}}{\sin{\theta_1}} \int_{-\infty}^{+\infty} d\theta [2p+1]_\theta \\
&+\sum_{\substack{p \in \mathbb{N} \\ 0<2p< h}} \frac{\sin{( 2p \theta_1 )}}{\sin{\theta_1}} \int_{-\infty}^{+\infty} d\theta [2p]_\theta \biggr)\\
&=\frac{\Lambda^2}{8 \pi \sin{\theta_1}} \frac{\g^2}{h} m^2_a \sum_{\substack{p \in \mathbb{N} \\ 0<p< h}} \sin{\bigl( p \theta_1\bigr)} \int_{-\infty}^{+\infty} d\theta [p]_\theta.
\end{split}
\end{equation}
From this last equality, we note that the one-loop correction to the squared of the mass of the particle $a$ is proportional to $m^2_a$ with a proportionality coefficient independent by $a$. This coefficient is easily computed noting that
\begin{equation}
\label{building_block_integrated}
    \int^{+\infty}_{-\infty} d\theta \ [p]_\theta = \bigl( 2 \pi -2 \theta_1 (p-1) \bigl) \sin{\Bigl( \theta_1 (p-1) \Bigr)}-\bigl( 2 \pi -2 \theta_1 (p+1) \bigl) \sin{\Bigl( \theta_1 (p+1) \Bigr)}.
\end{equation}
Substituting~\eqref{building_block_integrated} in~\eqref{universal_mass_renormalization_simply_laced_theory_1} we end up with
\begin{equation}
\label{universal_mass_renormalization_simply_laced_theory_2}
\begin{split}
\delta m_a^2&=\frac{\Lambda^2}{4 \sin{\theta_1}} \frac{\g^2}{h} m^2_a  \sum_{\substack{p \in \mathbb{N} \\ 0<p< h}} \Bigl( \sin{(p \theta_1)} \sin{( (p-1) \theta_1)} - \sin{((p+1) \theta_1)} \sin{( p \theta_1)} \Bigr)\\
&+\frac{\Lambda^2 \theta_1}{4 \pi \sin{\theta_1}} \frac{\g^2}{h} m^2_a  \sum_{\substack{p \in \mathbb{N} \\ 0<p< h}} \Bigl( (p+1) \sin{(p \theta_1)} \sin{((p+1) \theta_1)}- (p-1) \sin{((p-1) \theta_1)} \sin{(p \theta_1)}  \Bigr)
\end{split}
\end{equation}
By replacing $p$ with $p+1$ in the first term of~\eqref{universal_mass_renormalization_simply_laced_theory_2} we see that the contribution in the first-row of~\eqref{universal_mass_renormalization_simply_laced_theory_2} is zero.
The second row is instead different from zero and performing the substitution $p \to p+1$ in the last term of~\eqref{universal_mass_renormalization_simply_laced_theory_2} we obtain
\begin{equation}
\label{universal_mass_renormalization_simply_laced_theory_3}
\begin{split}
\delta m_a^2 &=\frac{\Lambda^2 \theta_1}{4 \pi \sin{\theta_1}} \frac{\g^2}{h} m^2_a  \sum_{p=0}^{h-1}   \sin{(p \theta_1)} \sin{((p+1) \theta_1)}\\
&=\frac{\Lambda^2 \theta_1}{8 \pi \sin{\theta_1}} \frac{\g^2}{h} m^2_a  \sum_{p=0}^{h-1}  \Bigl( \cos{\theta_1}- \cos{((2p+1) \theta_1)} \Bigl)=\frac{\Lambda^2 \theta_1}{8 \pi} \g^2 m^2_a  \cot{\theta_1}
\end{split}
\end{equation}
The last equality in~\eqref{universal_mass_renormalization_simply_laced_theory_3} is obtained by noting that the sum of $\cos{((2p+1) \theta_1)}$ over all $p \in \{0, 1, \dots ,h-1\}$ is null.

We have determined the one-loop corrections to the masses associated with the `white' roots in simply-laced Dynkin diagrams.
Note that each Dynkin diagram can be split into `white' and `black' roots in two different manners: for a given choice of `white' and `black' roots, there exists another legitimate choice determined by exchanging black roots and white roots.
For example, we can exchange white and black roots in the Dynkin diagram in figure~\ref{Example_of_Dynkin_diagrams_with_balck_and_white_roots}. All the previous discussion is valid no matter the choice we adopt.
For this reason, relation~\eqref{universal_mass_renormalization_simply_laced_theory_3} is valid for any index $a \in \{ 1, \dots, r\}$ and all the masses renormalise with the same multiplicative factor.
Following the notation in~\eqref{delta_m_square_j_does_not_depend_on_j}, it holds that
$$
\lambda= 1 +\frac{\Lambda^2 \theta_1}{8 \pi} \g^2  \cot{\theta_1}.
$$
All the surviving contributions of inelastic amplitudes of simply-laced affine Toda theories can therefore be cancelled by renormalizing the couplings as in~\eqref{coupling_integrable_renormalization}. This corresponds to the following transformation of the mass scale and roots of the affine Dynkin diagram
\begin{equation}
\label{root_system_renormalization_simply_laced}
    m^2 \to \lambda^{3 - 2 \rho} \cdot m^2 \hspace{4mm},\hspace{4mm}  \alpha_i \to \lambda^{\rho-1} \cdot \alpha_i.
\end{equation}
In~\cite{Braden:1990qa} it was observed that the choice $\rho=1$ produced $2$-to-$2$ one-loop elastic S-matrices with residues at simple poles that were in agreement with the results bootstrapped in~\cite{Braden:1989bu}.
With this choice, the root system does not renormalise and the masses and couplings are all scaled with the same factor 
$$
m_a \to \lambda m_a \hspace{5mm},\hspace{5mm} C_{a_1 \dots a_n}^{(n)} \to \lambda C_{a_1 \dots a_n}^{(n)} \, ,
$$
leading to the following renormalized Lagrangian
\begin{equation}
\label{Toda_theory_renormalized_lagrangian}
\mathcal{L}^{(\text{Toda})}=\frac{1}{2} \partial_\mu \phi_a  \partial^\mu \phi_a - \frac{\lambda m^2-\g^2 \delta m_{\text{tad}}^2}{\g^2} \Bigl( \sum_{i=0}^r n_i e^{\g \alpha^a_i  \phi_a} -h \Bigr) \, .
\end{equation}
In~\eqref{Toda_theory_renormalized_lagrangian} we omitted to write the contributions arising by the renormalization of the fields (i.e. the coloured contributions in~\eqref{renormalised_Lagrangian_due_to_standard_renormalisation_procedure_plus_integrable_corrections}), which should also be included: they can be easily included through the field redefinition defined in~\eqref{mass_renormalization_tab_definition}.

\section{Conclusions}
\label{Concluion_section}

A method to generate one-loop amplitudes in terms of tree-level amplitudes has been investigated and applied to integrable models with Lagrangians of type~\eqref{eq0_1} and satisfying properties~\ref{Condition_tree_level_elasticity_introduction} and~\ref{Non_degenerate_mass_condition}. In these models, it is proven that all the renormalized one-loop inelastic amplitudes are equal to their corresponding tree-level amplitudes in which the masses of the external particles and propagators are corrected by one-loop bubble diagrams, as shown in equation~\eqref{final_formula_we_need_to_prove_second_formulation}. 
It is proven that all the bosonic affine Toda field theories belong to this class of models.
Despite we analysed inelastic processes with two incoming particles and $n\ge 2$ outgoing particles, it is possible to show that formula~\eqref{final_formula_we_need_to_prove_second_formulation} is valid in all inelastic processes. Particularly interesting are the processes in which a single particle, say $c$, decays into two particles, $a$ and $b$.
In this case, \eqref{final_formula_we_need_to_prove_second_formulation} becomes
\begin{equation}
\begin{split}
    M_{c \to a b}&= M^{(0)}_{c \to a b}\Bigl|_{m_j^2-\Sigma^{(0)}_{jj}} + \ O\Bigl((\Sigma^{(0)}_{jj})^2 M^{(0)}_{c\to ab}\Bigl|_{m_j^2}  \Bigr).
\end{split}
\end{equation}
The three-point tree-level amplitude on the r.h.s. of this relation is just a coupling $-iC^{(3)}_{c \bar{a} \bar{b}}$ and does not depend on the masses. Therefore, it holds that
\begin{equation}
\label{1_to_2_inelastic_amplitude}
\begin{split}
    M_{c \to a b}=-iC^{(3)}_{c \bar{a} \bar{b}} + \ O\Bigl((\Sigma^{(0)}_{jj})^2 C^{(3)}_{c \bar{a} \bar{b}} \Bigl).
\end{split}
\end{equation}
For physical values of the external particles, this amplitude, to one-loop order, is identical to a three-point coupling. By property~\ref{Condition_tree_level_elasticity_introduction}, this coupling is zero any time $m_c>m_a+m_b$ otherwise the decay $c \to a+b$ would be allowed at the tree level. The amplitude~\eqref{1_to_2_inelastic_amplitude} is therefore null for physical values of the external particles. This observation generalises  results that were obtained in~\cite{Braden:1991vz} on a case-by-case study performed over different models.

We remark that the one-loop inelastic amplitudes reproduced by relation~\eqref{final_formula_we_need_to_prove_second_formulation} are nonzero in general and counterterms need to be added to the Lagrangian (in addition to the standard counterterms required by the renormalization procedure) to preserve integrability at one loop. 
We evaluated these counterterms in models preserving the mass ratios at one loop. 
We showed that the class of simply-laced affine Toda field theories belong to this class of models and therefore are one-loop integrable, in the sense that all one-loop inelastic amplitudes are zero. 

It would be interesting to study theories which do not preserve the mass ratios at one loop, as the non-simply-laced affine Toda models.
Also in this case, formula~\eqref{final_formula_we_need_to_prove_second_formulation} reproduces inelastic amplitudes; however, the Lagrangian corrections needed to cancel these amplitudes have not been investigated so far. Using~\eqref{final_formula_we_need_to_prove_second_formulation} it is possible to understand how to determine these corrections and hopefully observe the flowing between dual pairs of non-simply-laced theories at the Lagrangian level, as conjectured in~\cite{Delius:1991kt} and further investigated in~\cite{Corrigan:1993xh,Dorey:1993np,Oota:1997un}. In those papers, it was suggested that each dual-pair of Dynkin diagrams, i.e. diagrams connected by a transformation
$$
\alpha_i \to 2 \frac{\alpha_i}{\alpha_i^2} \, ,
$$
corresponds to a unique quantum field theory. The non-perturbative S-matrix of this quantum theory should interpolate between the tree-level S-matrices emerging from the Lagrangians~\eqref{Toda_theory_lagrangian_defined_in_terms_of_roots} associated with the pair of dual Dynkin diagrams in the limit of large and small coupling $\g$.

While in simply-laced models the counterterms necessary for the absence of inelastic processes at one-loop are introduced by the transformation~\eqref{root_system_renormalization_simply_laced}, which changes the root system only through an overall scale, it is expected that in non-simply-laced models the root system renormalises in a non-trivial way. This renormalization should correspond to a flow connecting dual pairs of Dynkin diagrams. In particular, it would be interesting the check if the flipping rule used in~\cite{Dorey:2021hub} to prove the cancellation of all inelastic processes at the tree level, is preserved at one-loop order in perturbation theory for the class of non-simply-laced models.

Another direction which should be pursued is the generalisation of the results obtained in this paper to elastic amplitudes. A formula to obtain one-loop amplitudes in terms of tree-level amplitudes of integrable models was proposed in~\cite{Bianchi:2014rfa}; however, as already mentioned in section~\ref{section_on_the_cutting_method}, if the tree-level S-matrices are purely-elastic, the formula proposed in~\cite{Bianchi:2014rfa} predicts that any time $M^{(0)}_{ab \to ab}$ has a pole of order one then $M^{(1)}_{ab \to ab}$ has a pole of order two. This statement cannot be true in general and the Bullough–Dodd model provides a simple example of a theory in which that formula fails. This could be due to some rational terms not captured by the unitarity cut method used in~\cite{Bianchi:2014rfa} or to the regularisation prescription adopted in that paper to avoid ill-defined cuts. It would be interesting to investigate elastic amplitudes for the simple class of models studied in this paper using the cutting approach described in section~\ref{section_on_the_cutting_method}; differently from unitarity cuts, this method corresponds to a different manner to write sums of one-loop diagrams and reproduces complete results for the amplitudes. It would be interesting to derive a universal formula for elastic amplitudes using this technique and compare the result with the formula obtained in~\cite{Bianchi:2014rfa} to check possible discrepancies. 

Another natural direction is the study of integrable models with polynomial-like interactions containing fermions. 
Examples of these models are provided by supersymmetric affine Toda theories, whose non-perturbative S-matrices were bootstrapped in~\cite{Delius:1990ij,Delius:1991sv} and have a mass spectrum which behaves well under one-loop corrections~\cite{Grisaru:1990sv}.

\vskip 20pt
\noindent

{\bf Acknowledgments}\\[3pt]
%
DP thanks Patrick Dorey, Giuseppe Mussardo, Alessandro Sfondrini and especially Ben Hoare for the many discussions on the construction of loop amplitudes from tree-level amplitudes which were the motivation for the work developed in this paper; the author thanks also Patrick Dorey for the suggesting idea of investigating if the flipping-rule is preserved at one-loop, mentioned in the conclusions, and Patrick Dorey and Alessandro Sfondrini for comments on an earlier draft of this work.
DP is grateful to the Kavli Institute for Theoretical Physics in Santa Barbara for the hospitality during the Integrable22 workshop, where part of this work was carried out.
This work has received funding from the  European Union's  Horizon  2020  research  and  innovation programme under the Marie  Sk\l odowska-Curie  grant  agreement  No.~764850 \textit{``SAGEX''}.
It was also supported in part by the National Science Foundation under Grant No.\ NSF PHY-1748958 and by 
STARS@UNIPD, under project ``Exact-Holography''.

\vskip 20pt

\appendix

\section{Counterterms for different particles with equal masses}
\label{Appendix_on_counterterms_for_different_particles_with_equal_masses}

In this appendix, we generalise the results of section~\ref{section_to_explain_the_mass_and_field_renormalization} to the case in which property~\ref{Non_degenerate_mass_condition} is not satisfied. 
As discussed in section~\ref{section_to_explain_the_cut_method},  given two particles of different types $c \ne d$ with $m_c=m_d$, property~\ref{Condition_tree_level_elasticity_introduction} implies $\Sigma^{(0)}_{cd}=0$ while property~\ref{Non_degenerate_mass_condition} implies $\Sigma^{(0)}_{cd}=\Sigma^{(1)}_{cd}=0$.
Therefore, the violation of property~\ref{Non_degenerate_mass_condition} allows for $\Sigma^{(1)}_{cd}$ to take values different from zero.
Due to this fact, the second definition in~\eqref{definition_of_mass_renormalization_and_tab_to_diagonalise_the_mass_matrix} needs to be modified to make equation~\eqref{renormalisation_condition_on_propagators} satisfied;
below we explain how to determine the coefficient $t_{cd}$ in the degenerate case in  which $c\ne d$ and $m_c=m_d$.

Labelling by $m$ the common mass of particles $c$ and $d$ and expanding the propagator $G_{cd}(p^2)$ near the pole $p^2=m^2$ we obtain
\begin{equation}
\label{degenerate_propagator_for_different_particles_with_the_same_mass}
    G_{cd}(p^2)= \frac{i \Sigma^{(1)}_{cd}}{p^2 -m^2} +  \frac{i t_{\bar{d} \bar{c}}}{p^2-m^2} + \frac{i t_{c d}}{p^2-m^2} +\dots 
\end{equation}
where the ellipses contain finite contributions at the pole.
Due to~\eqref{crossing_simmetry_on_loop_propagator} and the fact that $m_c=m_d$ it holds that $\Sigma^{(1)}_{cd}=\Sigma^{(1)}_{\bar{d} \bar{c}}$.
The choice cancelling the pole in~\eqref{degenerate_propagator_for_different_particles_with_the_same_mass} is therefore
\begin{equation}
\label{tab_for_particles_with_the_same_mass}
\begin{split}
    t_{cd}&\equiv-\frac{1}{2} \Sigma^{(1)}_{cd}+\lambda_{cd},\\
    t_{\bar{d} \bar{c}}&\equiv-\frac{1}{2} \Sigma^{(1)}_{\bar{d} \bar{c}}+\lambda_{\bar{d} \bar{c}},
\end{split}
\end{equation}
with
$$
\lambda_{\bar{d} \bar{c}}=-\lambda_{c d}.
$$
With this choice, the singular part of the propagator cancels at the pole and~\eqref{renormalisation_condition_on_propagators} is satisfied.
In the following, we discuss how inelastic amplitudes are affected by the presence of these nonvanishing coefficients $t_{cd}$ and $t_{\bar{d} \bar{c}}$.

Compared with the analysis presented in section~\ref{section_to_explain_the_mass_and_field_renormalization}, the only possible differences come from Feynman diagrams containing counterterms coloured red in figures~\ref{Image_all_propagators_and_two_point_vertices_of_renormalized_theory} and~\ref{Image_all_propagators_and_multiple_point_vertices_of_renormalized_theory}. Examples of these Feynman diagrams are depicted in figures~\ref{Image_cancellation_of_red_contributions_in_diagrams} and~\ref{Image_cancellation_of_red_contributions_in_external_legs} and were computed in section~\ref{section_to_explain_the_mass_and_field_renormalization}. The sum of pictures in
figure~\ref{Image_cancellation_of_red_contributions_in_diagrams} is still zero. 
Instead, for the pictures in figure~\ref{Image_cancellation_of_red_contributions_in_external_legs} the discussion changes slightly. If we write~\eqref{External_legs_off_diagonal_counterterms_diagram_two} as
\begin{equation}
\label{External_legs_off_diagonal_counterterm_D5_two_eq_masses}
D^{(5)}= -i \sum_{\substack{k=1 \\ k\neq j}}^r t_{j k} C^{(n+1)}_{a_1 \dots a_{n} k} -i \sum_{\substack{k=1 \\ k\neq j}}^r t_{\bar{k} \bar{j}} \frac{p^2 -m^2_j}{p^2 - m^2_k} C^{(n+1)}_{a_1 \dots a_{n} k}
\end{equation}
we note that the first term on the r.h.s. of~\eqref{External_legs_off_diagonal_counterterm_D5_two_eq_masses} cancels the contribution~\eqref{External_legs_off_diagonal_counterterms_diagram_one} as expected, but the second term is nonzero in general for $p^2=m^2_j$. In the second sum in~\eqref{External_legs_off_diagonal_counterterm_D5_two_eq_masses} certain particles $k$ may have the same mass as the external particle $j$.
For this reason, setting $p^2=m^2_j$ we obtain
\begin{equation}
\label{External_legs_off_diagonal_counterterm_D4_plus_D5_two_eq_masses}
D^{(4)}+D^{(5)}= \sum_{\substack{ k\neq j \\ m_k = m_j}} t_{\bar{k} \bar{j}}  \bigl(-i C^{(n+1)}_{a_1 \dots a_{n} k} \bigr)
\end{equation}
Due to this fact, Feynman diagrams containing counterterms coloured red in
figures~\ref{Image_all_propagators_and_two_point_vertices_of_renormalized_theory} and~\ref{Image_all_propagators_and_multiple_point_vertices_of_renormalized_theory} contribute to the amplitude associated with process~\eqref{2_to_nminus2_production_process_Introduction} with
\begin{equation}
\label{red_contribution_production_amplitude_with_degenerate_masses}
\begin{split}
    M^{(\text{red})}_{a_1 a_2 \to a_3 \dots a_n}&= \sum_{\substack{ k\neq a_1 \\ m_k = m_{a_1}}} t_{\bar{k} \bar{a}_1} M^{(0)}_{k a_2 \to a_3 \dots a_{n-1} a_n}\\
    &+ \sum_{\substack{ k\neq a_2 \\ m_k = m_{a_2}}} t_{\bar{k} \bar{a}_2} M^{(0)}_{a_1 k \to a_3 \dots a_{n-1} a_n}\\
    &+ \sum_{\substack{ k\neq \bar{a}_3 \\ m_k = m_{a_3}}} t_{\bar{k} a_3} M^{(0)}_{a_1 a_2 \to \bar{k} \dots a_{n-1} a_n}\\
    &\dots\\
    &+ \sum_{\substack{ k\neq \bar{a}_n \\ m_k = m_{a_n}}} t_{\bar{k} a_n} M^{(0)}_{a_1 a_2 \to a_3 \dots a_{n-1} \bar{k}}.
\end{split}
\end{equation}
If $n\ge4$ the expression above is zero since the sum on the r.h.s. is performed over tree-level production amplitudes, which are all zero by condition~\ref{Condition_tree_level_elasticity_introduction}. The case $n=4$ is a bit special and will be considered separately in the next section.

\subsection{Two-to-two inelastic processes with degenerate masses}
\label{Appendix_on_counterterms_for_different_particles_with_equal_masses_particular_process}

Let us consider the inelastic process
\begin{equation}
\label{degenerate_process_renormalization_of_fields_appendix}
a(p)+b(p')\to c(p)+b(p'),
\end{equation}
with $a \ne c$ and $m_a=m_c$. 
In this case, expression~\eqref{red_contribution_production_amplitude_with_degenerate_masses} becomes
\begin{equation}
\label{inelastic_process_degenerate_red_contribution}
\begin{split}
     M^{(\text{red})}_{a b \to c b}&= t_{\bar{c} \bar{a}} M^{(0)}_{c b \to c b}+ t_{a c} M^{(0)}_{a b \to a b}\\
     &=-\frac{1}{2} \Sigma^{(1)}_{ac} \bigl(M^{(0)}_{a b \to a b}+M^{(0)}_{c b \to c b}\bigl) +\lambda_{ac} \bigl(M^{(0)}_{a b \to a b}-M^{(0)}_{c b \to c b}\bigl).
\end{split}
\end{equation}
All the tree-level inelastic amplitudes that should appear on the r.h.s. of~\eqref{inelastic_process_degenerate_red_contribution} have been omitted since they are all zero by condition~\ref{Condition_tree_level_elasticity_introduction}.
We note that equation~\eqref{inelastic_process_degenerate_red_contribution} is nonzero in general since it contains two tree-level elastic amplitudes.
For $a$ and $c$ fixed, there are $r$ different processes of the type in~\eqref{degenerate_process_renormalization_of_fields_appendix}, one for each $b \in \{1, \dots, r\}$. For this reason, we have
$r$ quantities of the type in equation~\eqref{inelastic_process_degenerate_red_contribution}, each contributing to a different inelastic amplitude. 
Since in general the two combinations of amplitudes in parenthesis in~\eqref{inelastic_process_degenerate_red_contribution} should be independent functions of the on-shell momenta $p$ and $p'$, to have these $r$ contributions cancelling all together it should hold
$$
\lambda_{ac}=\Sigma^{(1)}_{ac}=0.
$$
While we are free to set $\lambda_{ac}=0$, we are not free to choose the value of $\Sigma^{(1)}_{ac}$, which depends on the theory under consideration. 
The vanishing of this quantity 
is a necessary condition for the universal validity of relation~\eqref{final_formula_we_need_to_prove_second_formulation} for all inelastic processes.
Indeed, formula~\eqref{final_formula_we_need_to_prove_second_formulation} receives a correction~\eqref{inelastic_process_degenerate_red_contribution} when it is applied to the process~\eqref{degenerate_process_renormalization_of_fields_appendix}. 

In all bosonic affine Toda theories condition~\ref{Non_degenerate_mass_condition} is satisfied and $\Sigma^{(1)}_{ac}=0$ whenever $a$ and $c$ are different particles with the same mass. This guarantees that contributions of the type in~\eqref{inelastic_process_degenerate_red_contribution} are always zero for these models.

\section{Potential ill-defined contributions in single cuts}
\label{Potential_ill_defined_contributions_in_single_cuts}

In section~\ref{section_to_explain_two_to_two_inelastic_processes}, we studied inelastic processes of type~\eqref{generic_scattering_process_ab_into_cd} and we showed that in the degenerate case in which $m_a=m_c$ and $m_d=m_b$ potential singularities can appear in~\eqref{4_point_one_loop_from_tree_double_cut_contribution}.
These singularities were avoided by introducing a regulator $x$ of the form in~\eqref{definition_of_my_regulator_x}, of small length $\mu$ and real rapidity $\theta_x$, and requiring~\eqref{regularization_through_x_introducing_outgoing_mass_deformations}. Now we focus on the same type of inelastic processes, with $m_a=m_c$ and $m_d=m_b$, and we show that the same regulator prevents singularities 
also in the integrands of~\eqref{4_point_one_loop_from_tree_single_cut_contribution}.
In this case, potential singularities are generated by cutting a pair of propagators, $e$ and $f$, of the same mass $m$, as shown in figure~\ref{relevant_terms_in_single_cut_t_channel_diagram}. 
We will show that collections of cut diagrams contributing to figure~\ref{relevant_terms_in_single_cut_t_channel_diagram} are not only finite but are zero. 

Following the same approach of section~\ref{section_to_explain_two_to_two_inelastic_processes}, we require~\eqref{regularization_through_x_introducing_outgoing_mass_deformations} with $a$ and $b$ on-shell; in this manner it holds that
\begin{subequations}
\label{r_and_rprime_written_in_terms_of_k_and_x}
\begin{equation}
\label{r_written_in_terms_of_k_and_x}
r=k-x,
\end{equation}
\begin{equation}
\label{rprime_written_in_terms_of_k_and_x}
r'=k+x
\end{equation}
\end{subequations}
where $k$ is the on-shell momentum of the cut propagators $e$ and $f$ in figure~\ref{relevant_terms_in_single_cut_t_channel_diagram}.
\begin{figure}
\begin{center}
\begin{tikzpicture}
\tikzmath{\y=1.4;}

\draw[directed] (4.6*\y,0.5*\y) -- (5.2*\y,1.1*\y);
\draw[directed] (5.2*\y,1.8*\y) -- (4.6*\y,2.4*\y);
\draw[directed] (8.8*\y,0.6*\y) -- (8.2*\y,1.2*\y);
\draw[directed] (8.2*\y,1.8*\y) -- (8.8*\y,2.4*\y);
\filldraw[color=gray!60, fill=gray!5, very thick](5.6*\y,1.5*\y) circle (0.4*\y);
\draw[directed] (6.5*\y,2.1*\y) -- (6*\y,1.7*\y);
\draw[directed] (7.4*\y,1.7*\y) -- (6.9*\y,2.1*\y);
\draw[directed] (5.95*\y,1.3*\y) arc(240:300:1.5*\y);
\filldraw[color=gray!60, fill=gray!5, very thick](7.8*\y,1.5*\y) circle (0.4*\y);

\filldraw[black] (4.6*\y,0.3*\y)  node[anchor=west] {\scriptsize{$a(p)$}};
\filldraw[black] (4.6*\y,2.5*\y)  node[anchor=west] {\scriptsize{$c(q)$}};
\filldraw[black] (8.2*\y,0.3*\y)  node[anchor=west] {\scriptsize{$b(p')$}};
\filldraw[black] (8.2*\y,2.5*\y)  node[anchor=west] {\scriptsize{$d(q')$}};
\filldraw[black] (6.15*\y,2.2*\y)  node[anchor=west] {\scriptsize{$e(k)$}};
\filldraw[black] (6.75*\y,2.2*\y)  node[anchor=west] {\scriptsize{$e(k)$}};
\filldraw[black] (6.4*\y,0.9*\y)  node[anchor=west] {\scriptsize{$f(r)$}};


\draw[directed] (9.6*\y,0.5*\y) -- (10.2*\y,1.1*\y);
\draw[directed] (10.2*\y,1.8*\y) -- (9.6*\y,2.4*\y);
\draw[directed] (13.8*\y,0.6*\y) -- (13.2*\y,1.2*\y);
\draw[directed] (13.2*\y,1.8*\y) -- (13.8*\y,2.4*\y);
\draw[directed] (12.45*\y,1.7*\y) arc(60:120:1.5*\y);
\draw[directed] (10.8*\y,1.2*\y) -- (11.2*\y,0.7*\y);
\filldraw[color=gray!60, fill=gray!5, very thick](10.6*\y,1.5*\y) circle (0.4*\y);
\draw[directed] (12*\y,0.7*\y) -- (12.4*\y,1.2*\y);
\filldraw[color=gray!60, fill=gray!5, very thick](12.8*\y,1.5*\y) circle (0.4*\y);
\filldraw[black] (9.6*\y,0.3*\y)  node[anchor=west] {\scriptsize{$a(p)$}};
\filldraw[black] (9.6*\y,2.5*\y)  node[anchor=west] {\scriptsize{$c(q)$}};
\filldraw[black] (13.2*\y,0.3*\y)  node[anchor=west] {\scriptsize{$b(p')$}};
\filldraw[black] (13.2*\y,2.5*\y)  node[anchor=west] {\scriptsize{$d(q')$}};
\filldraw[black] (11.4*\y,2.1*\y)  node[anchor=west] {\scriptsize{$e(r')$}};
\filldraw[black] (11*\y,0.6*\y)  node[anchor=west] {\scriptsize{$f(k)$}};
\filldraw[black] (11.7*\y,0.6*\y)  node[anchor=west] {\scriptsize{$f(k)$}};

\end{tikzpicture}
\caption{Pair of single cuts which are singular when $m_c=m_a$, $m_d=m_b$, $m_f=m_e$ and we are on the branch of the kinematics $\{q=p, q'=p'\}$. Each blob corresponds to a tree-level amplitude.}
\label{relevant_terms_in_single_cut_t_channel_diagram}
\end{center}
\end{figure}
The contribution associated with the combination of cut Feynman diagrams in figure~\ref{relevant_terms_in_single_cut_t_channel_diagram} is
\begin{equation}
\label{single_cut_relevant_contribution_u_channel_not_yet_expanded}
\begin{split}
&\frac{1}{8\pi}  \int d\theta_k \ M^{(0)}_{ae \to c f}(p,k,q,r) \ \Pi^{(R)}_f(r) \ M^{(0)}_{b f \to de }(p', r, q', k) \ +\\
&\frac{1}{8\pi}  \int d\theta_k \ M^{(0)}_{ae \to cf}(p, r', q, k) \ \Pi^{(R)}_e(r') \  M^{(0)}_{bf \to de }(p', k, q', r').
\end{split}
\end{equation}
Using the fact that $k^2=m^2$, the retarded propagators in~\eqref{single_cut_relevant_contribution_u_channel_not_yet_expanded} can be written as
\begin{equation}
\label{expansion_of_propagators_in_u_channel}
    \begin{split}
        &\Pi_f^{(R)}(r)= \frac{i}{-2 k \cdot x +\mu^2}= -\frac{i}{2 k \cdot x}-\frac{i \mu^2}{4 (k \cdot x)^2} + O(\mu),\\
        &\Pi_e^{(R)}(r')= \frac{i}{+2 k \cdot x +\mu^2}= +\frac{i}{2 k \cdot x}-\frac{i \mu^2}{4 (k \cdot x)^2}+ O(\mu).
    \end{split}
\end{equation}
The $i \epsilon$ factors in the denominators of propagators in~\eqref{expansion_of_propagators_in_u_channel} have been omitted since, as explained in section~\ref{section_to_explain_two_to_two_inelastic_processes}, $r$ and $r'$ cannot be on-shell for $\mu$ positive and $\theta_x$ real. Plugging~\eqref{expansion_of_propagators_in_u_channel} into~\eqref{single_cut_relevant_contribution_u_channel_not_yet_expanded} we obtain
\begin{equation}
\label{single_cut_relevant_contribution_u_channel_propagators_expanded}
\begin{split}
\frac{i}{16\pi}  \int \frac{d\theta_k}{k \cdot x} \ \Bigl( &-M^{(0)}_{ae \to c f}(p, k, q, r) \ M^{(0)}_{b f \to de }(p', r, q', k)\\
&+M^{(0)}_{ae \to cf}(p, r', q, k) \  M^{(0)}_{bf \to de }(p', k, q', r') \Bigr)\\
-\frac{i}{32 \pi} \int \frac{d\theta_k \mu^2}{(k \cdot x)^2} \ \Bigl( &+M^{(0)}_{ae \to c f}(p, k, q, r) \ M^{(0)}_{b f \to de }(p', r, q', k)\\
&+M^{(0)}_{ae \to cf}(p, r', q, k) \  M^{(0)}_{bf \to de }(p', k, q', r') \Bigr).
\end{split}
\end{equation}
The amplitudes appearing 
in the integrand of~\eqref{single_cut_relevant_contribution_u_channel_propagators_expanded} should be expanded around $\mu = 0$. 

In the general case in which $a\ne c$ and $ b \ne d$ all the tree-level amplitudes appearing in~\eqref{single_cut_relevant_contribution_u_channel_propagators_expanded} are inelastic and therefore are zero at the leading order in $\mu$ (i.e. when they are evaluated on-shell). For example, if we expand $M^{(0)}_{ae \to c f}(p, k, q, r)$ around $\mu=0$ we obtain
$$
M^{(0)}_{ae \to c f}(p, k, q, r)=M^{(0)}_{ae \to c f}(p, k, p, k) + O(\mu)= 0 + O(\mu).
$$
The same argument applies to all the tree-level amplitudes appearing in~\eqref{single_cut_relevant_contribution_u_channel_propagators_expanded} and the expansion of~\eqref{single_cut_relevant_contribution_u_channel_propagators_expanded} takes the form
\begin{equation}
\label{single_cut_relevant_contribution_u_channel_propagators_expanded_leading_order_in_mu}
\frac{i}{16\pi}  \int \frac{d\theta_k}{k \cdot x} \ O(\mu^2)
-\frac{i}{32 \pi} \int \frac{d\theta_k \mu^2}{(k \cdot x)^2} \ O(\mu^2).
\end{equation}
Since 
$$
k \cdot x =2 \mu \ m \cosh{(\theta_x - \theta_k)}
$$ 
is of order $\mu$, the first contribution in~\eqref{single_cut_relevant_contribution_u_channel_propagators_expanded_leading_order_in_mu} is of order $\mu$ while the second contribution is of order $\mu^2$. Therefore, both contributions are zero in the limit $\mu \to 0$.

If we consider the more degenerate situation in which $a=c$ and $b \ne d$, for $e=f$ \eqref{single_cut_relevant_contribution_u_channel_propagators_expanded} becomes
\begin{equation}
\label{single_cut_relevant_contribution_u_channel_propagators_expanded_degenerate_case}
\begin{split}
\frac{i}{16\pi}  \int \frac{d\theta_k}{k \cdot x} \ \Bigl( &-M^{(0)}_{ae \to a e}(p, k, q, r) \ M^{(0)}_{b e \to de }(p', r, q', k)\\
&+M^{(0)}_{ae \to ae}(p, r', q, k) \  M^{(0)}_{b e \to de }(p', k, q', r') \Bigr)\\
-\frac{i}{32 \pi} \int \frac{d\theta_k \mu^2}{(k \cdot x)^2} \ \Bigl( &+M^{(0)}_{ae \to a e}(p, k, q, r) \ M^{(0)}_{b e \to de }(p', r, q', k)\\
&+M^{(0)}_{ae \to a e}(p, r', q, k) \  M^{(0)}_{b e \to de }(p', k, q', r') \Bigr).
\end{split}
\end{equation}
and contains certain tree-level elastic amplitudes. In this case, due to property~\ref{Non_degenerate_mass_condition}, $M^{(0)}_{b e \to de }$ is zero at all orders in $\mu$ and~\eqref{single_cut_relevant_contribution_u_channel_propagators_expanded_degenerate_case} is null.
The discussion is analogous if we consider the case $\{b=d, a\ne c\}$.

We evaluated the two collections of diagrams in figure~\ref{relevant_terms_in_single_cut_t_channel_diagram} taking a particular \textit{off-shell} limit of the type discussed in section~\ref{section_to_explain_single_cut_contributions}; we could be interested in the sum of the two pictures in figure~\ref{relevant_terms_in_single_cut_t_channel_diagram} when we consider an \textit{on-shell} limit as the one considered in section~\ref{section_on_properties_of_tree_level_inelastic_amplitudes}. In this case, we should choose different momenta $\tilde{k} \ne k$ for the two particles associated with the cut propagator and perform the expansion of~\eqref{single_cut_relevant_contribution_u_channel_propagators_expanded} in the limit $\tilde{k} \to k$. 
Even in this case, it is easy to prove that the result is zero.
Indeed, despite the choice of variable we use for the expansion, the propagators in figure~\ref{relevant_terms_in_single_cut_t_channel_diagram} contribute with poles of order one in that variable. In contrast, each blob contributes with a zero of order one. The total order of the zeros is therefore always higher than the order of the poles and the contribution in figure~\ref{relevant_terms_in_single_cut_t_channel_diagram} is null. 

We conclude that in the situation in which $m_a=m_c$ and $m_b=m_d$, if the process~\eqref{generic_scattering_process_ab_into_cd} is inelastic, diagrams contributing to figure~\ref{relevant_terms_in_single_cut_t_channel_diagram} are well defined and generate the same result both performing an on-shell and an off-shell limit. In both cases, the result is zero. This is in agreement with our assumption that $\hat{R}_{abe \to cde}$ is the same in equations~\eqref{on_shell_limit_of_a_3_to_3_non_elastic_amplitude_infinities_removed} and~\eqref{off_shell_limit_of_a_3_to_3_non_elastic_amplitude}. 
While this conclusion is correct for all the two-to-two inelastic processes, it turns out to be false in the case $c=a$ and $b=d$ (i.e. when the process is elastic). In this case, any time $e=f$ formula~\eqref{single_cut_relevant_contribution_u_channel_propagators_expanded} generates a nontrivial contribution to the amplitude. While it is immediate to check that this contribution is finite, it is nontrivial to prove that it does not depend on the rapidity $\theta_x$ (i.e. on the direction we follow when we send $\mu \to 0$). A detailed study of these contributions and how they differ when we move to the elastic scattering configuration by performing an on-shell and an off-shell limit should highlight a general procedure to obtain one-loop elastic amplitudes from tree-level amplitudes in bosonic models with classically integrable Lagrangians of the form~\eqref{eq0_1}.

\end{document}